\documentclass[preprint,10pt]{aastex}

\usepackage{graphicx}
\usepackage{natbib}
\usepackage{txfonts}
\usepackage{dynamics}

\def\ggtau{GG~Tau}
\def\ggtaua{GG~Tau~A}

\begin{document}
\title{Dynamics of Circumstellar Disks~III: The case of \ggtaua}

\author{Andrew F. Nelson}
\affil{XCP-2, Mailstop T082, Los Alamos National Laboratory, Los
Alamos NM, USA, 87545}
\email{andy.nelson@lanl.gov}

\author{F. Marzari}
\affil{Universit\`a di Padova, Dipartimento di Fisica, via Marzolo 8, 35131
Padova, Italia}
\email{francesco.marzari@pd.infn.it}

\begin{abstract}

We present 2-dimensional hydrodynamic simulations using the Smoothed
Particle Hydrodynamic (SPH) code, VINE, to model a self-gravitating
binary system. We model configurations in which a circumbinary
torus$+$disk surrounds a pair of stars in orbit around each other and
a circumstellar disk surrounds each star, similar to that observed for
the \ggtaua\ system. We assume that the disks cool as blackbodies,
using rates determined independently at each location in the disk
by the time dependent temperature of the photosphere there. We assume
heating due to hydrodynamical processes and due to radiation from the
two stars, using rates approximated from a measure of the radiation
intercepted by the disk at its photosphere. We simulate a suite of
systems configured with semi-major axes of either $a=62$~AU (`wide')
or $a=32$~AU (`close'), and with assumed orbital eccentricity of
either $e=0$ or $e=0.3$. Each simulation follows the evolution for
$\sim$6500-7500~yr, corresponding to about three orbits of the torus
around the center of mass. Our simulations show that strong, sharply
defined spiral structures are generated from the stirring action of
the binary and that, in some cases, these structures fragment into 1-2
massive clumps. The torus quickly fragments into several dozen such
fragments in configurations in which either the binary is replaced by a
single star of equal mass, or radiative heating is neglected. The
spiral structures extend inwards to the circumstellar environment as
large scale material streams for which most material is found on
trajectories which return it to the torus on time scale of 1-200~yr,
with only a small fraction accreting into the circumstellar
environment. The spiral structures also propagate outwards through the
torus, generating net outwards mass flow and eventually losing
coherence at large distances from the stars. The torus becomes
significantly eccentric in shape over most of its evolution. In all
configurations, accretion onto the stars occurs at a steady rate of a
few $\times10^{-8}$\msun/yr, with the net result that, without
replenishment, the disk lifetimes would be shorter than $\sim10^4$~yr.
Our simulations show that only wide orbit configurations are able to
retain circumstellar disks, by virtue of accretion driven from the
robust material streams generated in wide configurations, which are
very weak in close configurations. In wide, eccentric orbit
configurations, accretion is episodic and occurs preferentially onto
the secondary, with rates strongly peaked near binary periapse. Based
on our results, we conclude that the \ggtaua\ torus is strongly self
gravitating and that a major contribution to its thermal energy input
is the shock dissipation associated with spiral structures generated
both by self gravitating disturbances and by the stirring action of the
binary. We interpret the sharply defined features observed in the torus
as manifestations of such spiral structures. We interpret the low
density disk surrounding it as an excretion disk created by the outward
mass flux generated by the spiral arms as they propagate outwards.
Typical eccentricities calculated for the shape of the tori modeled in
our simulations are large enough to account for the supposed
$\sim20$\degr\ mutual inclination between the stellar orbit plane of
\ggtaua\ and its surrounding torus, through a degeneracy between the
interpretation of inclination of the torus and its eccentricity. We
therefore interpret the observations in favor of a coplanar system with
an eccentric torus. Because accretion onto the disks occurs at rates
sufficient to sustain them only in wide orbit configurations, we
conclude that the gas currently resident in the circumstellar disks of
the \ggtaua\ system has been accreted from the torus within the past
few thousand years. Although circumstellar disks will persist over time
spans long enough to permit planet formation, the overall environment
remains unfavorable due to high temperatures and other conditions.
Given, the presence of circumstellar disks, robust accretion streams,
and our interpretation of the \ggtaua\ stellar orbit plane as coplanar
with the torus surrounding it, we conclude that the \ggtaua\ system is
in an eccentric, $a\sim62$~AU orbit, resolving questions in the
literature regarding its orbit parameters.

\end{abstract}

\keywords{Stars:Formation, Stars:Binary, Accretion Disks, Numerical
Simulations, Hydrodynamics}

\section{Introduction}\label{sec:intro}

In the early stages of the formation of a star \citep[see, e.g., the
review paper of][]{SAL87}, a cloud of gas and dust collapses and forms
a protostar with a disk surrounding it.  Later on, while the accretion
from the cloud continues, the star/disk system also begins to eject
matter into outflows whose strength varies in time. Finally, accretion
and outflow cease and over the next few million years the star loses
its disk and evolves onto the main sequence. A major refinement of
this paradigm has been an exploration of the parameter space for which
multiple objects form from a single collapse \citep[see
e.g.][]{PP4_BBKB,MONAL}. The dynamical evolution of multiple systems,
once they are formed, remains insufficiently well understood however.

Observations \citep{PP4_MGJS} have shown that in fact more stars in
low mass star forming regions (e.g. the Taurus star forming region)
are found in binary or higher order multiple systems than is the case
for older systems like the field stars in the solar neighborhood.
Further, nearly all of these stars show evidence for circumstellar
disks. The existence of such multiple stars in forming systems
suggests that the evolution of circumstellar material in orbit around
or infalling onto those stars will be significantly altered from the
evolution in a single star$+$disk system.

Several observed binary star$+$disk systems do in fact show evidence
of mutual interactions of various strengths
\citep[e.g.][]{UY-Aur,L1551_Rod}. In the L1551~IRS~5 system for
example, in which the observed angular separation on the sky
corresponds to a $\sim$50~AU physical separation, disks around each
component of the binary are truncated at about 10~AU, a size which is
far smaller than is typically observed in single systems or in
binaries with larger orbital separations (e.g. that observed in the
HK~Tau system; \citet{HKtau-c}). In addition to disk truncation, tidal
interactions due to the companion star can significantly alter the
evolution of the circumstellar disks by generating large scale shocks
and other internal heating. These factors act to accelerate mass
transport through the disk, significantly reducing its lifetime unless
that mass is replenished from a circumbinary disk on a similar
timescale. Such heating and transport may negatively influence the
formation of planets in binary systems, as described in \citet{N00}.

The situation becomes significantly more complex when the binary
system is also surrounded by a circumbinary torus and/or disk. This
additional player may exchange mass and angular momentum with each
star and its disk, affecting both the binary orbit and the mass
evolution of the circumstellar disks. For example, their dispersal due
to accretion onto the star, enhanced by the binary perturbations, may
be slowed by continued replenishment from circumbinary material.
If this reservoir is massive enough, the feeding mechanism may extend
the lifetime of the inner disks so that their dissipation timescale is
not significantly different from to single stars (\citet{PRASI}, see
also \citet{MONAL} for a review). Evidence for such replenishment has
recently been observed with using high resolution interferometry on
ALMA \citep{dutrey14}.

\subsection{The \ggtau\ system}\label{sec:ggtau}

The total number of observed circumbinary disks is not large. Of
those, most are found in binaries with a small separations, such as
GW~Ori, DQ~Tau, AK~Sco, V4046~Sgr \citep{MONAL}. Circumbinary disks in
wider binaries have been imaged in only a few cases, such as
L1551~IRS~5, Sr~24~N, UY~Aur and \ggtau. \ggtau\ is particularly
interesting because of the wealth of observational data available for
it, which can be compared to theoretical models.   

\ggtau\ was first determined to be a multiple system in the early
study of \citet{CoKu79}. Later observations of \citet{leinert91},
using speckle interferometry, confirmed this observation and set the
projected separation between the components to be $\sim10$\arcsec. The
same observations also resolved both the primary and secondary
components into a binaries. Imaging by \citet{Whiteetal99}, refined
the distance measurements, setting projected separations of
0.25\arcsec\ and 1.48\arcsec\  between the components of the A and B
binaries respectively, and 10.1\arcsec\ between the hierarchical
binary pairs. At an assumed 140~pc distance of Taurus, these
separations correspond to distances of $\sim35$~AU and $\sim207$~AU,
for the binaries, and a distance of $\sim1400$~AU, for the
hierarchical pairs.

The much more massive `primary' component of the system, \ggtaua, has
received extensive attention since its discovery, and a progressively
more detailed picture of the system has emerged. Measurements have
been made of the stellar masses, luminosities and spectral types,
their age, accretion rates, their relative motions and orbit
characteristics, the mass and mass distribution of the circumbinary
ring and disk and its rotation curve, and estimates of the
circumstellar disk configurations and substructure \citep[an
incomplete list of references
includes][]{sg92,dutrey94,roddier96,gds99,Whiteetal99,silber00,itoh02,krist02,tamazian02,duchene04,krist05,pietu11,beck12,dutrey14,difolco14}. 

Among the most detailed models of the various components of the
\ggtaua\ system is that described in \citet[][hereafter
GDS99]{gds99}, which we summarize here. In their model, the
circumbinary material is composed of two components, consisting of a
ring or torus and a disk with a combined mass of $\sim0.12$\msun, of
which some 70\% of the total is contained in the torus. Both the torus'
inner and outer edges are quite sharp, with edges $\lesssim10$~AU in
width, and are located at distances of 180 and 260~AU from the
apparent system center, respectively. Assuming that the torus is
circular, its inclination is determined from the observed projected
image to be 37 degrees. The disk component is fit to a rather steep
power law in density with exponent $s=2.75$, and a similarly steep
power law in temperature with exponent $q=0.9$.

The disk's outer edge is not as well resolved, but appears to extend
to approximately 800~AU, a distance consistent with tidal truncation
due to the B components of the combined quadruple system. Kinematic
measurements of the circumbinary material constrain the total mass of
the A components to be 1.28$\pm$0.07\msun. Unresolved emission
originating from the circumstellar environment sets a lower limit on
the mass of such material to be $\sim0.8\times10^{-4}$\msun. 

The wealth of observations of \ggtau\  is not without inconsistencies,
based both in the data themselves, and in attempts to fit theoretical
models to it. For example, while the combined mass of the `A'
components has been determined (as noted above) to be
$\sim1.28$~\msun, the masses of the two stars have been estimated from
their spectral signatures to be $\sim0.78$\msun\  and $\sim0.68$\msun\
for the primary and secondary, respectively, based on spectroscopic
measurements and fits to evolutionary tracks \citep{Whiteetal99}. Much
of the question regarding the system masses may be resolved with the
recent discovery of \citet{difolco14}, that the secondary, GG~Tau~Ab,
is itself a close binary with separation of $\sim4.5$~AU and masses
estimated to be 0.38 $M_{\odot}$ amd 0.25--0.35 $M_{\odot}$,
respectively.

Similarly, the orbital parameters of the stars have been established
from fits to their relative motions, and are estimated to define an
orbit with semi-major axis $a\sim32.4$~AU and eccentricity $e=0.34$
\citep{BEDU06}. These parameters appear to be inconsistent with the
expectations of theoretical models which predict a semi-major axis
$>50$~AU \citep{ArtLu94} based on the positions of various orbital
resonances and the $\sim$180~AU inner diameter of the circumbinary
disk. \citet{BEDU06} explored this discrepancy in an analysis of their
results, noting that with a slightly increased error bar on the
system's observed inclination, an alternate fit to the data yield a
semimajor axis of 62~AU and an eccentricity of 0.35. This orbit is
consistent with the dynamical constraints implied by the inner disk
edge location and with the observed projected separation of
$\sim36$~AU \citep[see, e.g.][]{difolco14}.

\subsection{Goals of this work}\label{sec:goals}

Our goals in this work are to understand the dynamical influences that
the disks and stars in a forming multiple star system have on each
other, what their observational consequences are and to make
inferences regarding the physical characteristics and behavior of the
system based on the outcomes of numerical simulations. To those ends,
we will present a series of two dimensional numerical simulations
configured to model systems like \ggtaua, using the numerical code
VINE. The \ggtaua\ system is interesting because of the depth and
quality of the observations made of it, which permit it to serve as
representative of a larger class of objects in which there are
significant dynamical interactions between the various components. Our
strategy will be largely qualitative in nature and will focus on
comparisons between the observations and simulations modeling a range
of plausible physical conditions for the system. Comparisons of
features seen in the simulations with those in the observations will
then permit us to rule out or confirm various physical properties of
the system, by virtue of being consistent (or not) with those
observations, and to assign their origin to specific physical
phenomena active in the system. 

The presence and characteristics of the binary will have influences on
the circumstellar and circumbinary materials at all scales. As noted
above however, the orbital parameters of the \ggtaua\ binary are not
yet uniquely defined by observations, so the characteristics of the
influence are likewise not uniquely determined. Therefore, a logical
starting point for our explorations will be to study the spectrum of
outcomes that different orbital parameters may produce in the behavior
of the gaseous material in the system. With this in mind, our strategy
will be to run a suite of simulations in which each member consists of
the same circumbinary initial condition and physical model, but stars
with different orbit parameters. Each member of the suite will then
define a unique numerical experiment to display the consequences that
a particular set of binary orbit parameters have for the system
behavior, and which characteristics are common to all members of the
suite. Given the simulation results, we can make assessments in terms
of comparisons of the appearance of the various gaseous components to
the observations. In particular, we will address three broad
categories of questions appropriate for the evolution at different
spatial scales. First, what are the influences of the binary on the
large scale structure of the circumbinary material and what structures
develop from independently of the binary? Second, what are the
influences of the binary on the circumstellar material and the
material `in transition' between the circumbinary and circumstellar
regions? Third, what are the likely orbit parameters of the binary?

We begin in section \ref{sec:binphys} with specifications of the
initial conditions, the physical models and the numerical methods used
to simulate their evolution. Then, in section \ref{sec:evo-circumbin},
we introduce the suite of models that form the basis of our study and
describe the evolution of each of the systems at large scales. In
section \ref{sec:evo-circumstar}, we revisit many of these same
simulations, but turn our focus towards the smaller scale
characteristics of the circumstellar environment. We will find that
many of the results described in these two sections are relevant to
more than one of the questions just noted. We therefore limit our
initial discussion to the various characteristics of the simulations
themselves, with limited reference to particular observations. Then,
in section \ref{sec:compare}, we compare our results to the observed
\ggtaua\ system. There, we focus first on assessing their consistency
with the observations in order to rule out less likely configurations,
and then on assessing the results for their physical significance,
i.e., what the models may reveal about the physical state and the
evolution of the system. Finally, in \ref{sec:discussion}, we
summarize our results, comparing them to previous work and discussing
a number of questions which may be of interest to address in the
future.

\subsection{An overview of our numerical
experiments}\label{sec:overview}

As a guide to the reader, we briefly introduce the various numerical
experiments and the goals we set for them. We introduce the main suite
of simulations and begin our characterization of the evolution at
large spatial scales in sections \ref{sec:eccentric} and
\ref{sec:circular}. A similar introduction to many of the same
simulations appears in section \ref{sec:circstar-conf}, where we
concentrate on the smaller scale, circumstellar evolution. These
sections will focus on exploring the morphology of large and small
scale structures developing under the tidal force of the central
stellar system.

At large scales, the presence of a binary affects the evolution of the
torus by continuously stirring the material, generating morphological
features in response, and by heating it. Also, in the specific example
of the \ggtaua\ torus, where the observed temperature and mass imply
that it is susceptible to gravitational instabilities, we expect that
features due to such instabilities to develop as well. As noted above,
\ggtaua\ is known to harbor assymmetries and other features within its
torus. Given these features, our first analyses will be qualitative:
are any similar features identifiable in our simulations? We address
this question in section \ref{sec:torus-features}. 

After finding qualitative correspondence between features in the
observed and simulated systems, we turn to investigations to couple
the physical processes that are active in the system with the features
they are responsible for producing. The most striking feature of the
\ggtaua\ circumbinary material is that it is composed of two very
distinct components: a torus and a disk, with a sharp boundary
distinguishing them from each other. In section \ref{sec:transport},
we test the possibility that these two components actually have a
common origin as a single primordial torus, out of which the disk
forms as an `excretion' feature generated by the outward mass
transport driven by the propagation of spiral arms. We run simulations
initially configured to include only a torus, in order to compare the
behavior of such systems to the torus$+$disk systems already studied,
focusing specifically on the relative efficiency of outward mass
transport and the resulting circumbinary mass distribution.

We continue with a study to determine the relative importance of
various sources of heating in the torus. This question is important
because of its bearing on the susceptibility of the torus to fragment.
Depending on its characteristics, if indeed any fragmenation occurs,
such fragmentation may lead to the formation of additional planetary,
brown dwarf or stellar components of the system. In section
\ref{sec:heatoff}, we describe the behavior of two simulations in
which we suppress, respectively, the radiative heating from the stars
and the internal heating due to the dissipation of shocks generated
both by the tidal action of the binary and self gravitating
instabilities. We will expect changes in the behavior of the evolution
if in fact either of these heating sources are important.

The physical significance of our results depends critically on having
both an accurate inventory of the important physical processes and an
accurate numerical realization of the evolution. When fragmentation
driven by gravitational instabilities plays a role, accurate numerical
simulations require that certain minimum resolution conditions be
observed over the lifetime of the simulation. We assess the physical
and numerical significance of the fragmentation behavior seen in our
simulations in \ref{sec:fragmentation}. Other properties of our
simulations may also be affected by numerical issues, with one notable
case being that they provide only very coarse resolution of the
circumstellar disks. An important consequence may be that numerical
dissipation there is unphysically high, leading to unphysically high
accretion rates onto the stars. As a component of our investigation of
the accretion in section \ref{sec:masstrans}, we describe simulations
run with two very different particle counts, thereby resolving the
circumstellar accretion disks to different extents, in order to test
whether or not this numerical issue affects our results.

Our suite of simulations cover a parameter space within which we
expect the actual \ggtaua\ system to reside, though its exact state is
currently uncertain. In order to set more stringent constraints, in
section \ref{sec:quantitative} we will make a number of quantitative
comparisons between our simulations and the observations. In
particular, we fit ellipses to the torus to determine its size and
shape over time and thereby infer other properties of the system, such
as the binary orbit parameters and the system's inclination.

Other system constraints can be derived by considering the influence
that the binary orbit has on the characteristics of both the
circumstellar disks and material in the gap between the circumstellar
disks and the torus. In section \ref{sec:masstrans}, we calculate the
amount of mass transferred from the circumbinary torus onto the
circumstellar disks in our simulations, to test how much this mass
flux depends on the orbit parameters of the binary. We also evaluate
the rates at which mass accretes from the circumstellar disks onto the
stars in different system configurations. The net rate of accretion
into and out of the circumstellar disks determines their lifetimes.
Given long enough lifetimes, planet formation may be enhanced. We
discuss this possibility in section \ref{sec:plan-form}.

\section{Physical Assumptions and Numerical Model}\label{sec:binphys}

The initial conditions for the simulations performed in this work are
based on the observed configuration of the \ggtaua\ system, as
discussed above, except that we have not incorporated the newly
discovered binary nature of \ggtaua b into our models. In order to
investigate various aspects of the evolution, we vary the initial
conditions over the plausible parameter space outlined by both
observational data and theoretical analyses, exploring uncertainties
in that data. The details of the physical models and the methods used
to set up the initial configurations are similar to those used in
\citet[][hereafter \dynone\ and \dyntwo, respectively]{DynI,DynII} and
we refer the reader to those works for a more complete exposition. In
this section, we shall summarize the models described previously, and
describe the details of the set up specific to the present study here.

\subsection{The Numerical Method}\label{sec:numerics}

We use the Smoothed Particle Hydrodynamics (SPH) code `VINE'
\citep{vineI,vineII}, to evolve our simulations forward in time. SPH
is a particle based, Lagrangian method for solving the hydrodynamic
equations, and so is very well suited to the complex geometry present
in the multi-component systems we study. Particles naturally follow
the flow, concentrating resolution specifically in regions where mass
itself concentrates, and removing it and the accompanying
computational cost, from uninteresting regions where little material
exists. This property permits the evolution of both the circumbinary
torus/disk and the circumstellar disks to be modeling at the same time
in the same simulation, despite their very large difference in spatial
scales. It also permits detailed studies of the the mass and momentum
exchange among the different components of the system, such as between
the circumbinary material and the circumstellar disks. 

\subsection{Initial Conditions}\label{sec:bininit}

We consider a simplified model of the \ggtau\ system in which the `B'
component of the system is not included. This approximation will be
reasonably accurate unless the orbit of \ggtau~B has a pericenter
smaller than $\sim$800~AU. Even when such conditions exist, we expect
it to affect only the outer portions of the circumbinary disk
\citep{BEDU06}, at distances well beyond the regions of interest for
our investigations. We further simplify our simulations by neglecting
the recent observation of \citet{difolco14} that the GG~Tau~Ab
component is also a multiple. Instead, we model the \ggtaua\ system as
two stars, each harboring a circumstellar disk, and a distribution of
circumbinary material, as described by GDS99 as a combined torus/disk
configuration in orbit around the binary. Following this description,
we configure the initial conditions for each of the three components
of the system independently, then combine them together to create a
synthesized, complete initial condition for each simulation. We
discuss the set up of each component in turn.

\subsubsection{Circumbinary Material}\label{sec:circumbin}

\begin{figure*}
\includegraphics[angle=0,scale=0.75]{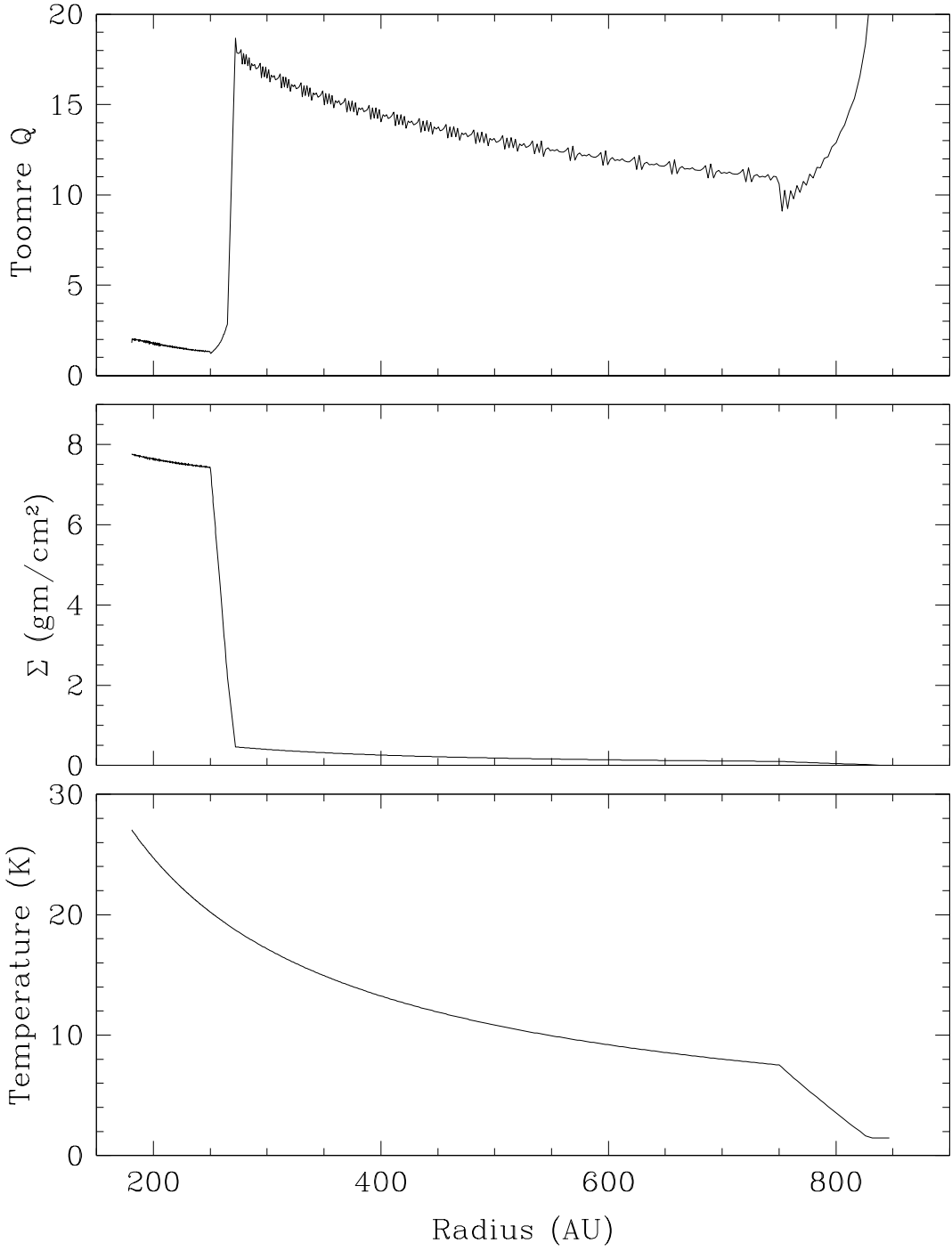}
\caption{\label{fig:circumbin-IC}
The initial condition profile in the circumbinary torus. The top panel
shows the local value of the Toomre $Q$ parameter. The middle panel
shows the surface density and the bottom panel shows the temperature.}
\end{figure*}

We model the mass distribution of the circumbinary material as a
torus$+$disk combination. We set the dimensions of the combination
such that its inner and outer edges each define a circle 180~AU and
800~AU in diameter, respectively. The outer boundary of the torus is
set to 260~AU. The center of mass of the circumbinary material is set
to coincide with the system barycenter. The initial conditions of the
circumbinary material are summarized graphically in figure
\ref{fig:circumbin-IC}.

SPH particles are initially laid out on a series of concentric rings
centered on the origin and extending from the inner torus/disk edge at
a radius of 180~AU outwards to a radius of 800~AU. The particle
distribution follows a surface density power law:
\begin{equation}\label{eq:bindenslaw}
\Sigma(r) = S_T \Sigma_T  + S_D \Sigma_0 \left[ 1 +
       \left({r\over r_c}\right)^2\right]^{-{p\over{2}}},
\end{equation}
where the values of $\Sigma_T$ and $\Sigma_D$ are adjusted to give the
desired mass ratio between the torus and disk masses. GDS99 fit a power
law to the volume density in the disk with a proportionality
$r^{-2.75}$. Assuming a constant ratio of scale height to radius
($H/r$), this translates to a surface density power law with exponent
$p=1.75$. This exponent is broadly consistent with that expected for
most circumstellar disk configurations and, for the sake of simplicity,
we choose the exponent $p=3/2$, used in our previous work for both the
circumbinary disk and for the circumstellar disks, discussed below.

The core radius $r_c$ for the power laws is set to $r_c$=1AU, a value
insignificant for the circumbinary material, but important for the
circumstellar disks described below. In the suite of simulations
described in this work, we explore both torus$+$disk configurations and
torus-only configurations. For the torus$+$disk models, we define the
torus mass such that it contains 70\% of the total, consistent with
GDS99. The two factors, $S_T$ and $S_D$, are linearly decreasing
functions near the torus or disk boundary, respectively. They are each
defined by:
\begin{equation}\label{eq:sdens-alter}
S = \cases{  1                          & for $r<R-\delta$ ;\cr
             1 - {{(r-(R - \delta))}\over{2\delta}}
                                        & for $R-\delta<r <R+\delta$;\cr
             0                          & for $r>R+\delta$ ;\cr }
\end{equation}
where $R$ plays the role of the defined torus or disk edge, for $S_T$
and $S_D$, respectively. With this definition, the outer torus and
disk edges are smoothed in the region within a distance $\delta$
inward and outward of their nominally defined radii.

We smooth both the outer edge of the disk and the torus/disk
interface. This smoothing both improves consistency with the observed
width of the interface, where data are available, and side-steps the
unphysically large pressure gradients that would otherwise perturb the
initial condition if the interface were perfectly sharp. Consistent
with the observed width as discussed in GDS99, we have chosen the
width of the smoothed torus/disk interface to be $\delta=10$~AU. No
observational constraints are readily available for the disk's outer
edge, and we choose a wider smoothing parameter of $\delta=50$~AU for
the outer edge of the disk. As a practical matter, the inner edge of
the circumbinary material is unsmoothed in order to permit a more
evenly distributed initial particle configuration throughout the
torus. In consequence, some small shear instabilities develop over the
course of the first few hundred years of the simulations. These features 
are quickly overwhelmed by the larger scale evolution of the system,
and so are not significant on any longer time scale.

We specify the initial thermodynamic state of the disk assuming that
the specific internal energy, or equivalently, temperature, in the
torus and disk follows a single power law profile, given according to:
\begin{equation}\label{eq:circumbin-temp}
u(r) = u_D\left[1 + \left({r\over r_c}\right)^2\right]^{-{q\over{2}}}.
\end{equation}
The exponent, is set to $q=0.9$, a value chosen to reproduce the
fitted value obtained by GDS99 and the value $u_D$ is chosen to
reproduce, in combination with the equation of state, the observed
temperature of $T\approx20$~K at a distance of 300~AU from the origin.
As for the surface density law, the core radius, $r_c$, is set to
$r_c$=1AU, which will not significantly alter the profile for the
circumbinary material, but will influence the inner edge of the
circumstellar disks defined below. At the outer edge of the disk, we
modify the temperature profile following the same smoothing procedure
as used for the surface density, such that it falls to a minimum of 3~K
at the outermost extent. 

Matter is set up on initially circular orbits around the origin
assuming rotational equilibrium. Radial velocities are set to zero.
For purposes of generating an initial state only, we replace the two
binary components with a single point mass at the origin. It's mass,
of 1.3\msun, is the sum of that assumed for the two stars used in the
actual simulations. Gravitational and pressure forces are balanced by
centrifugal forces by setting
\begin{equation}\label{eq:coolrotlaw}
\Omega^2(r) = { {GM_*\over{r^3}} + {1\over{r}}{
       {\partial\Psi_D}\over{\partial{r}}} + {1\over{r}}{
       {{\bf\nabla}{P}}\over{\Sigma} } },
\end{equation}
where $\Psi_D$ is the gravitational potential of the disk and the
other symbols have their usual meanings. The magnitudes of the
pressure and gravitational forces are small compared to the stellar
term, therefore the disk is nearly Keplerian in character.

The circumbinary material, at a mass of one tenth the combined stellar
masses, contains sufficient mass to be marginally self-gravitating.
The extent to which this is true is quantified by the well known,
Toomre $Q$ parameter defined as: \begin{equation}\label{eq:toomreq} Q
= {{\kappa c_s}\over{ \pi G \Sigma}}, \end{equation} where $\kappa$ is
the local epicyclic frequency and $c_s$ is the sound speed. Together,
the conditions specified above combine to determine the radial profile
for $Q$ shown in figure \ref{fig:circumbin-IC}.
Within the torus values range from $\sim1.5-2$, and we may therefore
expect significant signatures of self gravitating flows. Further out,
in the disk, values increase to $Q\sim12-18$, and we expect this
region will be largely insensitive to self gravitating disturbances.
Note that the $Q$ profile shown includes small amplitude
`ringing' oscillations, particularly in the disk, which are artifacts
of the numerical derivatives (of $\Omega$ with respect to radius) that
are used to determine the epicyclic frequency. They are of little
significance and may be disregarded.

\subsubsection{The stars and circumstellar
material}\label{sec:circumstellar}

The two stellar components are each independently defined as point
masses, and each is orbited by a circumstellar disk. The stars have
masses of $M_*=0.7$ or $M_*=0.6$\msun\ for the primary and secondary,
respectively. These values are consistent with the total system mass
determined from the rotation curve of the circumbinary material
(GDS99) and from orbital motions \citep{tamazian02}. It lies slightly
below the `best' values derived from their luminosities and fits to
evolutionary tracks \citep[see, e.g.,][where values of 0.78\msun\ and
0.68\msun\ are fit for the primary and secondary,
respectively]{Whiteetal99}, but remain within the error bars imposed
by those models. As discussed in section \ref{sec:energy-gen}, below,
we incorporate a simple model of the radiative heating on the disks,
supplied by the stars. Following \citep{roddier96}, we specify their
luminosities to be $L_p=1.0$\lsun\ and $L_s=0.76$\lsun, for primary
and secondary, respectively. Gravitational forces due to the stars are
softened, with a softening radius of $r_s=0.3$~AU and, as the systems
evolve, mass may be accreted onto the star if its trajectory takes it
inside this radius. The mass and momenta of accreted particles are
added to that of the star.

We assume that the mass of the disk around each star is
$M_D=10^{-3}M_*$, consistent with the observed lower limit on the mass
of $\gtrsim10^{-4}$\msun\ imposed by millimeter observations of the
circumstellar emission (GDS99). This assumption is also consistent
with the masses inferred from the more recent observations of
\citet{pietu11}, of $\sim1.5\times 10^{-3}$\msun. As for the
circumbinary material, we assume the mass distribution follows a
surface density power law using equation \ref{eq:bindenslaw}, with an
exponent of $p=3/2$. For the circumstellar disks, we omit the
contribution in equation \ref{eq:bindenslaw} due to the torus. We set
the spatial dimensions of the disks to be approximately consistent
with the maximum radius dynamically permitted by the orbital
parameters we assume for a given simulation. Specifically, for
simulations with $a=62$~AU, we assume the initial outer radius of each
disk to be 10~AU, and for simulations with $a=32$~AU, we set the outer
radius to be 4~AU. In each case, we smooth the outer edge, over a
range $\pm1$~AU of the nominal edge location, following the same
prescription used above for the circumbinary material.

No observations of sufficient detail exist, which may be used to
determine an initial thermal state for the circumstellar disks with
any precision. We are therefore free to choose conditions that provide
a simple and convenient starting point, from which the disks may
evolve as the various components of the system interact. Therefore, as
for the circumbinary disk, we set the initial specific internal
energy distribution to be a power law, identical in form to equation
\ref{eq:circumbin-temp}. Here again, we choose an exponent of $q=0.9$
for the power law, and a radial scale of $r_c=1$~AU. We set the
initial scale of the power law, $u_D$, to be such that the disks are
assumed to have initial minimum Toomre stability \qmin=1.5. 

In practice, and as one of us found in previous work on binary systems
\citep{N00}, we will find that the initial thermodynamic properties of
the disks matters very little, since they readjust both their mass and
temperature distributions within a few orbits around their primary, as
the system evolves. Similarly, the mass distribution also changes in
response to the evolution, though on a slower ($\sim 1000$~yr)
timescale.  

\subsubsection{Combining the Components}\label{sec:combined-system}

Once specified as isolated systems, the two star/disk systems defined
in the last section are combined to form a binary system, assuming a
Keplerian orbit around the system barycenter, defined to be at the
origin. For systems with an initial eccentricity, the initial
positions of the stars are such that the system is at apoapse, so that
the interactions between the components are at their weakest levels.
We then embed this combined system into the circumbinary environment
defined in section \ref{sec:circumbin}, replacing the single point
mass originally used to as a place-holder to configure the
circumbinary system in isolation. Once configured, the combined
system, including the orbital elements of the binary, is free to
evolve in response to gravitational forces between all of the
components.

\subsection{Thermal Energy in the System}\label{sec:energy}

The thermodynamic properties of the system are very similar to those
presented in \dyntwo, and interested readers should consult that work
for a detailed description. Briefly summarized, we assume that the
disk matter can be approximated by an ideal gas and that radiative
cooling from the disk surfaces is the only thermal energy sink.
Thermal energy is generated by internal sources (shocks and viscous
turbulence), implemented by means of an artificial viscosity term in
the numerical method. In addition, for this work, we also employ a
radiative heating model, with the two stars as the sources of radiant
heating, as described below.

\subsubsection{The Equation of State}\label{sec:eos}

The hydrodynamic equations are solved assuming a vertically integrated
gas pressure and a single component, ideal gas equation of state given
by:
\begin{equation}\label{eq:ideal-eos}
P=(\gamma - 1)\Sigma u
\end{equation}
where $\gamma$ is the ratio of specific heats, $P$ is the vertically
integrated pressure and $u$ is the specific internal energy of the
gas. The value of $\gamma$ assumed for these calculations is the same
as that in \dyntwo, of $\gamma=1.53$, for the hydrodynamical evolution
and $\gamma=1.42$, for the three dimensional calculation of the disk
vertical structure.

\subsubsection{Thermal Energy Dissipation}\label{sec:energy-diss}

As in \dyntwo, thermal energy in the disks is permitted to escape from
the system using a simple model of the disk's vertical structure to
determine a `photosphere temperature' based on local conditions at
each time step. The temperature is then used as an input into a
blackbody cooling rate. The blackbody temperature is derived for each
particle and at each time, using the surface density, specific
internal energy and a distance as input parameters for a calculation
of the temperature and density structure as a function of altitude,
$z$, above the midplane. This calculation assumes the disk is
instantaneously vertically adiabatic, plane parallel and self
gravitating. From the newly derived $(\rho(z),T(z))$ structure, we
derive the photosphere temperature by integrating the optical depth,
$\tau$, downwards from $z=\infty$ to the altitude at which
\begin{equation}\label{eq:tau-a}
\tau = 2/3 = \int_\infty^{z_{phot}}\rho(z)\kappa(\rho,T)dz.
\end{equation}
In our previous work, the opacities were derived from a combination of
the tabulations of \citet{pmc85} and \citet{AF94}. For this work, we
have updated our model to use the Rosseland mean opacities from the
work of \citet{semenov03}. In optically thin regions, for which
$\tau<2/3$ at the midplane, we assume the photosphere temperature is
that of the midplane. As in \dyntwo, we multiply the opacity by a
factor $R=0.005$ in regions where the midplane temperature is above
that of the grain destruction temperature, in order to crudely model
the effect of grain destruction and reformation on the opacity.

The present configuration is considerably more complex than that
described originally in \dyntwo, where the disk's vertical structure
was defined assuming only the gravitational influence of a single star
contributed to the vertical structure. In this work, of course, two
stars contribute, each with a different relative magnitude depending
on the distance of a given location from those stars. To simplify the
specification of the vertical structure, we define three distinct
regions, in which one or another limiting case holds and only one
point mass can be assumed to contribute the only significant external
contribution to the disk's vertical structure. For each of these
limiting cases, we create separate tabulations of the vertical
structure. The regions are, respectively, locations where we may
assume that the gravitational forces of either the primary or
secondary alone dominate the vertical structure, and locations where
the combined effect of both stars contribute to the vertical
structure. We assume arbitrarily that all material at distances
greater than 150~AU from the system center of mass falls into the
latter case. For purposes of defining the disk's vertical structure in
this region, the mass used to define the gravitational influence of
the two stars is set to the sum of the masses of both stars, and its
location is set artificially to the system center of mass. For
material inside this 150~AU limit, we determine the distance to each
star and define the material as `belonging' to the closer of the two.

Given the blackbody temperature, each SPH particle loses thermal
energy at a rate defined by
\begin{equation}\label{eq:dudt-radcool}
{{du_i}\over{dt}} = {{ -2\sigma_R T_{eff}^4 }\over{\Sigma_i} }
\end{equation}
where $\sigma_R$ is the Stefan-Boltzmann constant, $u_i$ and
$\Sigma_i$ are the specific internal energy and surface density of
particle $i$ and $T_{eff}$ is its photosphere temperature. The surface
density $\Sigma_i$ serves the dual purposes of first, relating the
blackbody flux (quantified as an energy flux `per unit area') to the
specific internal energy (quantified as an energy `per unit mass')
used in our simulations and, second, of disambiguating the `surface
area' of a single SPH particle, which would otherwise suffer from
confusion because of the overlap of SPH particles with each other.
The factor of two accounts for the two surfaces of the disk. In
regions where $\tau<1$ we multiply the derived flux by the optical
depth, $\tau$, in order to more correctly model the cooling of
optically thin regions.

\subsubsection{Thermal Energy Generation}\label{sec:energy-gen}

Thermal energy in the disk is generated from bulk mechanical energy
via both reversible and irreversible processes during the course of
the system's evolution. Thermal energy is generated reversibly as
spiral arms and similar dynamical structures become compressed and
$PdV$ work is done on the gas. Thermal energy is generated
irreversibly in viscous processes and shocks. Both of the latter
processes are modeled through the use of an artificial viscosity, as
is common in many implementations of hydrodynamic codes. Two time and
space dependent artificial viscosity terms are included. As discussed
in \dyntwo, the first is proportional to $\nabla\cdot{\bf v}$ (the
`bulk viscosity') and corresponds to unresolved viscous turbulence in
the system. The second is proportional to $|\nabla\cdot{\bf v}|^2$
(the `von~Neumann-Richtmyer' viscosity) and corresponds to shock
dissipation.

Because of the time and space dependence, an exact equivalance
between the artificial viscosity and \citet{SS73} `$\alpha$' models
does not exist. However, a correspondence can be drawn between the
bulk viscous term in the code and the local value of the
\citealt{SS73} $\alpha$ \citep{Murray96}, which we can use to
calibrate the magnitude of internal disspation in our simulations in
comparison to the model. In 2D, the correspondence can be expressed
for particle $j$ as 
\begin{equation}\label{eq:Balsara}
\alpha^j_{SS} = {{f_j \bar\alpha_j h_j\Omega_j}\over{8c_j}},
\end{equation}
where $f_j$ is the coefficient to reduce numerical shear viscosity for
SPH simulations defined by \citet{B95}, $\bar\alpha$ is the artificial
bulk viscosity coefficient, $\Omega_j$ is the orbit frequency, 
$c_j$ the sound speed and $h_j$ the smoothing length of particle $j$.
Averaged over many particles in some suitable radial range, equation
\ref{eq:Balsara} provides a useful estimate of the magnitude of the
artificial viscosity.

The calculations of \dyntwo\ showed that the artificial viscosity
produced a dissipation comparable in magnitude to an $\alpha$ equal to
a few $\times 10^{-3}$, but that the dissipation was resolution
dependent so that the physical dissipation of the real system was
probably lower than this value. The circumbinary torus/disk
configuration used in this work is both different than the
circumstellar disks in \dyntwo and is resolved with far more
particles, so we do not expect dissipation at comparable levels to
that work. Indeed, using the relationship quoted in \citet{Murray96},
we find energy dissipation rates in typical regions of the torus to be
an order of magnitude or more smaller, comparable to models with
$\alpha\sim 10^{-4}$ or less. In the case of the circumstellar disks,
which are configured similarly to \citet{N00}, we estimate the energy
generation rates to correspond to those in $\alpha$ models where
$\alpha\sim10^{-2}$.

It is important to note in these estimates that while the magnitude of
the dissipation in our simulations does correspond to that in a
\citet{SS73} model with a given parameter, the physics as realized in
the simulation does not corresponds directly to the physics underlying
such models. Specifically, the turbulent eddies out of which the
dissipation is assumed to arise are absent in our simulations, because
we model only two dimensions. Rather, the presence of such dissipation
is effectively an assumption that such turbulent eddies are present
and are the only source of internal energy generation in the fluid. As
discussed in \dyntwo, this includes the assumption that the flow is
reasonably smooth, and that strong shocks are not present.

We shall find this assumption is violated in the work presented here,
as it was in our earlier work on binary systems \citep{N00}. Strong
interactions between the binary components produce large amplitude
spiral structures in the circumstellar disks, including strong shocks
that inevitably accompany such structures. The circumbinary material
also produces large amplitude spiral structures and shocks generated
by the stirring motion of the binary itself. In both cases, additional
heating of the disk matter occurs. In comparison to the bulk
viscosity, these heating events generate internal energy at several
times the smooth `background' rate in the torus, though only in a much
smaller fraction of the total volume, where the spiral structures
themselves are present. 

In early simulations, performed only with internal heating sources, we
found that temperatures in the circumbinary torus and disk fell to
values well below those observed for the \ggtaua\ system and, in the
outer regions of the disk, even to the lower limit of 3~K we have
implemented in the code itself. The rate of thermal energy generation
due to internal processes is less than is required to balance the
radiative cooling in the circumbinary material. We conclude that a
correct model of the system requires that additional, external heating
sources be included in the inventory of physical processes describing
the system.

Although insufficient by themselves to heat the disk and balance
radiative cooling losses, we will show below (e.g. section
\ref{sec:heatoff}) that heating from dynamical processes does still
provide a significant source of energy input to the system. Therefore
a passive radiative heating model, such as described by \citet{CG97},
will also not be an appropriate ingredient to an accurate physical
model of the system. We have therefore developed a simple model for
the radiative heating of the disks by the stars, using a prescription
which is similar in spirit to the radiative cooling model described
above. Specifically, we assume that each star radiates isotropically
at a constant luminosity, and that all of the stellar radiation
impinging on the two disk surfaces (i.e. both above and below the disk
midplane) is absorbed. We estimate that energy absorption rate as
follows. By definition, a star's luminosity is the energy flux that it
emits per unit time, with its value per unit area of any spherical
volume centered on that star being $L_*/4\pi r^2$. This flux falls on
the two surfaces of the disk, each oriented at an angle to the
radially outward directed stellar flux. The flux absorbed per unit
area by the disk's surface will be diluted from the stellar flux due
to the mismatch between the normal directions of the stellar flux and
the disk surface. In order to quantify the energy absorbed correctly,
we must determine the size relationship between a surface element
oriented radially and one oriented normal to the disk surface.

Following GDS99, we assume that stellar flux impinges on the disk
horizontally and that the disk surface's normal vector is tilted at an
angle $\gamma$, as measured downwards from a vector pointed towards
the disk's north or south `pole'. Then, simple geometric arguments
provide the relationship we require: the flux per unit area of disk
surface will be diluted from the flux per unit area of stellar flux by
the dot product of the unit vectors defining their respective
orientations. That dot product reduces to the cosine of the angle
between them and, if we assume that the disk is not significantly
flared, it's value will be defined at every location by it's radial
distance from the star and the altitude of surface at which the disk
material becomes optically thick to stellar radiation:
\begin{equation}\label{eq:cosgamma}
\cos(\gamma) = {{H_{0}}\over{r}}.
\end{equation}
Then, the rate of energy absorption of the stellar flux at any given
location of the disk's surface will be given by the relation
\begin{equation}\label{eq:dudt-radheat}
{{du_i}\over{dt}} = { {2\cos(\gamma) L_* }\over{4\pi r^2 \Sigma_i} }. 
\end{equation}
The factor of two accounts for the fact that both the upper and lower
disk surfaces will be heated by the stellar flux. The surface density
term plays the same role in this equation that it does in equation
\ref{eq:dudt-radcool} above. Namely, to recast the energy flux per
unit area in terms of the specific internal energy (i.e. as a change
to the energy per unit mass), which is the conserved quantity employed
in VINE. 

The remaining unknown in equation \ref{eq:dudt-radheat} is the value
of $H_0$, through its presence in the cosine term in equation
\ref{eq:cosgamma}. In general, the value of $H_0$ will be different
than the altitude of the disk's photosphere however, because the
latter characterizes an optical thickness to its own internal
radiation, rather than to stellar photons. We therefore utilize a
measure of the disk's scale height, derived from the same vertical
structure calculation used to determine the cooling rates described in
in section \ref{sec:energy-diss}. 

As for the cooling calculations, and to simplify the model, we employ
the same three limiting regions as were used in the radiative cooling
model. At distances more than 150~AU from the system center of mass,
we the calculate vertical structure assuming a single star at the
origin whose mass is the sum of the two binary components. At closer
distances, the scale height is determined assuming that each SPH
particle belongs to the nearer of the two stars in the binary. 

\subsection{Limitations of the radiative heating and cooling models
relevant for our simulations}\label{sec:limits}

The radiative heating and cooling models used in this work are
deliberately simplified from the actual radiative transfer processes
present in the real systems. Here we explore the implications that
these simplifications make on our simulations, in order to provide
insights into the conclusions we may draw from them.

As noted above, we multiply the derived flux by the optical depth,
$\tau$, in regions where $\tau<1$, in order to more correctly model
the cooling of optically thin regions. The circumstellar disks and the
torus region of the circumbinary material are optically thick to their
own radiation throughout the calculation, so that in practice, this
modification has little effect. In the circumbinary disk region
however, the optical depth falls to near unity, making the disk only
marginally optically thick and endangering the assumptions made in
deriving the cooling rates there. Because the focus of our work is
mainly on the interactions between the stars, their circumstellar
material, and the torus, errors in the cooling rates at large
distances, and their consequences for the dynamical behavior, will be
comparatively small. 

For other regions, such as the gap between the circumstellar and
circumbinary material, material densities become even more tenuous,
with correspondingly smaller optical depths as well. In these regions,
the coupling of the material to the radiation becomes increasingly
tenous as well, and the radiative heating and cooling treatments used
in our work fail to provide accurate treatments of the energy flow.
Since we expect that the radiative energy fluxes into and out of the
material will be substantially diminished by the reduced coupling
efficiency, we simply disable both the radiative heating and cooling
entirely when such conditions are found in our simulations.
Specifically, when the surface density of any given SPH particle falls
to a value less than .05~gm/cm$^2$, we assume that no radiative
heating or cooling occurs for that particle.

As for radiative cooling, the accuracy of the radiative heating
suffers in the outer regions of the circumbinary disk. Because our
model assumes that thermal energy is deposited in an entire vertical
column of material, rather than only in a surface layer, we
overestimate the actual amount of heating there, since in fact most of
the column is shielded by material closer to the stars. Because we are
interested primarily in the dynamical behavior of the torus and
circumstellar disks where the effect is smaller, the over estimates
at larger radii will be of limited concern for interpreting our
results. 

\subsection{Relevant Physical Scales}

The dimensions of the \ggtaua\ system are typical of a number of
similar systems, as noted in section \ref{sec:ggtau}, so that our
models serve as prototypes for the behavior of that class. Of course,
the exact dimensions of each system vary and, with them, so do the
time scales important for their evolution. In order to provide a
calibration for our readers, we define here several relevant time and
distance scales for the initial conditions used in our simulations. 

Together with the masses of the binary components, the semi-major axis
assumed for the binary orbit defines its orbital period. The initial
conditions defined for our simulations include setups with semi-major
axes of $a=32$~AU and $a=62$~AU, which correspond to orbital periods
of 158~yr and 426~yr, respectively. The orbital period of material in
the torus provides another interesting timescale. Given its
initial inner boundary at 180~AU and the initial condition
specification above, we can estimate its orbit period to be $\sim
2100$~yr. These orbit radii, together with the inner torus edge,
define dimensionless ratios of $R_T/a\sim5.6$ for the 32~AU semi-major
axis, and $R_T/a\sim2.9$ for the 62~AU semi-major axis, where $R_T$ is
the torus' inner radius. 

We defined the initial configuration of the circumstellar disks to
have outer radii that are each a large fraction of the orbital
separation between the components making up the binary, consistent
with our expectation that such a configuration will be nearer to the
`asymptotic' long term configuration. As the simulations proceed, they
will grow or shrink according to the amount of accretion into the
circumstellar environment from the torus and the amount of accretion
out of the disks onto the star around which they orbit. In this
context, another relevant dimension for the system is the limiting
distance for stable orbits around the star to exist. \citet{HolWie99}
have calculated such criteria for a variety of system configurations
and, for the systems studied here, derive outermost stable radii of
$\sim10.5$ and $\sim15.5$~AU for the a$=$62~AU systems with
eccentricity of $e=0.3$ or $e=0$, respectively, and $\sim5.4$~AU and
$\sim8$~AU for the a$=$32~AU systems with $e=0.3$ and $e=0$,
respectively. These radii are slightly larger than the initial sizes
of the disks in our study. The difference does not imply any
inconsistency in the model because we anticipate (and as we shall
observe below, correctly so) that the disks will quickly reconfigure
themselves according to the dynamical evolution of the system, and
because the present calculations include hydrodynamical effects not
addressed in \citet{HolWie99}. 

\section{Evolution of the Circumbinary Environment}\label{sec:evo-circumbin}

Using the initial conditions and physical models described above we
have performed a number of simulations modeling systems similar to
\ggtaua. The simulations typically follow the evolution for 6500~yr,
corresponding to approximately three orbits of the inner edge of the
circumbinary disk. These times also correspond to $\sim16$ binary
orbits for the 62~AU series, and to $\sim41$ orbits for the 32~AU
series. Complete specification of the initial parameters of the
simulations are listed\notetoeditor{Note to copyeditor: it would be good if
the table were placed near here in the final text. For reasons beyond
my skill to repair, latex seems to want it at the very end of the text}
in table \ref{tab:bin-params}. In the discussion below, we will often
refer to the two classes of simulations as the `wide' (62~AU) binary
simulations and `close' binary (32~AU) simulations. We examine the
qualitative nature of our simulations first, then examine in detail the
structures which form and their physical origin.

\begin{deluxetable}{lcccccccc}
\tablecaption{\label{tab:bin-params} Initial Parameters For Simulations}
\tablehead{
\colhead{Name}    &
\colhead{Primary}  & \colhead{Secondary} & \colhead{Circumbinary} &
\colhead{Disk Radius}     & \colhead{Orbital}   & \colhead{Semi-Major}    &
\colhead{End Time}  & \colhead{Rotations} 
\\
\colhead{}        & 
\colhead{Particles}   & \colhead{Particles}    & \colhead{Particles}  &
\colhead{(AU)} & \colhead{Eccentricity} & \colhead{Axis (AU)}  &
\colhead{(yr)}   & \colhead{$T_{end}/T_{180AU}$ }
}

\startdata
Clos3elo  &  2289 & 1955 &\phn  421384 &   4 & 0.3     & 32      & 4400 & 2.09  \\
Wide0elo  &  2207 & 1892 &\phn  421384 &  10 & 0.0     & 62      & 6500 & 3.09  \\
Wide3elo  &  2207 & 1892 &\phn  421384 &  10 & 0.3     & 62      & 6500 & 3.09  \\
Clos0ehi  &  9995 & 8592 &     1861769 &   4 & 0.0     & 32      & 7000 & 3.33  \\
Clos3ehi  &  9995 & 8592 &     1861769 &   4 & 0.3     & 32      & 7500 & 3.57  \\
Wide0ehi  & 10134 & 8418 &     1861769 &  10 & 0.0     & 62      & 6500 & 3.09  \\
Wide3ehi  & 10134 & 8418 &     1861769 &  10 & 0.3     & 62      & 6500 & 3.09  \\
Singlehi  &    -  & -    &     1861769 &   - &  -      & -       & 2100 & 1.00  \\
Noheat3ehi  & 10134 & 8418 &     1861769 &  10 & 0.3     & 62      & 2100 & 1.00  \\
ClosT3elo &  2289 & 1955 &\phn  421081 &   4 & 0.3     & 32      & 4000 & 1.90  \\
ClosT0ehi &  9995 & 8592 &     1867987 &   4 & 0.0     & 32      & 6500 & 3.09  \\
ClosT3ehi &  9995 & 8592 &     1867987 &   4 & 0.3     & 32      & 6500 & 3.09  \\
WideT0ehi & 10134 & 8418 &     1867987 &  10 & 0.0     & 62      & 6500 & 3.09  \\
WideT3ehi & 10134 & 8418 &     1867987 &  10 & 0.3     & 62      & 6500 & 3.09  \\
\enddata
\end{deluxetable}

\subsection{Systems with Eccentric Orbits}\label{sec:eccentric}

The analyses of the proper motion data lead to two alternatives for
the orbital motion of the stars in \ggtaua, with $a=32$~AU and
$a=62$~AU. In both cases, the orbital eccentricity is calculated to be
$e\approx0.3$. In this section, we examine the results of two
simulations, each with eccentricity $e=0.3$, but with differing
semi-major axis, referenced in Table \ref{tab:bin-params} as
simulation {\it Wide3ehi} ($a=62$~AU) and {\it Clos3ehi} ($a=32$~AU). 

\subsubsection{Wide Separation}\label{sec:wide-morph}

\begin{figure*}
\includegraphics[angle=0,scale=0.87]{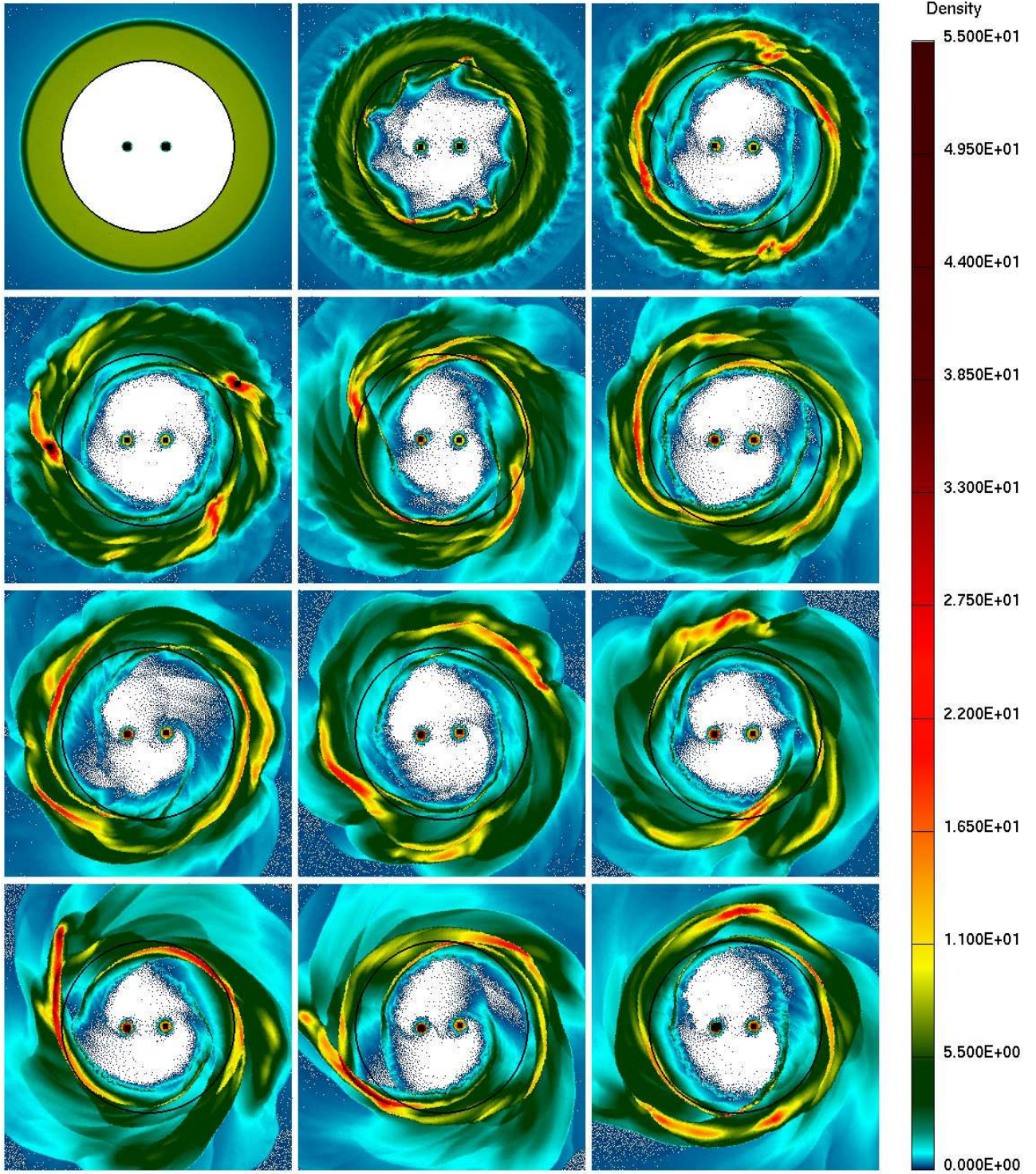}
\caption{\label{fig:wide-ecc-manyorb}
The morphology of the circumbinary material at successive apoapse
passages of the binary, for sumulation {\it Wide3ehi}. Each frame
defines a 600$\times$600~AU region centered on the system center of
mass. Ordered from top left, the first five frames are shown at time
intervals of two complete binary orbits, while the next seven frames
are separated intervals of one orbit. } 
\end{figure*}

In figure \ref{fig:wide-ecc-manyorb} we show twelve
images\footnote{Note that for ease of comparison, we will use
identical color maps, with identical scales and limits in this and all
other similar figures throughout the paper, although in a few images
the limits will be exceeded in small regions. Also, note that each
of our figures show the system as represented by the positions of each
SPH particle directly, rather than as interpolated onto a grid based
graphical representation. We believe that although the images may
appear somewhat granular using this manner of presentation, the viewer
will be left with a better sense of both the physical system state
and its numerical realization. } of the same simulation,
{\it Wide3ehi}, at the same orbital phase, spread over 16 orbits of a
binary orbiting with semi-major axis $a=62$~AU. The mass distribution
in the initial condition, seen in the first frame, is completely
axisymmetric and clearly shows the various components in the idealized
initial configuration we employ in all of our simulations. As the
simulation time increases and the system begins to evolve (frames
1-3), the mass density begins first to exhibit many small scale
disturbances at the interface between the torus and disk near
$\sim250-300$~AU, which penetrate into the low density disk
surrounding it. Larger scale disturbances are present in the mass
distribution at the torus's inner boundary as well. We attach little
physical significance to these early time features and instead
attribute them merely to the growth and relaxation of `start up
transients' related to our idealized initial condition for the actual
complex system. 

After the first several orbits of the binary, such small scale
instabilities have been largely overwhelmed by the growth of features
driven by the strong stirring action of the stars on the circumbinary
material. Initially, the torus condenses into a narrowly peaked
structure some 10-20~AU in radial extent, with several higher density
lumps approximately equally spaced in azimuth. Later, after 6-8 binary
orbits and as the results of interactions between the torus and binary
begin to accumulate, the density inhomogeneities become more and more
widespread. At this time, the lumps have largely dissipated, and the
ring itself has broken up into several filamentary spiral arms, each
of which contain still finer scale substructure. 

Over time, the original sharp inner edge of the torus becomes
progressively more ragged and diffuse as the binary interactions
perturb it. As it evolves, it both moves inwards towards the stars and
its shape changes, becoming significantly non-circular and producing
spiral streams that extend inwards towards the binary. As pictured,
the inner edge of the torus appears similar to an ellipse whose
semi-major axis is approximately perpendicular to the axis joining the
two stars. Because each of the images in the mosaic is generated at
the same orbital phase, the relative phase of the binary orbit and the
elliptical torus edge appears nearly constant. We will explore the
extent to which this is always true in section
\ref{sec:quantitative}, below.

At larger radii, and as they begin to overwhelm the start-up
transients, spiral structures become ubiquitous features throughout
the torus at all times in the simulation. The spiral structures are
generated by a combination of self gravitating instabilities and the
stirring action of the stars in their orbit. The finer scale
substructures in the torus are largely the remnants of streams of
material which propagate through the gap region and return to the
torus. The details of the size and shape of these streams vary from
one orbit to the next. For some orbits, long streamers of material are
visible extending inwards towards one or the other star, while in
others they are nearly absent. We will discuss the behavior of these
streams over the course of one binary orbit in detail in section
\ref{sec:circstar-conf}. Of importance for our discussion here however
is the fact that comparatively little material is actually accreted
into the circumstellar environment compared to the fraction that
returns to the torus. Instead, most material falls through the gap,
interacting with one or both stars only through a relatively distant
encounter during which only its trajectory is changed. Then, as it
continues on its new trajectory, material begins to propagate outward
again, returning to the torus from which it originated.

\subsubsection{Close Separation}\label{sec:close-morph}

\begin{figure*}
\includegraphics[angle=0,scale=0.87]{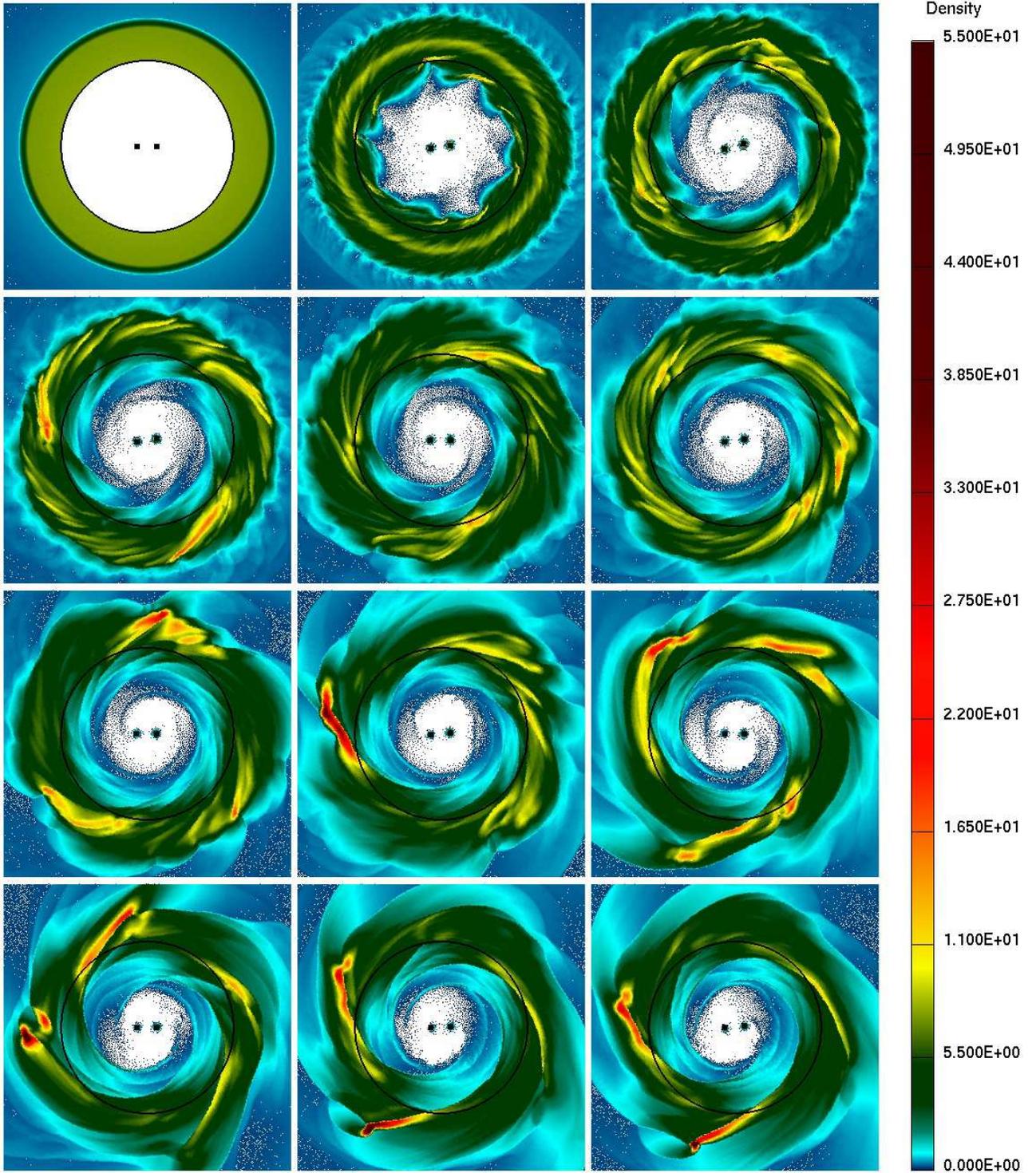}
\caption{\label{fig:close-ecc-manyorb}
The morphology of the circumbinary material at successive apoapse
passages of the binary for simulation {\it Clos3ehi}. Each frame
defines a 600$\times$600~AU region centered on the system center of
mass. Ordered from the top left, the first nine frames are shown at
time intervals of five binary orbits, while the last three frames show
intervals of three, three and one binary orbits each.}
\end{figure*}

In figure \ref{fig:close-ecc-manyorb}, we show twelve images of the
{\it Close3ehi} simulation, each at the same orbital phase, spread
over 47 orbits of a binary orbiting with semi-major axis $a=32$~AU. As
in the prototype wide orbit configuration in the last section, the
mass distribution in the circumbinary disk first begins to exhibit
small scale inhomogeneities which we attribute to start up transients
as the simulation begins to evolve away from its idealized initial
configuration. After $\sim1500$~yr of evolution (frame 3), larger
scale structures develop in the torus, which extend througout its
radial extent. Unlike the wide orbit case however, the structures are
more filamentary in character, with no features comparable to the
radially narrow, peaked structure noted in panel of figure
\ref{fig:wide-ecc-manyorb} developing, even at later times. Instead,
smaller amplitude spiral structures develop which, after $\sim5000$~yr
(panel 7-9), begin to organize into larger scale condensations. These
condensations continue to grow and coalesce over the next
$\sim2000$~yr and, in the final configuration of the simulation, only
one remains. We expect that if the simulation were to continue, this
fragment would continue to grow, perhaps to a size at which it would
begin to disrupt the disk itself. We will discuss what interpretation
should be given to such fragments in section \ref{sec:fragmentation}.

Over the lifetime of the simulation, the inner portion of torus
material migrates inwards towards the stars, with its inner edge
eventually reaching distances from the system center of mass of only
about half of its original distance. Unlike the wide orbit case, large
streamers of material do not appear in the gap region and, to the
extent that they are present at all, they do not change their
character from one orbit to the next. This behavior is consistent with
the increased distance from the stars to the inner edge of the torus
and the correspondingly smaller perturbing influence that they exert
on it. Even at late times, when the streamers appear most distinct,
they do not extend inwards to the positions of the stars or their
disks as much as is seen in the wide orbit case above. Instead, they
appear as extensions of the spiral structures further out, and we
associate their presence to the fact that precursors of the fragments
noted above have begun to influence the torus morphology.

\subsection{Systems with Circular Orbits}\label{sec:circular}

\begin{figure*}
\includegraphics[angle=0,scale=0.75]{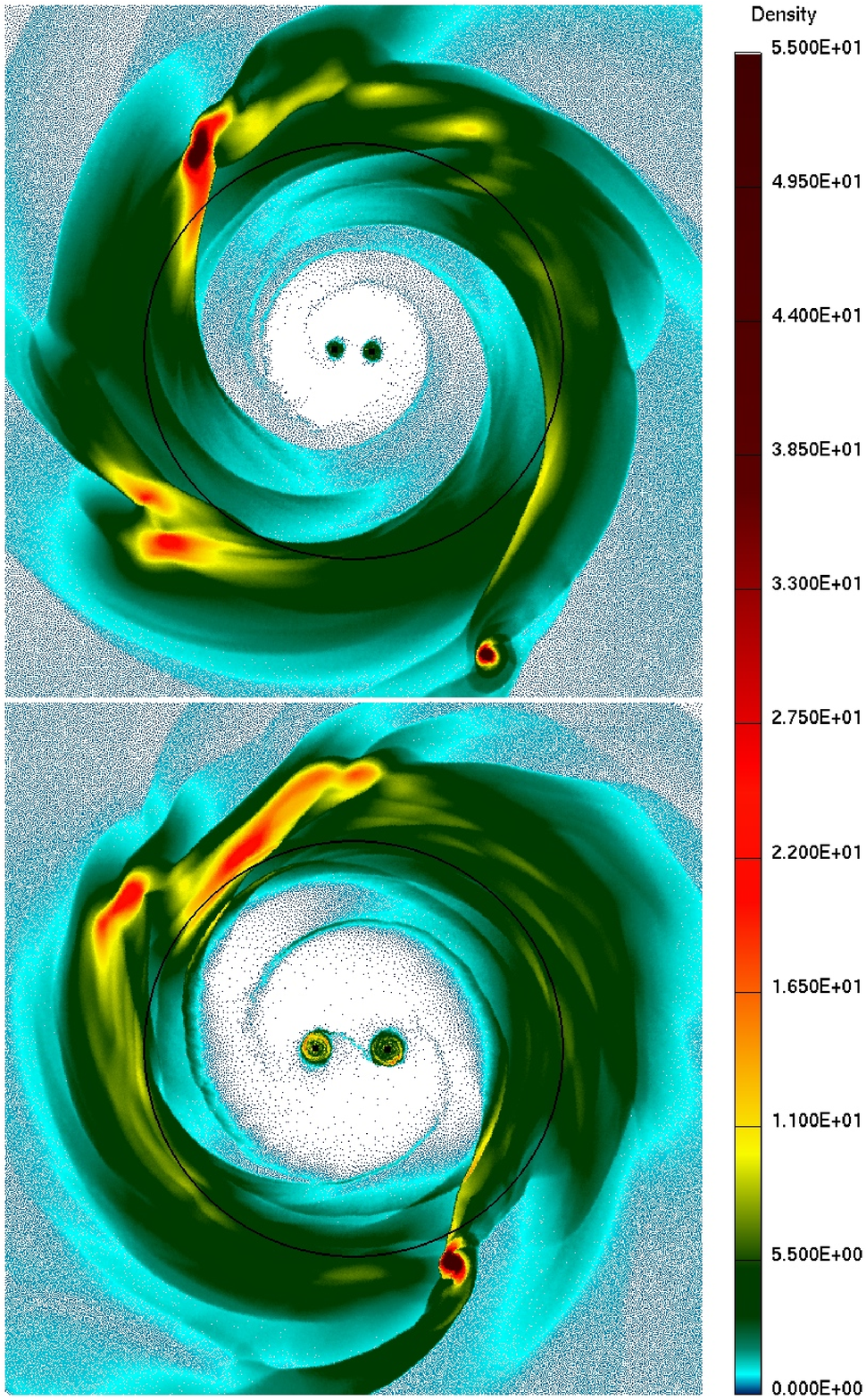}
\caption{\label{fig:e0-end}
The morphology of the circumbinary material at the end of each of the
two $e=0$ configurations in our study. In the top panel we show
the variant with the `close' binary orbit separation ($a=32$~AU, {\it
Clos0ehi}), while in the bottom panel, matching `wide' binary orbit
separation ($a=62$~AU, {\it Wide0ehi}) simulation. Each frame defines
a 600$\times$600~AU region centered on the system center of mass, shown
after evolving for a time of $t\sim3.3$ (top) and $t\sim3.1$ (bottom) 
orbits of the torus' inner edge, as noted in table \ref{tab:bin-params}.}
\end{figure*}

The behavior of the cirumbinary environment surrounding a system with
an eccentric orbit will of course differ from that of a circular
orbit, due to the presence or absence of time variations in the
strength of the gravitational forces exerted by the stars. We have
investigated these differences with a set of simulations whose
conditions are identical to those in sections \ref{sec:wide-morph} and
\ref{sec:close-morph}, except that we assume that the orbit of the
binary is defined by an eccentricity of $e=0$. We show the system
configurations at the end of each simulation in figure
\ref{fig:e0-end}, for the same two semi-major axes ($a=62$~AU and
$a=32$~AU) described in sections \ref{sec:wide-morph} and
\ref{sec:close-morph} above.

In both the wide and close cases, the inner edges of the tori have
moved inwards over the time scale of the simulation to radii of
120-140~AU from the system barycenter or less. Both models show
evidence of streams of material present in the gap region. Although
less pronounced than in the $e=0.3$ cases, they extend further inwards
towards the stars, likely because they are not disrupted to as great
an extent by close encounters with the stars. Relative to each other
the wide binary configuration shows more massive streams than the
close binary configuration, as was the case in the $e=0.3$ models, In
addition, the wide configuration also exhibits a small band of
material joining the two components, which appear to originate from
material pulled from the outermost extents of the circumstellar disks.
Closer to the stars, the circumstellar disks exhibit only minimal
spiral structure, with the highest densities found in the outer
extents of the disks rather than close to the stars.

The spiral structures that develop in each torus are much more diffuse
in character than in the corresponding $e=0.3$ simulations, and
exhibit much less internal structure. In consequence, the outer edges
of each torus are both more circular and less sharply defined than in
the eccentric binary cases, where well defined edge features form as
structures propagate through the tori to their outer limits. In spite
of the less pronounced spiral structure, the circumbinary tori in both
simulations have produced large fragments, some 10-20~AU in diameter.
In each case, the fragments have grown to such an extent that they
have begun to generate large amplitude spiral wakes in the torus,
independently of the influence of the binary's stirring action. We
will discuss the physical significance of these fragments in section
\ref{sec:fragmentation}.

\subsection{Angular Momentum in the System}\label{sec:transport}

The observed, two-component torus/disk configuration of the \ggtaua\
circumbinary material provides a challenge for simple models of star
formation to reproduce. While one can easily imagine mechanisms by
which the actions of the binary conspire to keep a sharply bounded
inner `gap' region clear, similar arguments are more difficult to make
at the outer edge of the torus. How did a sharply bounded torus form
with a surrounding low density disk, and how is the structure
maintained in the system over time? 

An important result of the sections above was to show that the stirring
action of the binary on the surrounding material is very strong, with
large scale spiral structures being continually generated. Such
stirring serves as a foundation for the model described in
\citet[e.g.][]{pringle91} for example, and one consequence he describes
is the formation of an `excretion' disk as a portion of the
circumbinary material gains angular momentum from the binary and moves
outwards in response. Is this mechanism the origin of the low density
disk in the \ggtaua\ system? In this section, we explore this question
first by simulating the evolution of a family of systems in which the
initial circumbinary configuration includes {\it only} a torus with the
same mass as the combined torus/disk configuration as above. Then, we
study behavior of the torques generated by the the stars and torus on
each other.

\subsubsection{Torus-only Initial Configurations}\label{sec:torus-only}

\begin{figure}
\includegraphics[angle=0,scale=0.90]{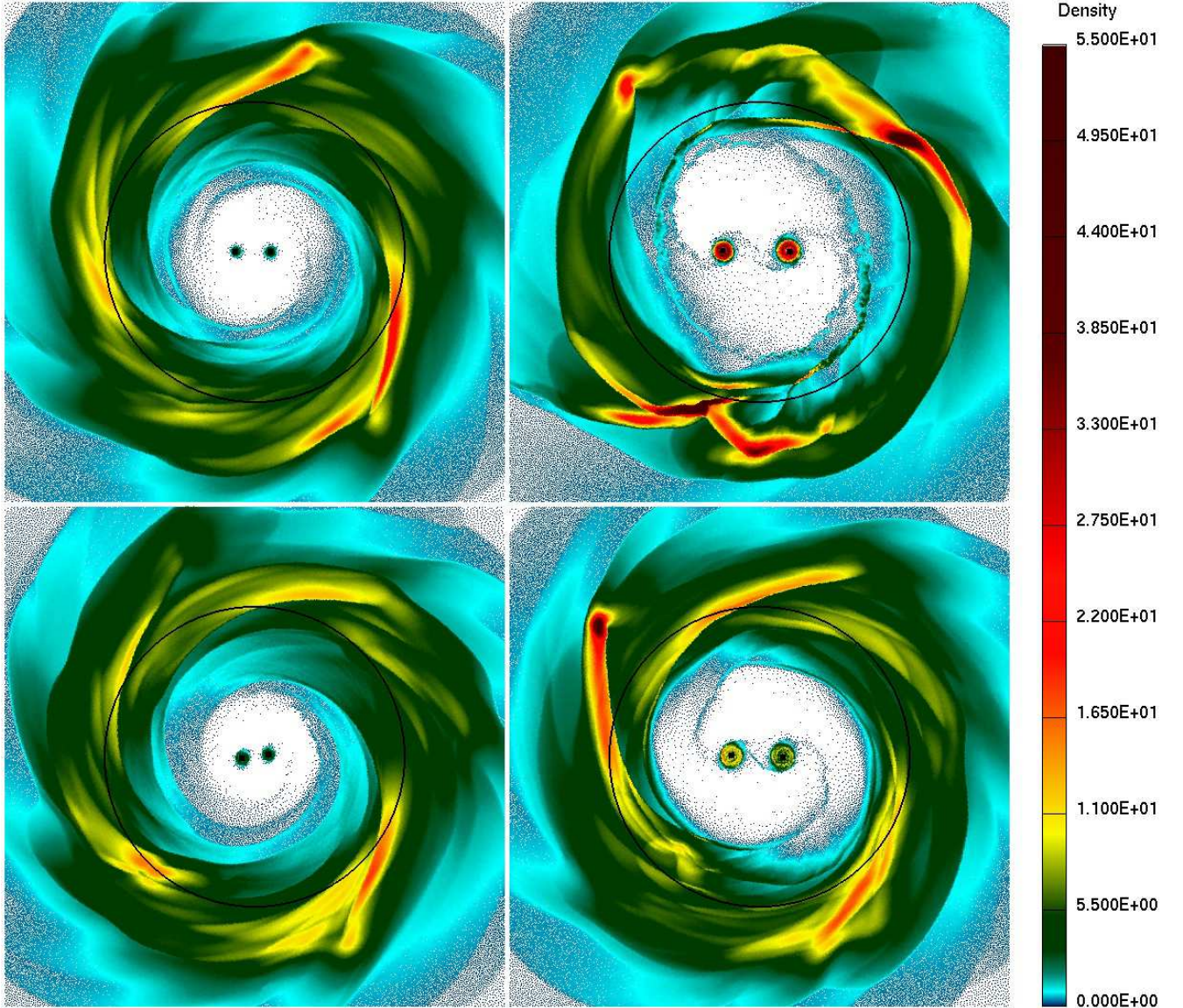}
\caption{\label{fig:torus-morphs}
System configuration of four simulations which assume all material in
the circumbinary environment is initially located in a torus. Each
panel shows the configuration at the binary's apoapse for the time
closest to $\sim6500$~yr of evolution, corresponding to $t\sim3.09$
torus orbit periods, such that the systems have evolved for a similar
amount of time. The top panels show the two simulations with the
assumed binary eccentricity set to $e=0.3$ ({\it ClosT3ehi} and {\it
WideT3ehi}), while the lower panels show the two configurations with
assumed binary eccentricity set to $e=0.0$ ({\it ClosT0ehi} and {\it
WideT0ehi}). The right and left panels show the $a=32$~AU and
$a=62$~AU semi-major axes, respectively.}
\end{figure}

Figure \ref{fig:torus-morphs} shows the counterparts of each of the
four torus/disk configurations described above, each after evolving
for $\sim6500$~yr. As in the torus/disk models discussed above, torus
material migrates inwards, partially filling the gap region between
the cirumstellar environment and the initial inner boundary of the
torus. The gap region in the close binaries is visibly smaller in size
than in the wide, and more in the $e=0$ cases than in the eccentric
orbit cases. In every case, and similarly to the torus/disk models
above, well defined spiral structures form in the tori and, in several
examples, they fragment into clumps. In all cases, the spiral
structures are driven by the stirring action of the stars and
propagate outwards, carrying material as they do so. Also as in the
torus/disk models, the edges of the spiral arms are sharply defined,
particularly in the two wide binary examples where the stars interact
more strongly with the torus. Although sharply defined during their
passage through the torus, they become less and less distinct at
greater distances from the system center, eventually reaching a lower
and smoother background density at distances beyond the original outer
edge of the torus.

\begin{figure}
\includegraphics[angle=-90,scale=0.65]{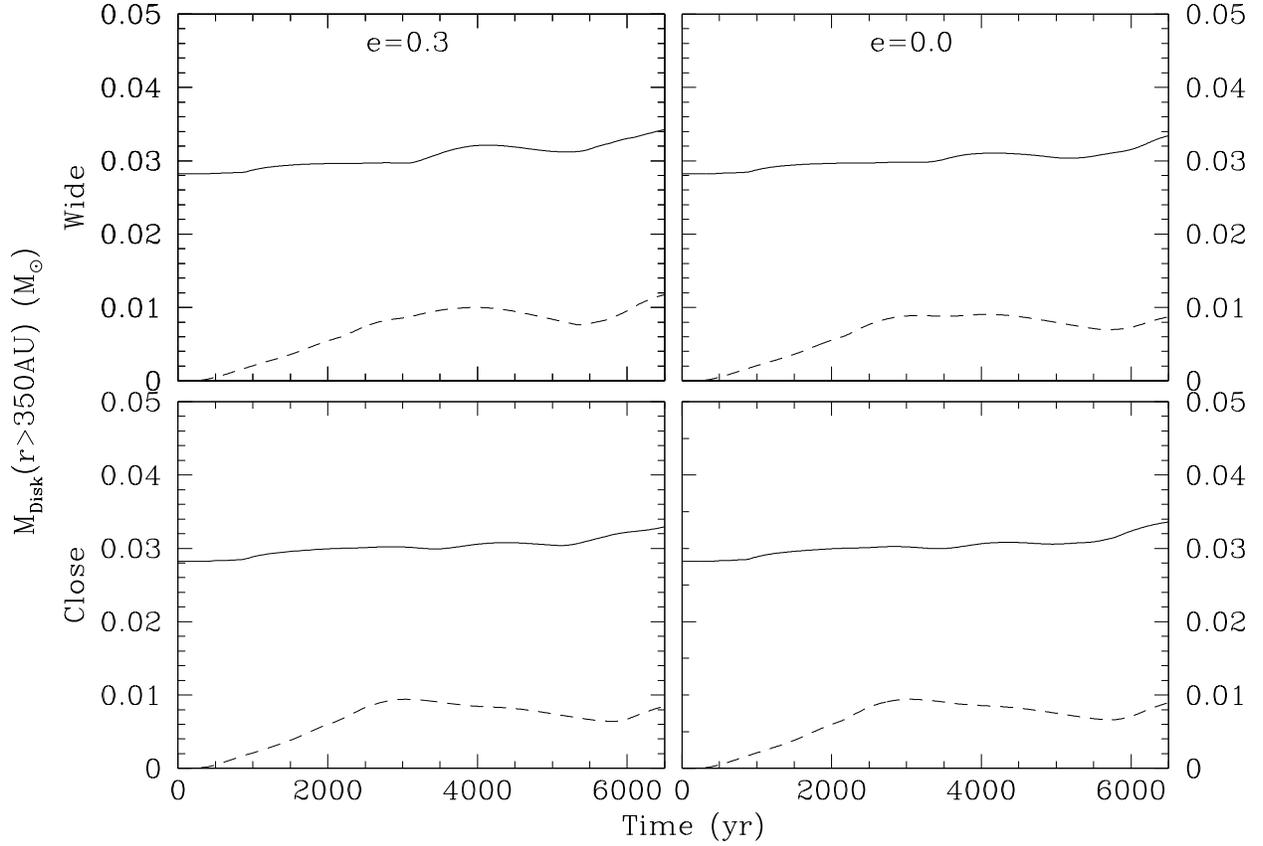}
\caption{\label{fig:mass-outside}
The mass further from the system barycenter than 350~AU as a function
of time, for each of the torus-only and torus$+$disk models in our study.
The top and bottom panels show the set of wide (62~AU binary orbit) and
close (32~AU binary orbit) runs, respectively. The right and left
panels show the circular and eccentric orbits, respectively. The dashed
lines in each panel show the masses for the simulations with torus-only
initial conditions. The solid lines show the masses for the simulations
with torus/disk initial conditions. }
\end{figure}

The presence of material at large separations is evidence for some
level of outwards mass transport in the system. Figure
\ref{fig:mass-outside} shows the amount of mass found there as a
function of time for each of the simulations. As defined in the
figure, the `outer disk mass' is defined as the amount of mass found
at radii larger than an arbitrarily defined critical distance from the
system barycenter. Guided by the results we will describe in section
\ref{sec:quantitative} below, we set a value of 350~AU for this
radius, which we choose as a compromise between capturing the true
mass of the disk component over time, and masking out transient changes
to it that are due to portions of the torus flexing such that parts of
it extend beyond our critical radius. 

In every case, the mass increases over time, showing that the process
itself is robust across a wide variety of configurations. The
magnitude and overall pattern of the mass transport varies however,
both from simulation to simulation and from time to time within the
same simulation. After $\sim3$ torus orbits, corresponding to 6500~yr
run time of the simulations, the total mass transport corresponds to
about 5\% of the total initial mass of the circumbinary material in
the torus$+$disk systems, and is $2-3$ times higher, at $\sim15$\%, in
the torus only systems. In the torus$+$disk runs, the outer disk mass
increases at a near constant rate, with small oscillations visible,
superimposed on the long term trend. In contrast, in the torus only
runs, the mass transport is almost entirely limited to the first
$\sim3000$~yr of the evolution, with the outer disk mass becoming
essentially constant thereafter or even decreasing slightly. Only in
the last $\sim500$~yr of the runs, does the outer disk mass begin to
increase again.

The duration of our simulations are long in the context of the
circumstellar environments and the number of binary orbit periods we
simulate. Unfortunately, they are short in comparison to the orbital
timescales of circumbinary material, where a time of 6500~yr
corresponds to only $2-3$ orbits. Therefore, due to the high cost of
running the simulations, we have not followed the evolution for the
long times required to determine whether or not the outer disk masses
return to their high rate of growth, or if the decrease is simply a
signature of the overall flexing action of the circumbinary structures
as they respond to the binary's stirring action. We are therefore
unable to conclude that the outward material migration will continue
at the same rate or decrease, or even if time variations within a
given simulation continue indefinitely. Nevertheless, the fact that
the outer disk mass increases in all of our models, is strong evidence
that the excretion process will continue in the initially torus-only
systems, such that they evolve to become robust torus$+$disk systems.

Given the rate of outward material movement, and the fact that the
disk in the \ggtaua\ system is observed (GDS99) to contain a mass of
$\sim0.04$\msun, the entire disk could be generated from such
transport activity within a few $\times10^4$~yr. We therefore conclude
that the disk material surrounding the \ggtaua\ torus most likely
originates as torus material which was thrown outwards by the stirring
action of the binary.

\subsubsection{Torques on the stars and gas}\label{sec:torques}

Gravitational torques between the stars and circumbinary material play
an important role in their mutual evolution. Here, we turn to a study
of these torques, in order to understand their importance in forming
multiple star systems like \ggtaua.

\begin{figure}
\includegraphics[angle=-90,scale=0.65]{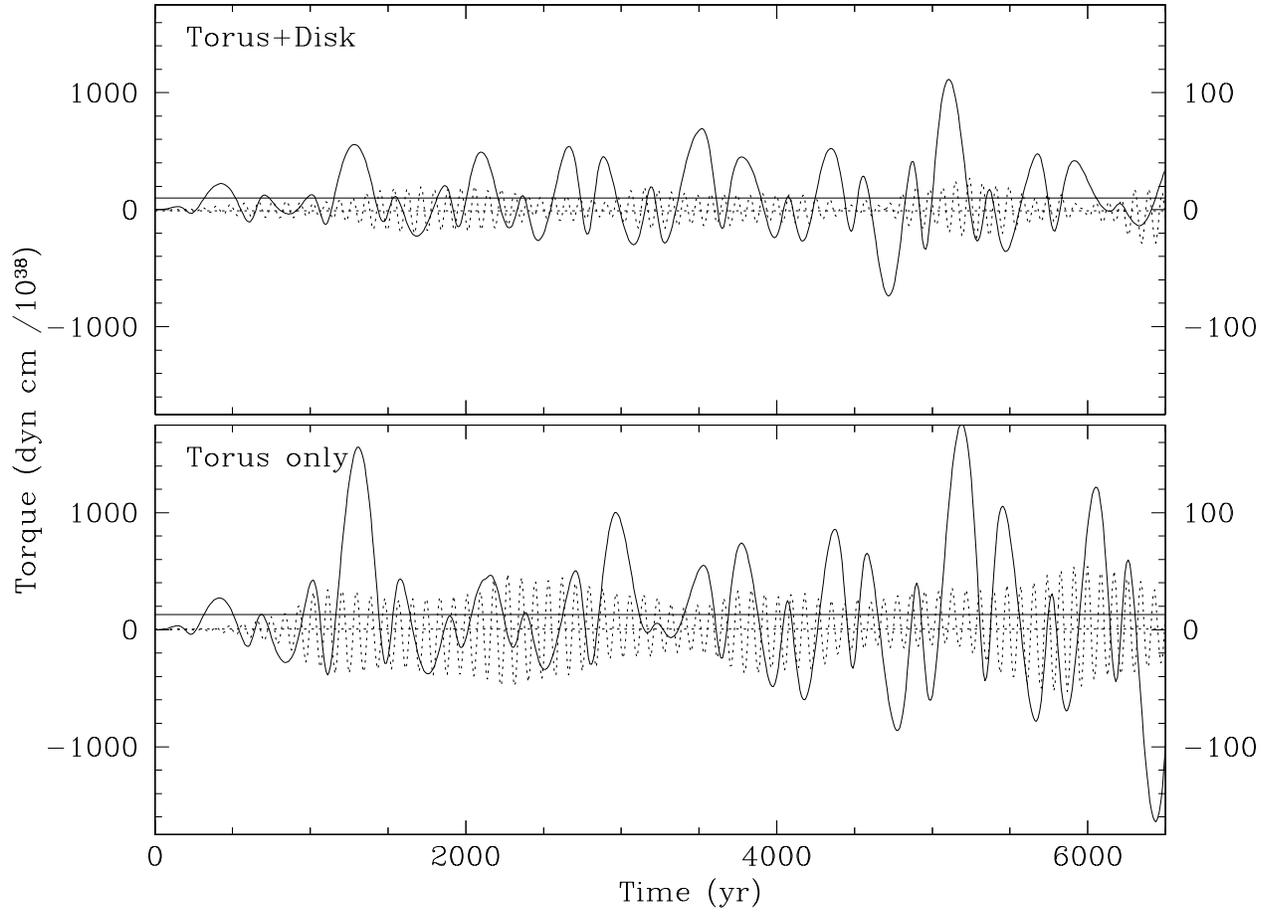}
\caption{\label{fig:sumtorq}
The total torque exerted on the circumbinary material as a function of
time by the stars, as calculated about the system barycenter.
Wide orbit simulations {\it Wide3ehi} and {\it WideT3ehi} are shown
with solid curves, while close orbit runs {\it Clos0ehi} and {\it
ClosT0ehi} are shown with dotted curves. The torus$+$disk models appear
in the top panel and torus-only models in the bottom panel. The left
axis labels correspond to the wide orbit models, while the right axis
labels (expanded by a factor ten) correspond to the close orbit
models. The horizontal lines define the time averaged torques for each
of the simulations shown. } 
\end{figure}

Torques between the stars and the torus will affect both and, if large
enough, cause measurable changes even over the comparatively short
duration of our simulations. Figure \ref{fig:sumtorq} shows the net
torque exerted by the stars on the circumbinary material. In all cases,
the torque magnitudes vary both in amplitude and sign on time scales
comparable to the orbit periods of the binaries. Also, while
oscillations are correlated with the binary orbit, consecutive peaks
and troughs are neither evenly spaced in time nor are their amplitudes
the same. The amplitude of the variations are larger in the torus-only
systems than in the torus$+$disk systems, consistent with the increased
mass concentration in the former case. With few exceptions, the
magnitude of the peaks in the wide orbit examples exceed the magnitude
of adjacent troughs, such that the average over many orbits is
positive. As seen, e.g., in section \ref{sec:wide-morph} above, the
torques from the binary drive strong spiral waves into the torus in
wide configurations, but are much less effective in the close
configurations shown later. Their absence leads to the question of what
process remains to drive the excretion flow discussed above? We note
that, while stellar torques are weak in such systems, torques due to
self gravity remain, and we conclude that they are important
contributors to the excretion flow. 

With a change in sign, the total torque on the circumbinary material is
the same as that on the binary. To what extent do torques of these
magnitudes alter its orbit over time? \citet{pringle91} makes estimates
using a diffusion approximation based analysis where the torques are
mediated by viscosity. Here, no such model is required, since we have a
direct quantification of the torques from the simulations themselves.
We therefore estimate the time scales using time averaged torques
determined directly from the curves shown in figure \ref{fig:sumtorq}.
While the torques in the wide orbit runs exhibit positive maxima of
$\sim5-10\times10^{40}$~dyn~cm and negative minima of
$\sim1-5\times10^{40}$~dyn~cm, their averages over time take on values
near $\sim1\times10^{40}$~dyn~cm. The circumbinary material thus gains
angular momentum at the expense of the binary, consistent with the
analysis in \citet{pringle91}. The close orbit torques are each more
than an order of magnitude smaller and are also more nearly symmetric
over their positive and negative extrema, with time averages consistent
with zero net torque.

Neglecting factors near unity, the net angular momentum of the stars
around the system barycenter is given by \citet{pringle91} as $J_* =
M\sqrt{GMa}/4$, with $M$ being the combined mass of the two stars. The
time scale for changes in the semi-major axis is $\tau = 2J_*/(dJ/dt)$.
In terms of the parameters of the wide and close orbit systems, the
angular momentum translates to numerical values of $\sim2.5\times
10^{53}$ or $\sim1.8\times 10^{53}$~gm~cm$^2$/s, respectively. Thus,
combined with the time averaged torques derived from figure
\ref{fig:sumtorq}, the timescale for significant changes to the
semi-major axis becomes $\tau \lesssim 1.4$~Myr for the wide systems
and $\tau \gtrsim 100$~Myr for the close systems. The long time scales
in the close orbit models, in Pringle's interpretation, correspond to a
system in an initial phase where the torus has yet to `find out' about
the presence of the binary. In contrast, the shorter time scales in the
wide orbit simulations correspond to the period in which the torus is
gradually expelled by the torques applied by the binary. They are also
comparable to the pre main sequence lifetimes of stellar systems
\citep{HLL01}, indicating that the final \ggtaua\ system configuration
has yet to be established.

\subsection{Comparison to the evolution of systems with reduced
heating}\label{sec:heatoff}

Before we may conclude that the action of the orbiting stars on the
circumbinary material is significant for the overall evolution of the
system, we must compare the results of our simulations to systems with
analogous configurations where the binary is not present. Similarly,
in order to conclude that the radiative heating from the stars is
important, we must compare to simulations in which that heating is
omitted. To those ends, we have configured and run a simulation in
which the radiative heating term has been neglected, and another in
which  we have replaced the two components of the binary with a single
star of the same mass, located at the origin. In order to provide a
similar level of heating to the system, this 'star' radiates with a
luminosity equal to the sum of the luminosities of the binary
components. 

\begin{figure}
\includegraphics[angle=0,scale=0.62]{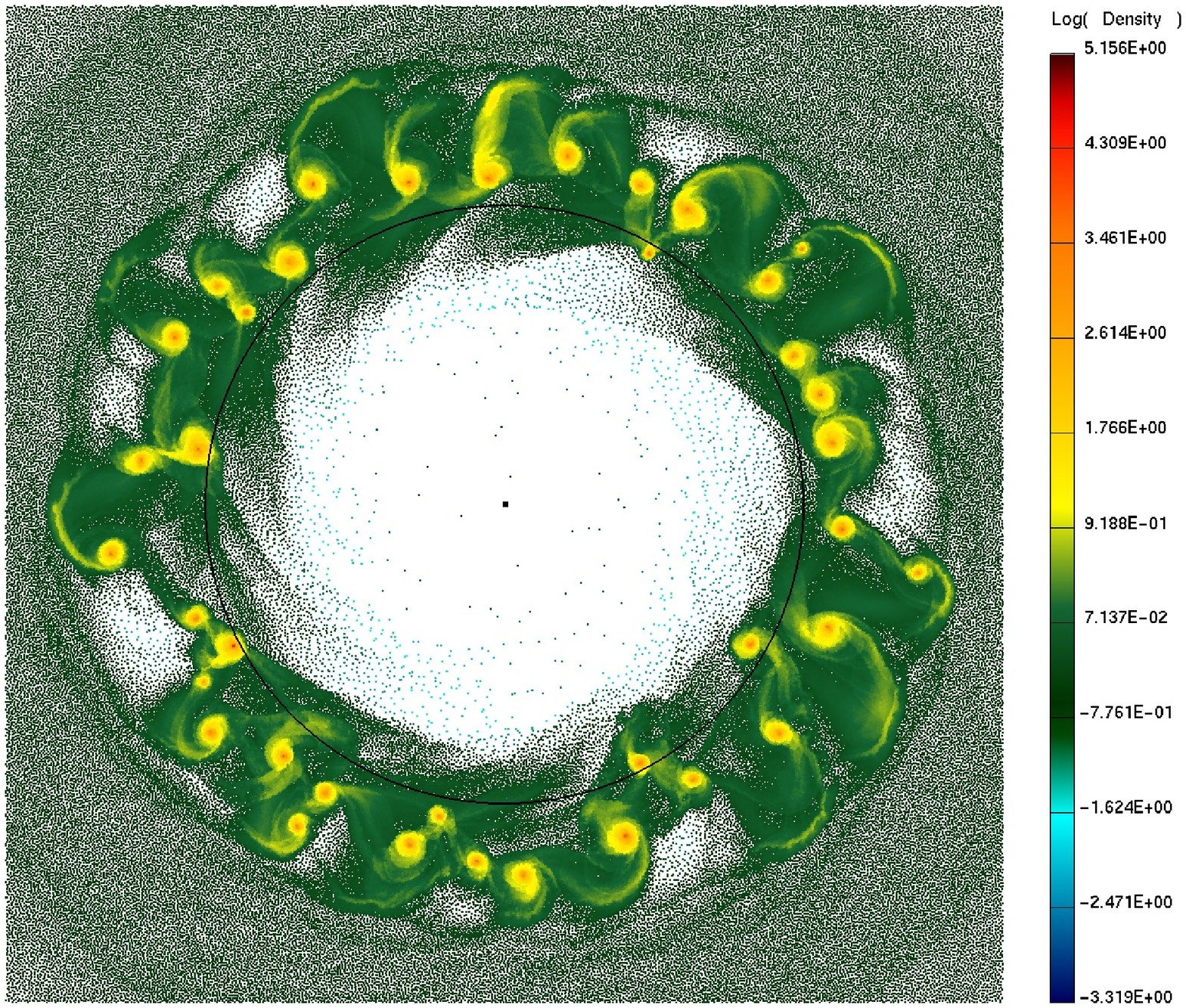}
\includegraphics[angle=0,scale=0.62]{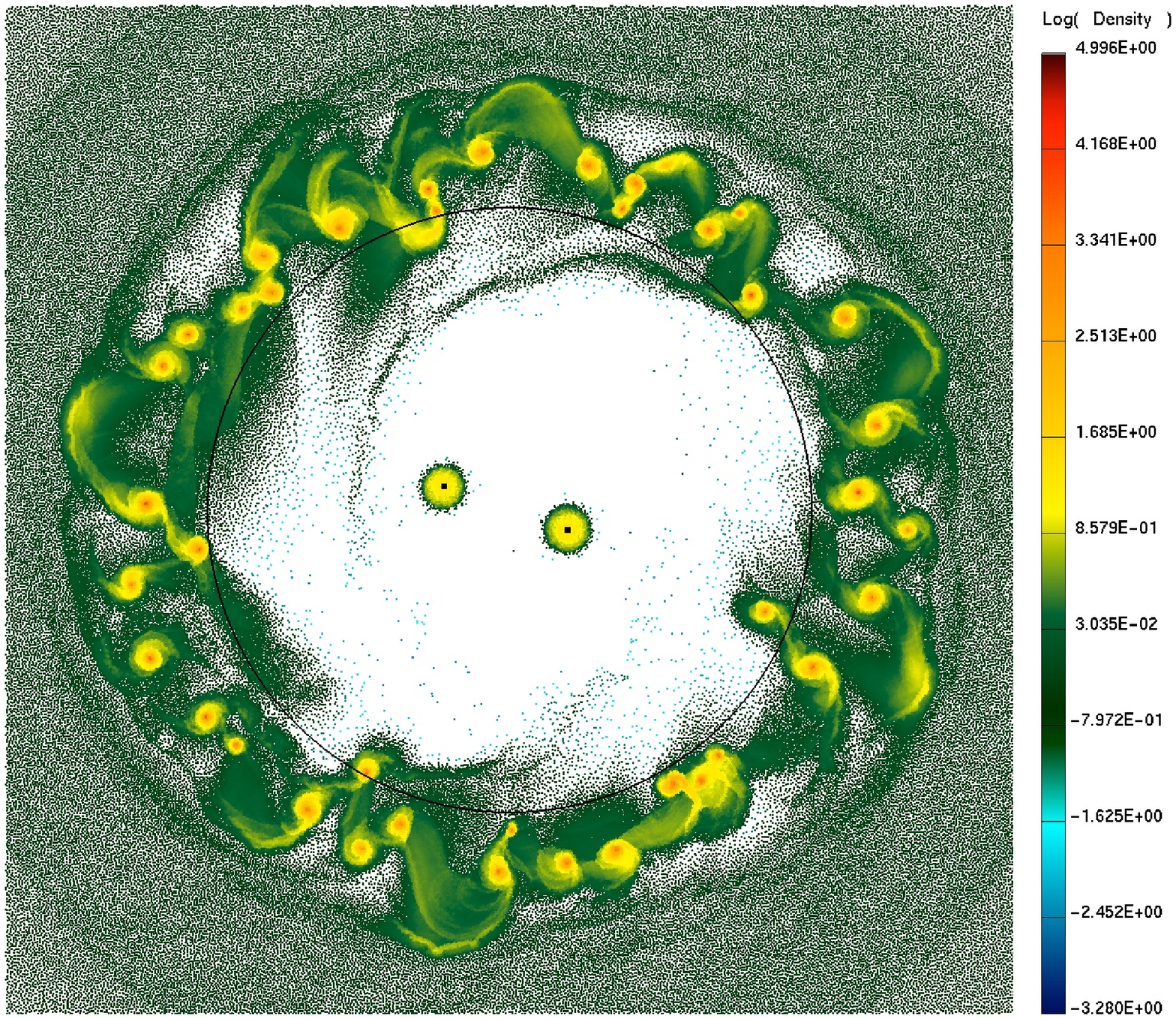}
\caption{\label{fig:heatoff-morph}
The surface density morphology of simulations {\it Singlehi} (top) and
{\it Noheat3ehi} (bottom), each after $\sim2100$~yr (approximately one
orbit of the torus) of evolution.}
\end{figure}

Figure \ref{fig:heatoff-morph} shows the results of these simulations
after they have evolved for $\sim2100$~yr, corresponding to one orbit
of the inner edge of the torus. The image shows that in each case,
approximately 38-39 separate clumps have formed from the initially
smooth tori during this comparatively short timescale. Although not
shown in the images themselves, these clumps initially appeared after
only $\sim1400$~yr of evolution and continued to develop further as
the simulations proceeded. We monitored the status of the simulations
for adherence to the numerical resolution requirements discussed in
\citet{N06} and found that these clumps are indeed sufficiently
resolved according to the criteria described there.

Evidence for self gravitating instabilities is present in the torus 
region of all of our other simulations, including several which
exhibit one or more clearly developed fragments. However, even the
simulations in which fragmentation was observed do not produce
fragments so quickly or in such large numbers. In this regard, the
configurations that develop here are unique in our suite of
simulations, and follow a clearly different evolutionary path than any
of the simulations that include both the binary components and
radiative heating. Since such ubiquitous fragmentation requires cold
torus material, and the action of the stellar irradiation and the
tidal interactions of the binary will be to increase the internal
energy of the material, we conclude that both the stellar irradiation
and the stirring action of the binary are critical contributors to the
overall thermodynamic state of the circumbinary environment of
\ggtaua.

\subsection{Quantitative metrics of the circumbinary
material}\label{sec:quantitative}

In the sections above, we have discussed the overall morphology of the
circumbinary material as it evolves using qualitative metrics of its
behavior. We now turn to more quantitative metrics to describe the
evolution. In particular, we have noted above that the torus appears
to flex both radially and azimuthally over time, with various
components moving both inwards and outwards as spiral waves pass
through them. Here, we quantify the changes in the shape of the torus,
by fitting its shape to a set of three ellipses, chosen as
approximations of its inner and outer edges and its midpoint. In the
following sections, we show example fits at two snapshots in time in
order to illustrate the fidelity of the fits to the underlying data,
and then as functions of time, in order to show their variations as
the system evolves.

For each ellipse, we fit for three important quantities: its
semi-major axis, its eccentricity and its orientation. We discuss the
fits for the two simulations {\it Wide3ehi} and {\it Clos0ehi}. We
choose the first because both the binary's wide orbit (semi-major
axis) and its eccentricity carry it closer to the torus than any
other. We choose the second for exactly the opposite reason: its
tighter orbit and zero eccentricity mean that it stays most distant
from the torus over the course of its orbit. We expect that these
system configurations will therefore bracket the parameters for which
the effect of the binary on the torus is the largest and smallest of
all the configurations in our study.

\subsubsection{Generating fits from the simulation
data}\label{sec:fitting}

SPH particles do not easily lend themselves to standard procedures for
generating fits to the ellipses we use to model the system shape.
Therefore, in order to generate fits, we first map the surface density
of the disk material onto an 1800$\times$1800 Cartesian grid, on which
the fitting can be accomplished straightforwardly. The grid has
spatial dimensions of 900$\times$900~AU and is centered on the origin,
such that each grid cell is 0.5~AU in linear extent. The mapping
implements the same interpolation procedure as is used in the SPH
algorithm to determine the surface densities for each particle in the
simulation itself. Unfortunately in the present context however, the
resolution afforded by our simulations and the mapping retains very
small scale features in the flow, including accretion streams and
other spiral structures only peripherally associated with the torus.
This level of detail frequently causes the fitting procedure outlined
below to produce poorer fits of the large scale torus morphology in
favor of matching these peripheral structures. In order to more
accurately capture only the large scale features of the torus, we
therefore convolve the density mapping with a Gaussian PSF with a
characteristic length scale (i.e., its half-width, half maximum) of
10~AU. 

With this modification of the density mapping, we compute an azimuth
averaged surface density and extract a maximum over all radii. This
maximum is used, in turn, as a critical value to extract the set of
all grid points lying on three distinct isodensity contours. We define
one such contour by the condition that the local surface density
matches the maximum azimuth averaged value, and two others by the
condition that the local density is 20\% of that maximum value, and
falls either radially inside of, or outside of, a circle centered on
the origin and 220~AU in radius. These contours correspond roughly to
the torus' inner and outer boundaries and to its high density core.
Finally, given the points defining the three contours, we use the
Numerical Recipes routine `{\tt mrqmin}' \citep{NumRec} to extract the
semi-major axis, the eccentricity and the axis of pericenter for the
best fit ellipses to each isosurface. In addition to the three fit
parameters, we also permit the origin of each ellipse to be offset
from the system barycenter in the $x$ and $y$ coordinates, taking the
final ellipse parameters from the offset for which the $\chi^2$ value
for the fit is minimized.

In order to avoid contamination of our fits by circumstellar material,
we mask out a circular region of the grid, centered on the origin.
Its radius is defined to be slightly larger than the largest extent
to which the circumstellar material extends at any time during its
orbit, which will occur when the binary is at apoapse and also
includes the combined radii of the two circumstellar disks.
Specifically, we define radius of the circumstellar mask region as:
\begin{equation}\label{eq:CSradius}
r_{mask} = {{a(1 + e)}\over{2}} + f(r_1 + r_2).
\end{equation}
With the identification of the semi-major axis and eccentricity of the
binary using their usual identifiers, $a$ and $e$, the first term
defines half the apoapse separation of the binary components. In the
second term, the quantities $r_{1}$ and $r_{2}$ are each defined using
the expression derived in \citet{HolWie99}, to approximate the largest
radius for which an orbit around the star is stable over a suitably
long duration. We multiply the combined radii by a scale factor,
$f=1.5$, to account for the fact that material may be present
intermittently at larger distances from each star for shorter durations
than are accomodated by the \citet{HolWie99} formalism.

\begin{figure*}
\includegraphics[angle=-90,scale=0.65]{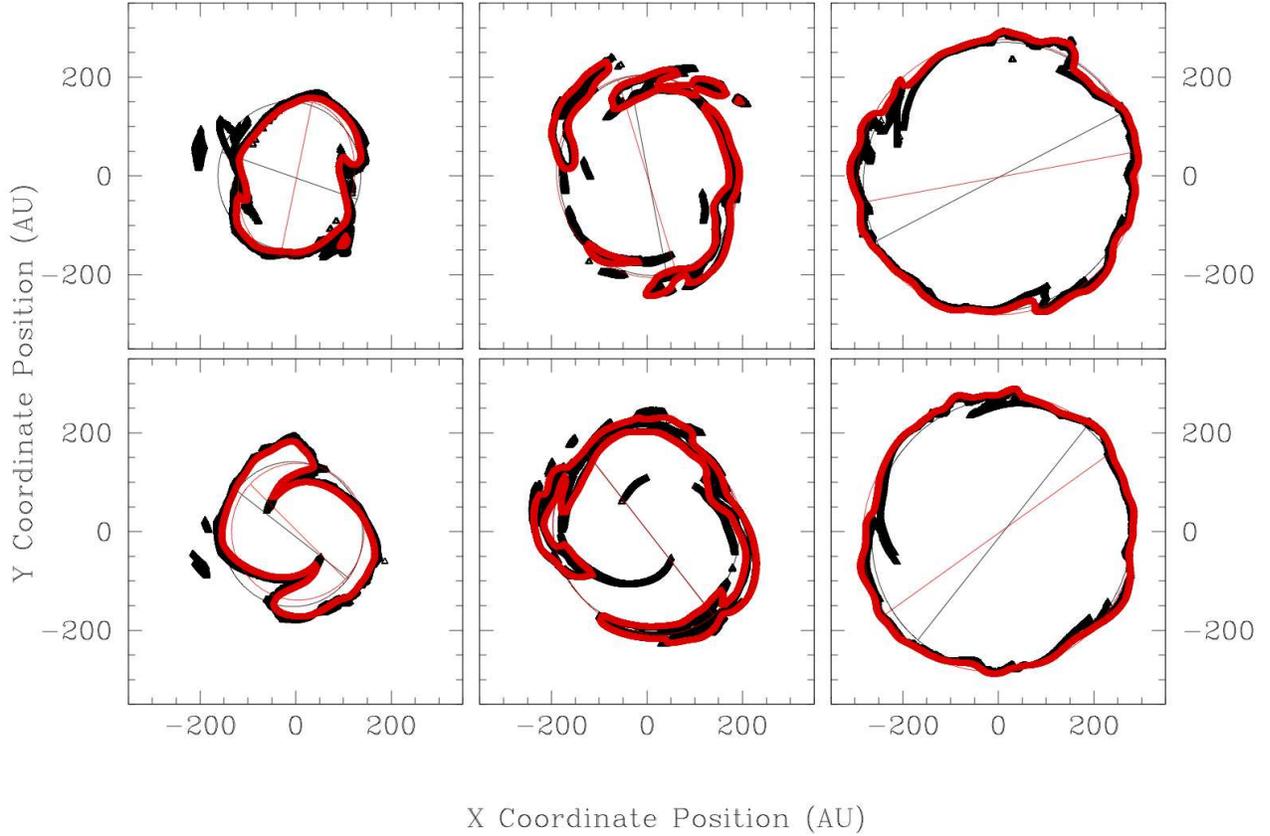}
\caption{\label{fig:torus-fits}
Iso-density curves and fits for the inner (left), maximum (middle) and
outer (right) contours as defined in the text. The top panels show a
snapshot of simulation at the binary's orbital apoapse, the bottom
panels show the immediately following periapse. Each panel shows
contours and fits in black, denoting the densities derived directly
from the simulation, and in red, denoting densities initially
convolved with a 10~AU Gaussian PSF. The lines bisecting each fit
ellipse defines the semi-major axis orientation.}
\end{figure*}

Figure \ref{fig:torus-fits} shows isodensity contours and fits for
simulation {\it Wide3ehi}, at each of the three locations defined
above, both as raw contours and as convolved with a 10~AU PSF. The
times at which the fits were obtained correspond to the same apoapse
and periapse passages shown in figure \ref{fig:wide-ecc-1orb}, below.
In each panel, the fitted ellipses are shown along with the contours
but are frequently obscured by the data, visually demonstrating the
fit quality. Although the differences between the raw and the
convolved contours are not large, they are nevertheless significant.
For example, in the case of the maximum isodensity contours (center
panels), the accretion streams are eliminated in the convolution.
Little change occurs in the fitted ellipses in these cases, but both
the inner and outer contours exhibit much larger differences,
particularly so in the case of the orientation. For example, the
difference is $\sim15-20$ degrees in each of the two snapshots of the
outer contour, a value comparable to the precession angle over the
timespan of the two fit times. The largest differences are seen in the
fits for inner contour at apoapse (upper left panel), where the
difference is nearly 90 degrees. The inner contours are also those for
which both the difference between the two times is most significant
and the fidelity of the fit to the underlying data is least accurate.
While the apoapse snapshot (top) requires significant eccentricity for
a good fit to both the raw and convolved data, the periapse snapshot
(bottom) is best fit to a much more circular shape. In this case,
although the contour is indeed intrinsically more circular, the fit
remains compromised by the incompletely obscured presence of the
accretion streams in the contours which contribute non-negligibly to
the fit. We conclude that while our fits to the middle and outer
contours are minimally affected by such features, fits to the inner
contour will be of lesser quality due to such contributions.

\subsubsection{The size of the torus}\label{sec:torus-size}

\begin{figure}
\includegraphics[angle=-90,scale=0.65]{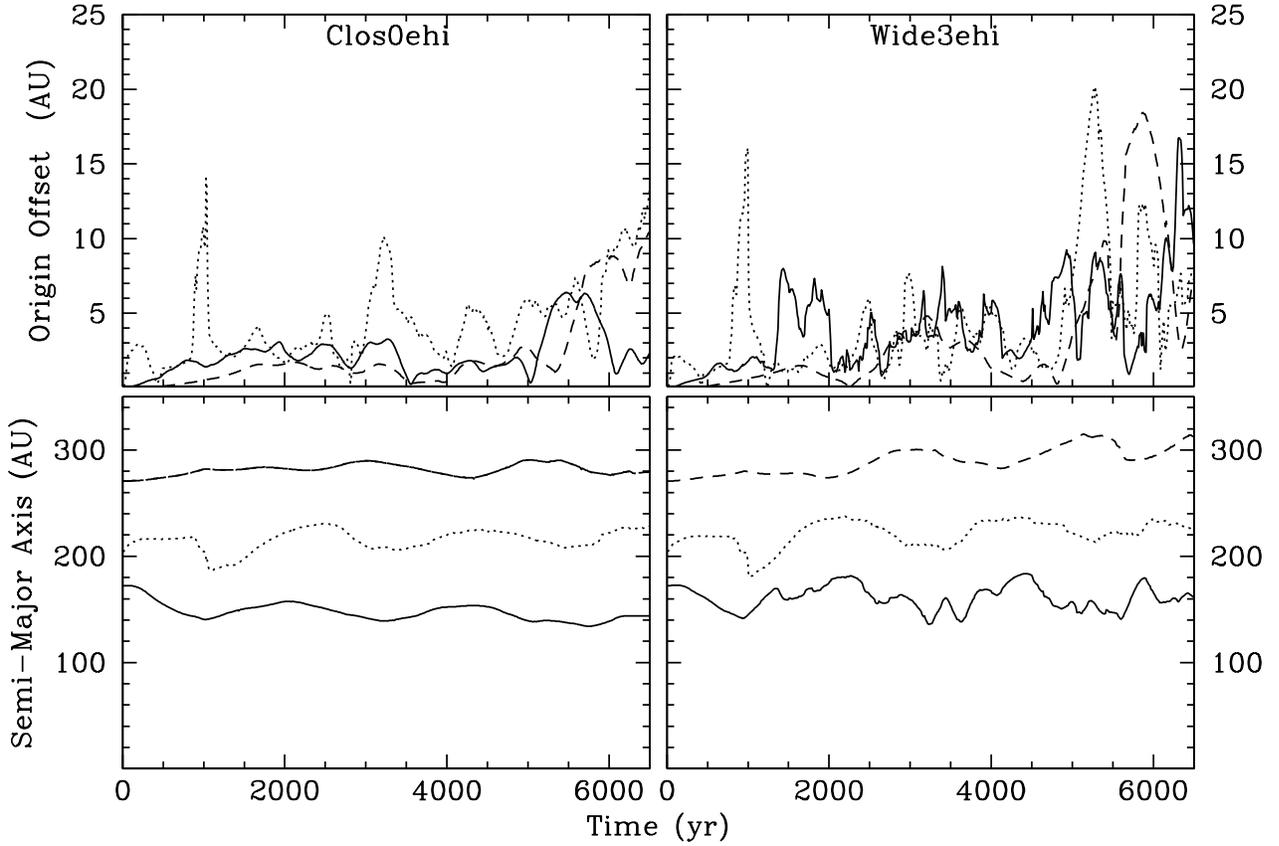}
\caption{\label{fig:torus-radii}
(bottom) The semi-major axes of the three ellipses described in the
text, defining the inner (solid), center (dotted) and outer (dashed)
limits of the torus, each as a function of time, for simulations {\it
Clos0ehi} and {\it Wide3ehi}, as labeled. (top) The offset of the
origin of each ellipse from the system barycenter.}
\end{figure}

Figure \ref{fig:torus-radii} shows fits for each of the three
semi-major axes defined above, as functions of time, along with the
offset of each ellipse from the actual system barycenter. In each case
the semi-major axis varies over time, with the inner ellipse featuring
the most rapid changes and the outer ellipse the slowest. The
amplitudes of the oscillations extend some $\sim20$~AU in either
direction from its long term averaged value. The inner and outer
ellipses also exhibit slow secular trends to smaller and larger
extents, respectively, as the torus itself becomes distorted by the
passage of spiral structures through it. Although the trends are
visible over the entire duration of the simulation, it is unlikely
that they will continue over much longer time spans. For the inner
ellipse, the trend will be constrained by the fact that the action of
the binary will tidally truncate the torus edge. The outer edge trend
will be limited by the fact that the spiral structures decrease in
amplitude as they move outwards, merging with the smooth background
disk. 

In each case, the offset of the best fit ellipse's origins from the
system barycenter fluctuates most rapidly of all, but with
comparatively small amplitudes of $\sim5$~AU, and often with evident
correlations between the set of offsets for the three ellipses at a
given time. At late times in the simulation ($\gtrsim5000$~yr), the
offsets each increase to between 10 and 20~AU, in response to the
larger inhomogeneities in the tori from which they are derived, as
seen in section \ref{sec:evo-circumbin}. Even these late time offsets
remain substantially smaller than the $\gtrsim30$~AU offsets derived
from the \ggtaua\ system by GDS99 and \citet{MDG02} however. In part,
we attribute this difference to the fact that the fits to the
observations assume a circular torus, while our ellipse fits include
additional degrees of freedom, which permit the ellipse origin to
relax toward the system's barycenter, while retaining a good fit to
the shape. More importantly however, the discrepancy serves to
highlight an important difference between the characteristics of the
observations and our simulations. Namely, that the fitted parameters
for our simulations do not trace exactly the same physical features as
fits of various torus parameters derived from observed systems do.
Here, they account only for density features in our 2D representation
of the flow, while in observed systems such as \ggtaua, the various
system characteristics include features that trace both the system's
3D geometry, such as the disk scale height, and its radiated output,
both from scattered and internal sources. Such characteristics may be
correlated with the density, but do not trace it directly. Therefore,
although the specific radii will be similar to those observed for
corresponding features in the \ggtaua\ torus, it is important to make
the distinction clear, in order to avoid confusion.

\subsubsection{The elliptical shape of the torus}\label{sec:torus-ellipse}

\begin{figure}
\includegraphics[angle=0,scale=0.75]{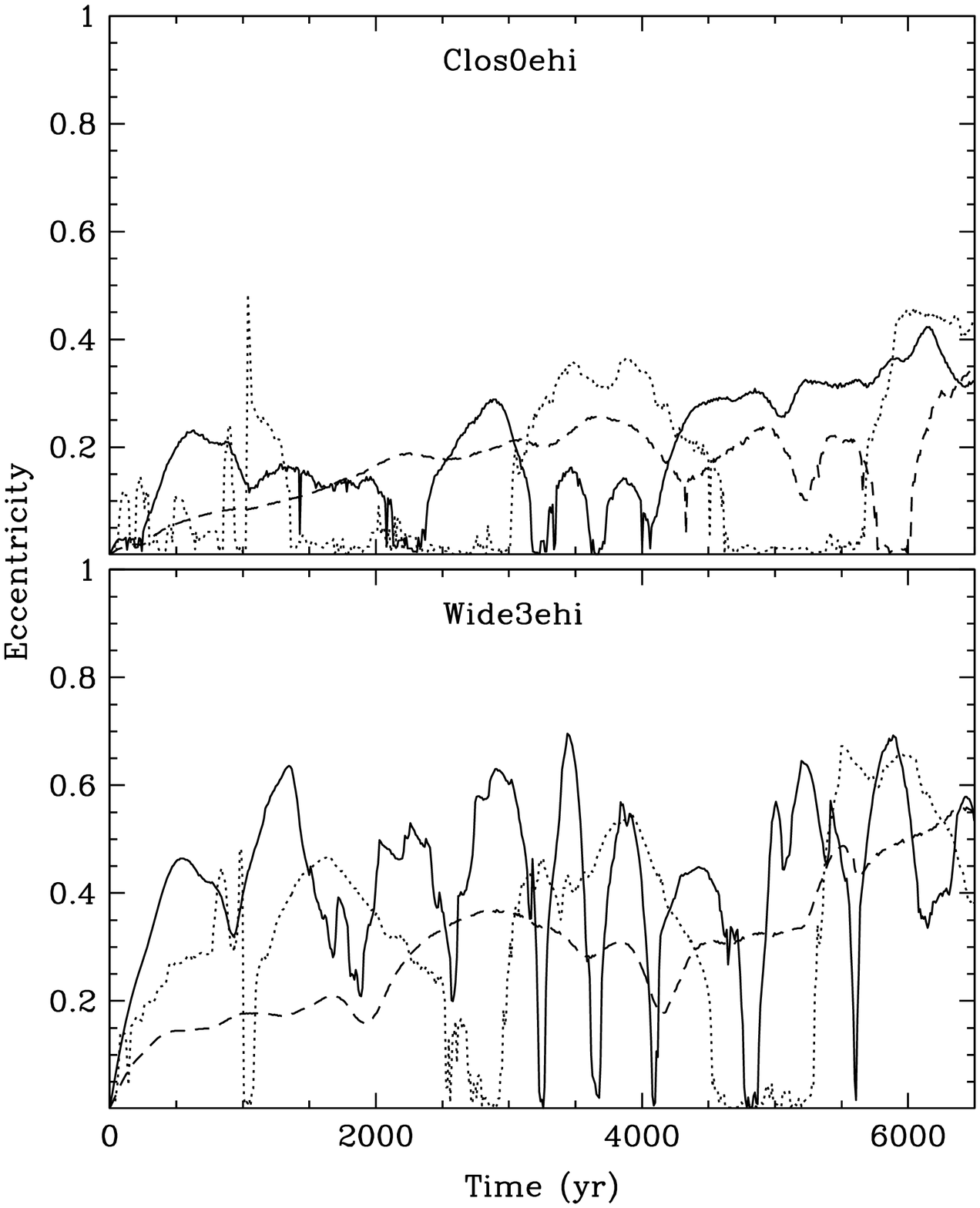}
\caption{\label{fig:torus-ecc}
The eccentricity of the three ellipses described in the text, defining
the inner (solid), center (dotted) and outer (dashed) limits of the
torus, each as a function of time, for simulations {\it Clos0ehi} and
{\it Wide3ehi}, as labeled.} 
\end{figure}

Figure \ref{fig:torus-ecc} shows the eccentricities as functions of
time, for the same fits shown in figure \ref{fig:torus-radii}. In both
simulations, and for all three fits, the tori quickly become
significantly eccentric. The eccentricities for simulation {\it
Clos0ehi} however, typically near $e=0.2-0.4$, fall well below those
for {\it Wide3ehi}, where they are found near $e=0.3-0.6$. The
difference is consistent with the fact that the torus is more strongly
perturbed in the wide binary configuration. For all three fits, the
eccentricities also rise to values typical of their longer term
evolution within only 1-2 orbits of the binary, but continue to vary
widely over the full duration of the simulation. This rise time is far
shorter than a single orbit of the torus' material defining the edge
itself, and demonstrates the importance of the stars on the torus
morphology. Consistent with its proximity to the perturbing effects of
the stars, the torus' inner edge suffers the largest variations from
its initial circular form. In the case of the wide orbit
configuration, the eccentricity undergoes repeated excursions to
values greater than $e=0.6$ and less than $e=0.1$, corresponding to a
nearly circular shape. Also, the variations appear to be nearly
periodic, perhaps arising as consequences of the binary's influence on
the inner torus. To the extent that they are quantifiable however, the
ocscillations do not correspond to the binary orbit period through any
simple relationship, arguing against any direct connection.

Similar variability is present in the fits derived from central torus
region, but the maxima do not extend to values as large as at the
inner edge. Also, in both simulations, the eccentricities of the
central ellipses remain near zero for long time spans, alternating
with periods in which their values are much higher. In contrast, both
the inner and outer ellipses typically retain their large
eccentricities during these periods. Direct examination of the
simulation morphology reveals that the systems have developed several,
more or less equally spaced, spiral structures at these times. In
consequence, though the non-axisymmetric character of the torus
remains, the best fit ellipse derived from it becomes more circular,
as the various features driving the fit to higher eccentricity cancel
each other out. Later, as the spiral structures evolve into more
asymmetric shapes, the eccentricity increases again.

The trend towards slower and smaller amplitude variations continues in
the eccentricity for the outer edge. For both configurations, the
outer ellipses require more time to increase to similar values as are
seen for the other two. In contrast to the inner and central ellipses,
the outer ellipses exhibit only comparatively small excursions away
from their long term averages and the variations occur most slowly
with time. They also exhibit none of the periods of near circular
shapes (low eccentricities) seen for the other two. Even at the end of
the simulations, the long term trend values continue to increase,
meaning that the outer edge has yet to evolve to a configuration for
which all artifacts in the results due to the initial conditions have
been fully erased.

\subsubsection{The orientation of the torus}\label{sec:torus-w}

\begin{figure}
\includegraphics[angle=0,scale=0.80]{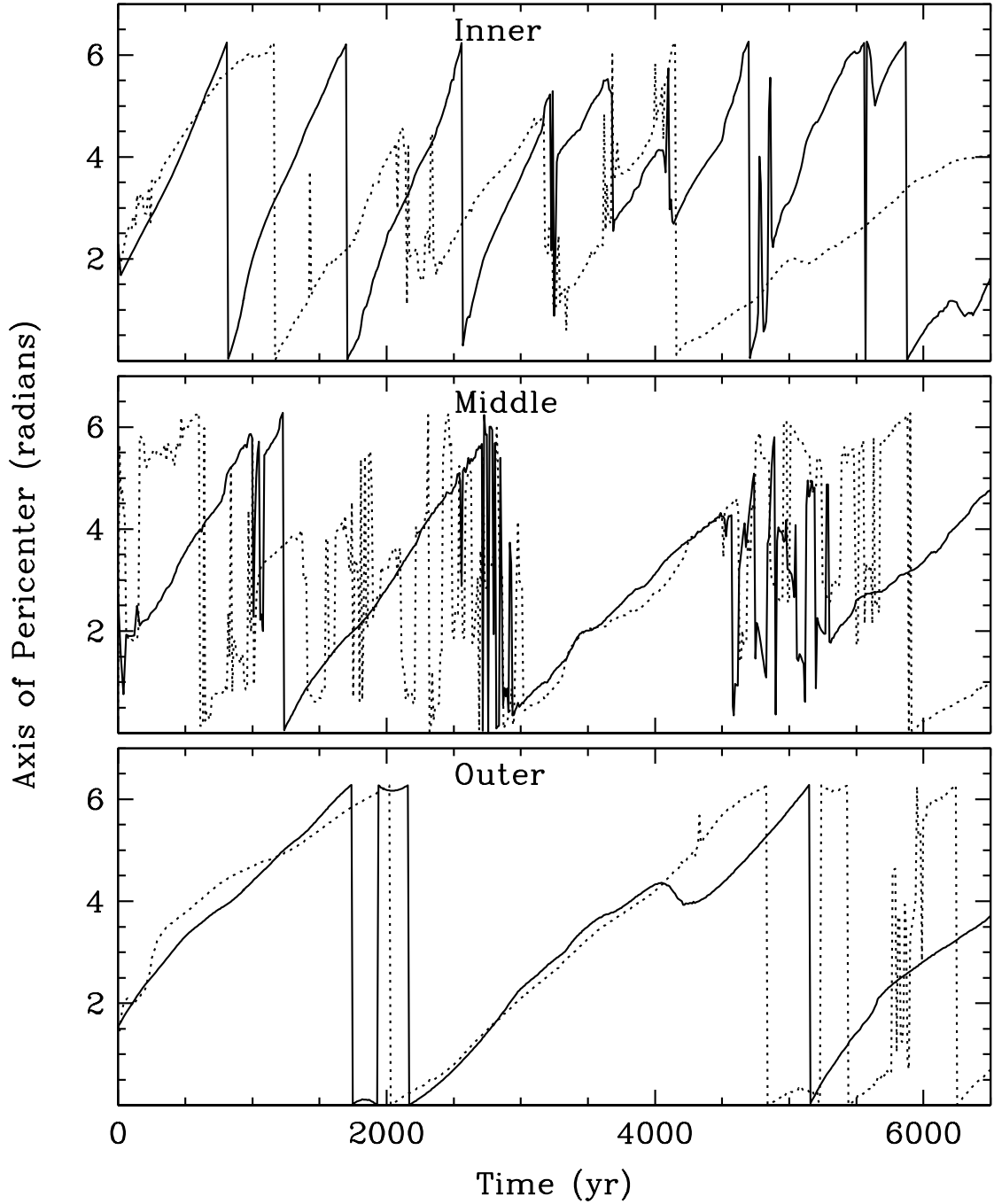}
\caption{\label{fig:torus-omega}
The axis of pericenter of the three ellipses described in the text,
defining the inner, middle, and outer limits of the torus as labeled.
Simulation {\it Wide3ehi} is shown with the solid curves, while simulation
{\it Clos0ehi} is shown with the dotted curves. } 
\end{figure}

Figure \ref{fig:torus-omega} shows the axis of pericenter angles
fitted for the tori over time. By definition, this parameter specifies
the angle along which the long axis of the ellipse points, with a zero
angle being defined here as the long axis of the stellar orbit
ellipse, for the eccentric binary case, with the same zero point used
for the for the circular binary simulation as well. In the case of the
eccentric binary configuration, we might expect the torus to remain
aligned along that same axis, since the binary's orbital motion
implies a strong forcing function to facilitate such alignment. This
expectation is clearly not met in our simulations. Instead, the tori
around both the eccentric and circular orbit binary appear to precess
over time, as indicated by their continuing increases in angle and
their periodic returns to zero as the motion sweeps around the
compass. 

The tori precess at different rates depending on distance from the
system barycenter, with both simulations producing similar rates for
the fits at the middle and outer edge of the torus, as measured by the
similar slopes of the curves and the time required to return to a
given orientation. In contrast, the precession behaviors for the inner
ellipses in the two simulations are not similar to each other.
Instead, the precession in the eccentric binary run corresponds
approximately to a periodicity of twice the binary orbit period (i.e.
$\sim860$~yr), over the course of many binary orbits, while for the
circular binary run, it is much longer. We speculate that the
precession in the first case is being driven by the forcing action of
the binary. In the circular binary case, and for the center and outer
tori, the influence of the stars on the tori is much weaker. In these
regions, the precession periods correspond loosely to the orbital
periods of the tori at each location, with the exception discussed
below for the center torus.

Although the correspondence itself is apparent, we do not believe a
direct, causal connection between the precession rates and the orbital
motion of the torus material can be established however because, in
general, the precession is clearly only approximately periodic. The
axis of pericenter angle neither changes at a uniform rate, nor does
it increase monotonically with time. Instead, occasional reversals
occur, as the fit adapts to details of the local flow, and sudden
discontinuities in the angle are also common, as the fit adjusts to
the growth or decay of some structure in the torus. Such events are
visible at $t\approx5800$~yr in fit for the outer torus edge in the
{\it Clos0ehi} simulation, for example, and much more frequently in
the center and inner torus fits. Jumps are particularly common in the
central torus, and occur at the same times for which we noted above
that the fitted eccentricity hovers near zero. Given this
correspondence, we attribute the jumps there to the same cause: i.e.
that the torus has evolved into a set of spiral structures of near
equal spacing and appearance, causing the fitted axis of pericenter to
become highly sensitive to small scale features of the torus as a
result.

Of some interest is that the phase of the precession in one simulation
is frequently similar to that in the other--the pericenter angles are
nearly the same at the same times for the outer and middle tori. We
believe that the similarity is not driven by any physical condition
common to the two configurations, but is rather a consequence of the
identical initial condition specified for the torus, which has not yet
been erased by its subsequent evolution. The inner torus edge follows
the same pattern as well, both for the correlation in angle and in
periodicity, but the correlation remains close for less than a single
orbit period there, before being lost. 

\section{Evolution of the Circumstellar Environment}\label{sec:evo-circumstar}

Observations of \ggtaua\ have conclusively demonstrated that a
significant quantity of material is present in very close proximity to
the stars themselves. Recent observations have even detected material
in locations where its dynamical lifetime cannot be exceed a few
hundred years \citep{beck12} before being accreted into the
circumstellar disks or otherwise cleared from the system. In this
work, we have assumed that all circumstellar material is initially
contained in circumstellar disks orbiting each star, and in this
section, we discuss the evolution of that material as it evolves. We
will investigate the processes by which material is transfered into
the circumstellar environment from the circumbinary environment, and
what configurations the cirumstellar disks evolve towards over time as
material moves through them into and out of them. 

\subsection{The configuration of the gap and the circumstellar
disks}\label{sec:circstar-conf}

\begin{figure*}
\includegraphics[angle=0,scale=0.87]{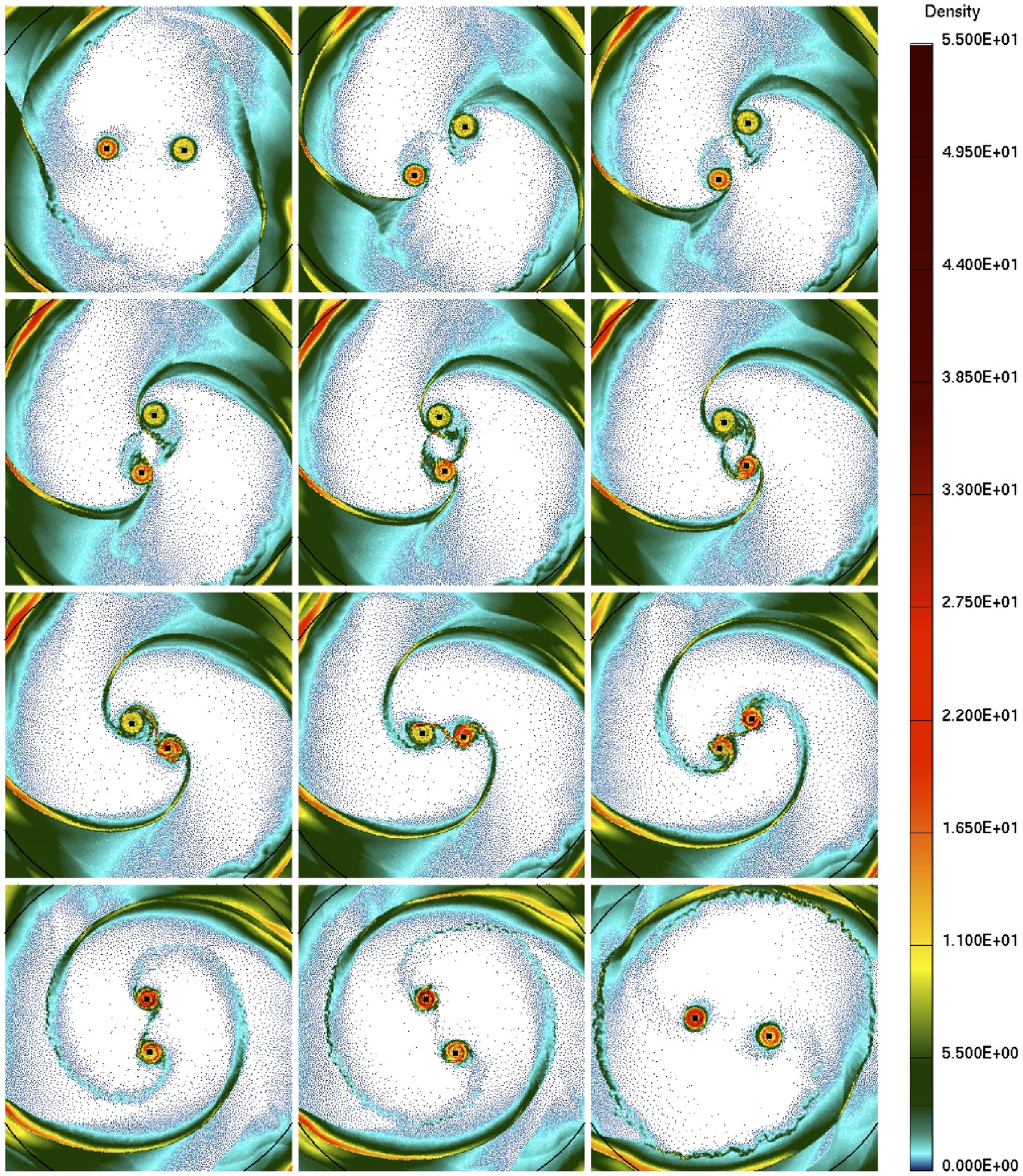}
\caption{\label{fig:wide-ecc-1orb}
The morphology of the circumstellar disks and the inner torus/gap
region at successive points along the eighth orbit of the binary from
the start of simulation {\it Wide3ehi}, corresponding to a time
$\sim3500$~yr after the beginning of the run. Each frame defines a
300$\times$300~AU region centered on the system center of mass. Time
increases in successive panels from the top left, to the lower right,
with the first frame chosen to be near apoapse of the orbit. The
primary and secondary are initially located on the right and left
sides of first frame, respectively, and their orbital motion is in a
counter-clockwise sense.}
\end{figure*}

The morphology of a system with a wide, eccentric orbit (simulation
{\it Wide3ehi}) is shown in figure \ref{fig:wide-ecc-1orb} for various
phases of its evolution through one binary orbit period. Details of
the orbital evolution vary between orbits, and we have chosen to show
the eighth orbit after the simulation's beginning (i.e. beginning at
$t\approx3400$~yr) because it exhibits the largest magnitude episode
of mass accretion of any single orbit during the simulation (see
section \ref{sec:acc-onto-disks} below). It therefore provides the
most easily visible illustrations of similar events that occur at
other times, but which are of smaller magnitude. In addition, rather
than show frames equi-distant from each other in time, we have chosen
frames selected to highlight various aspects of the flow as the binary
travels through a complete orbit, referencing the time since last
apoapse for various time intervals in the orbit in the text.

In the first frame of the mosaic, where the components are at apoapse,
the disks show little evidence of internal structure such as spiral
density waves. Consistent with its higher mass (see section
\ref{sec:acc-onto-disks} below), the mass densities in the secondary
disk (to the left in the image), are higher than in the primary. The
shape of the cavity between the stellar components and the
circumbinary material is significantly non-`round' and includes
fragmentary spiral structures that are remnants of previous
interactions between the components. These fragments are traveling
outwards towards the circumbinary torus, which they will soon rejoin.

As the binary components begin another orbit, falling towards each
other, large streams of material are pulled away from the torus and
follow the stars on similar inward trajectories. Due to its orbital
motion, the inbound material forms large scale spiral structures as it
travels through the gap region and does not directly impact either
circumstellar disk. By the time the stars have reached their 1/4 phase
(frames 2-4, some 80-140~yr after apoapse), the leading edges of the
spiral structures have overtaken the stellar components and wrapped
around them, with the close tidal interactions causing them to begin
to spread out again as they move away from the star. A portion of this
material may impact the other component of the binary, while the
remainder forms a temporary `bar' extending between the two binary
components. Material arriving slightly later (frames 4-7, 140-220~yr
after apoapse), does directly impact portions of the outer
circumstellar disks and the flow pattern becomes quite complex. 

\begin{figure*}
\includegraphics[angle=0,scale=0.62]{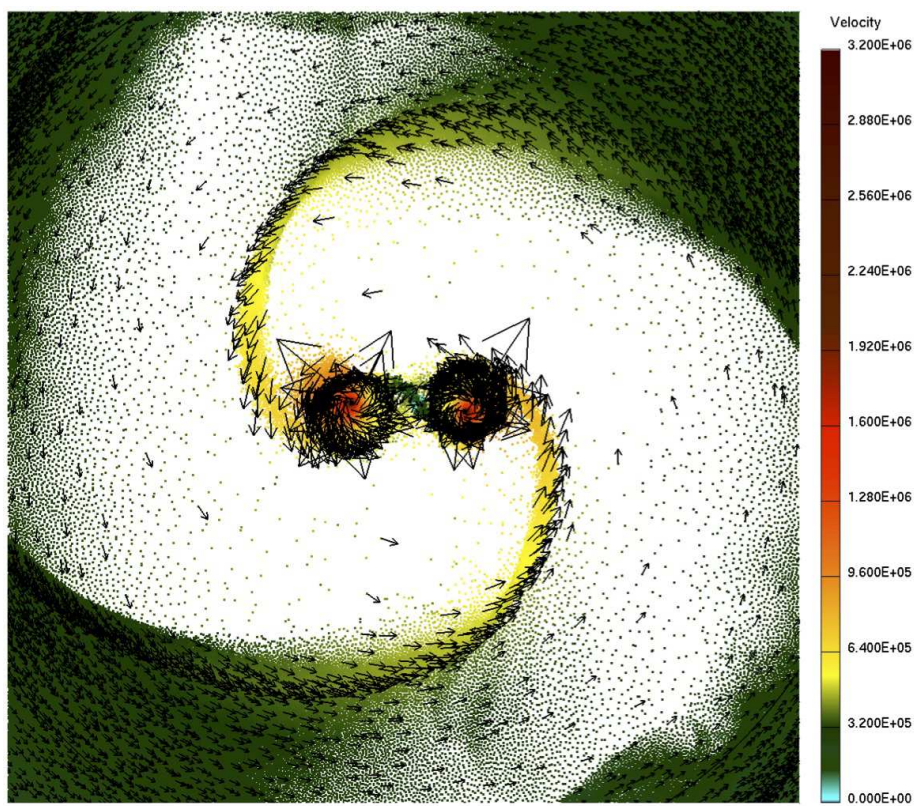}
\caption{\label{fig:velocity}
The velocity magnitude (in cm/s) relative to the system barycenter, at
the same time as the sixth panel show in figure
\ref{fig:wide-ecc-1orb}, for the same 300$\times$300~AU region
centered on the system center of mass. Vectors indicating direction
and relative magnitude scaled by an arbitrary constant factor is
attached to one out of every 64 particles. } 
\end{figure*}

The time interval shortly before and during periapse is,
unsurprisingly, the interval over which the most complex interactions
take place. The various interactions between the infalling streams,
the `bar' and disks cause substantial disruption in each. As a
consequence, the spatial extent of the circumstellar material
increases, with mass from all three sources contributing to the
extended disks for each star. The amount and distribution of
material in the accretion streams is near its maximum at this time as
well. It is noteworthy however, that only a small fraction of the
material in the streams actually accretes into the circumstellar
environment, with the remainder returning to the torus. Illustrating
this fact, figure \ref{fig:velocity} shows the velocity magnitude at
the same periapse passage shown in the sixth panel of the mosaic in
figure \ref{fig:wide-ecc-1orb}, along with velocity vectors on a small
set of representative particles as well. The latter show the motion of
the streams extending well around and away from both the stars and the
surrounding circumstellar material. Near the circumstellar disks,
velocity magnitudes in the streams increase to as much as 5-8~km/s,
comparable to the Keplerian motion of the material within the
circumstellar disks themselves, and well in excess of the 2-3~km/s
orbital velocities of the components themselves.

\begin{figure}
\includegraphics[angle=0,scale=0.90]{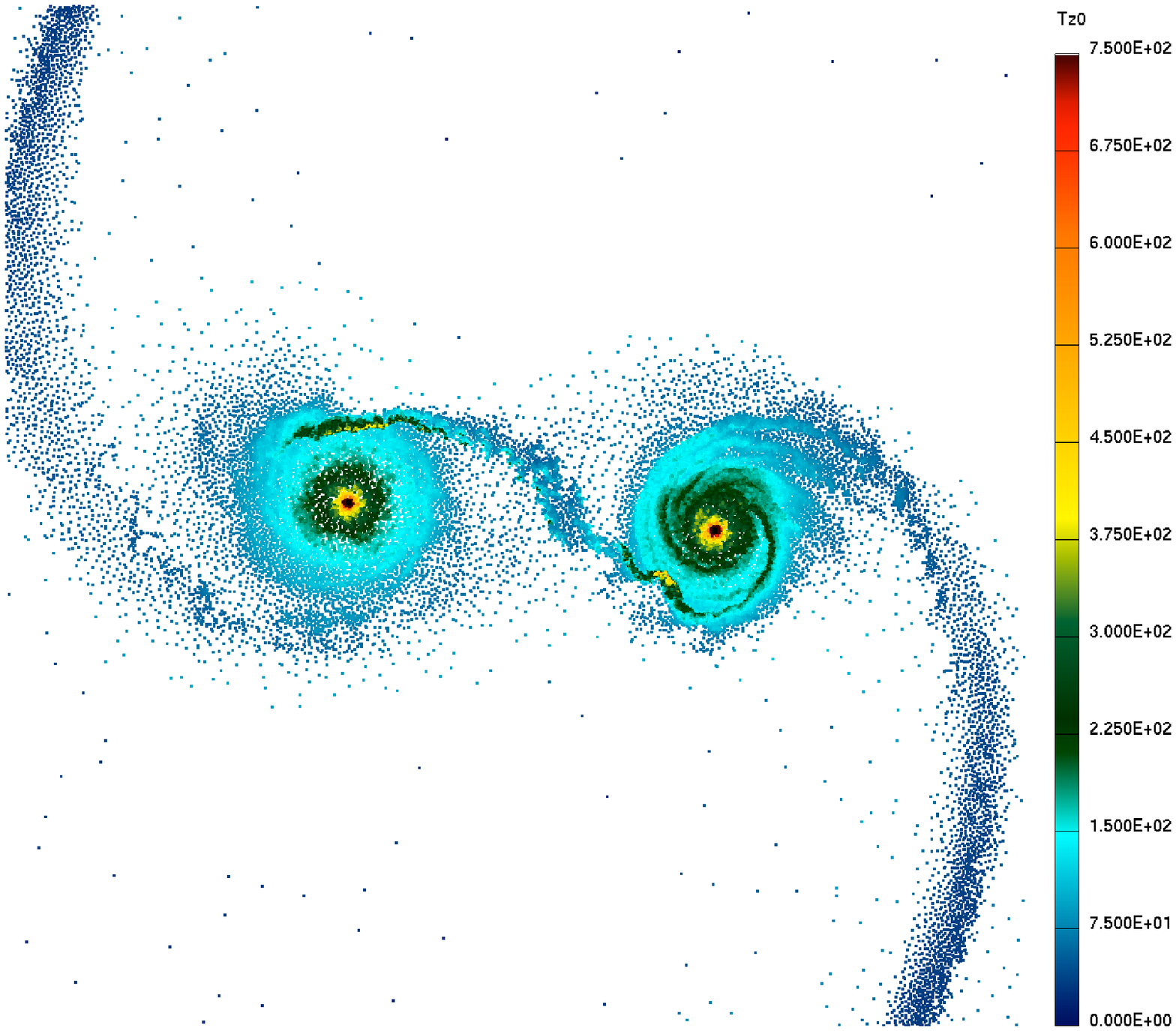}
\caption{\label{fig:disk-temps}
The midplane temperature distribution in the 120$\times$120~AU region
around the system barycenter at the time of the eighth periapse after
simulation {\it Wide3ehi} begins. The secondary is on the right.
Midplane temperature is determined for each SPH particle and at each
time using the procedure described in section \ref{sec:energy-diss}.}
\end{figure}

The high velocities generated as the streams move through the gap are
also important for conditions well within the circumstellar
environment itself. Figure \ref{fig:disk-temps} shows the midplane
temperatures of the material, again at periapse. While the accretion
streams themselves contain the coldest of all material in the
circumstellar environment, at temperatures below $\sim$50~K,
temperatures in small regions near the impact points of the streams
increase to 3-500~K as the kinetic energy of the infall is converted
to heat. These temperatures are well in excess of the $\sim150$~K
temperatures found in immediately adjacent regions of the disks, not
affected by the streams' impact. These $\sim150$~K temperatures at the
outer disk edges are the lowest found within the disks. At other
locations, temperatures increase to as high as 1000~K (extending to
values beyond the color bar range shown in the figure) within 1~AU of
each star. Non-azimuth symmetric structure is also visible,
particularly in the disk around the secondary in which temperatures
increase to as high as 300~K inside the spiral structures that
develop. We will discuss implications of these temperatures in section
\ref{sec:plan-form}, below.   

As the stars pass through periapse and begin to recede from each other
again (frames 7-9 of figure \ref{fig:wide-ecc-1orb}, 200-240~yr after
apoapse), the extended disk begins to shrink as the material it
contains begins to reintegrate into the main volume of the
circumstellar disks. At the same time, the disk structures closer to
the stars become more inhomogeneous, due to their interactions with
the highly eccentric flows of the infalling material that disrupt
their Keplerian motion. The disk structures generated by the
interactions near periapse decay to near axisymmetry over the next
1-200~yr, as the stars recede from each other and again approach
apoapse (frames 8-12, 250-420~yr after apoapse). 

During these later phases of the orbit, and at larger distances from
the stars, material in the tidal streams which was able to pass by
both circumstellar regions without impacting either, continues on
trajectories which returns it to the circumbinary torus from which it
came. As they travel through the gap, the spiral structures into which
material is organized begin to lose the sharp definition that
characterizes their appearance in the earlier phases of the binary
orbit. Such loss of coherence is most noticable in the innermost
portions of the streams, for which tidal interactions with the binary
were largest. Nevertheless, such structures remain coherent well after
their return to the main volume of the torus and, as discussed in
section \ref{sec:evo-circumbin} above, their existence in and
influence on the torus are of major importance for the overall density
and temperature structure well beyond its inner edge.

\begin{figure*}
\includegraphics[angle=0,scale=0.87]{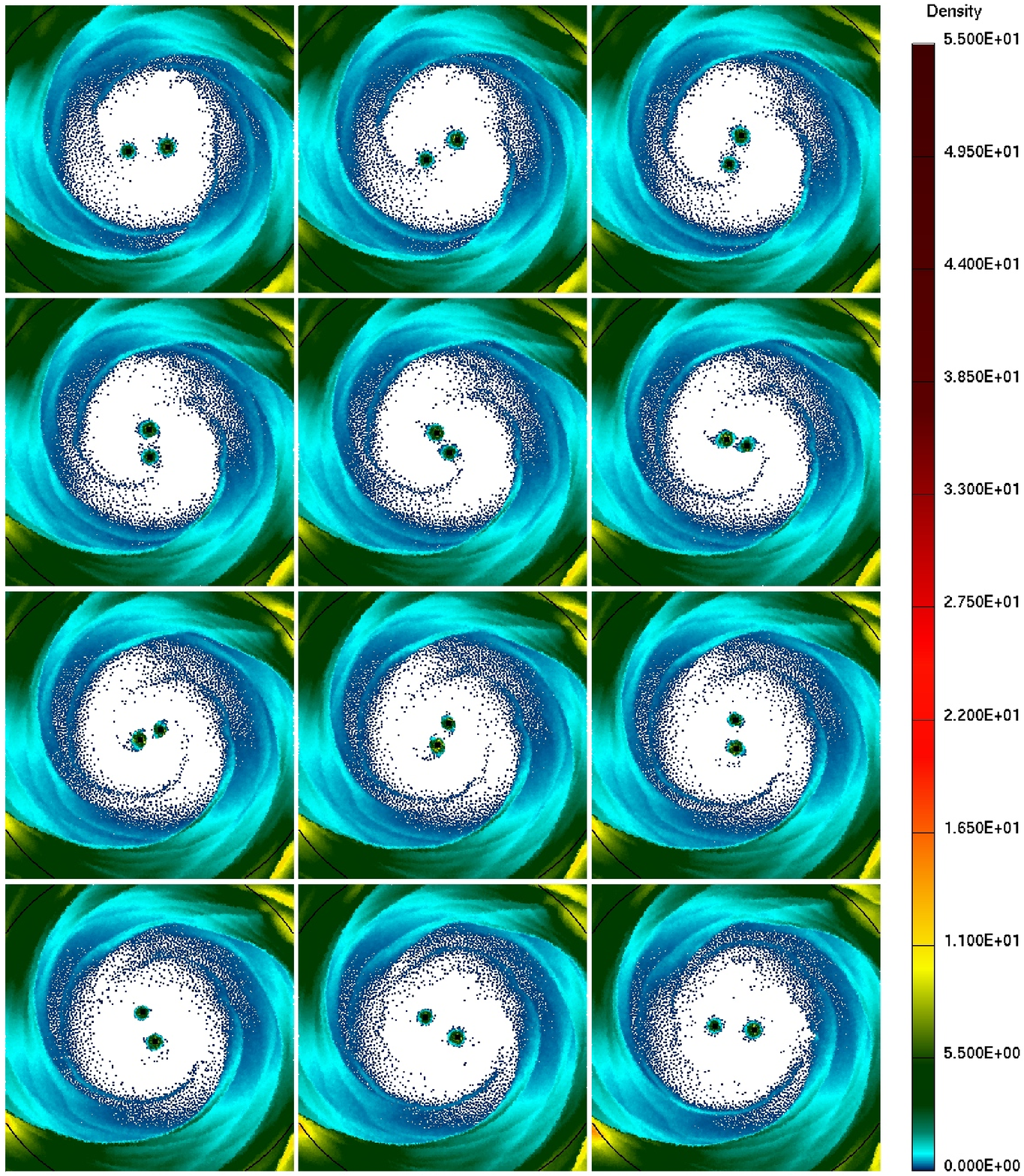}
\caption{\label{fig:clos-ecc-1orb}
The morphology of the circumstellar disks and the inner torus/gap
region at successive points along the 35th orbit of the binary
from the start of simulation {\it Clos3ehi}, corresponding to 
$\sim5500$~yr after the beginning of the run. Each frame defines a
300$\times$300~AU region centered on the system center of mass. Time
increases in successive panels from the top left, to the lower right.
The primary and secondary are initially located on the right and left
sides of first frame, respectively, and their orbital motion is in a
counter-clockwise sense.} 
\end{figure*}

A markedly different pattern of behavior occurs in the case of a
$a=32$~AU orbit for the binary. Figure \ref{fig:clos-ecc-1orb} shows
the evolution of the {\it Clos3ehi} simulation over the course of one
binary orbit. In comparison to the wide orbit case, a much larger
fraction of torus material has moved inwards into what was originally
the outer gap region. Despite this, and in contrast to the behavior
seen with the wide orbit model, the remaining gap region continues to
be clear of material over the entire binary orbit. Accretion streams
do develop, but are quite weak and remain distinct for only a fraction
of the distance inward to the stars. No bar develops between the
stars at periapse because the weaker accretion streams are unable to
feed the circumstellar region with enough material to produce one.
Also, due to the closer orbital separation, the circumstellar disks
are spatially more compact than in the wide orbit case and assymetries
that develop within them are also correspondingly less distinct. 

With small variations, the same behaviors appear during each binary
orbit in both the wide and close cases. We have not shown the $e=0$
models corresponding to the eccentric orbit models described here,
however we note that their behaviors largely mimic those of their
eccentric cousins. An important exception to this trend is the fact
that only the wide, eccentric model in figure \ref{fig:wide-ecc-1orb}
shows the strong accretion streams and episodic accretion. 

\subsection{Mass transfer into and out of the circumstellar
disks}\label{sec:masstrans}

We have seen that as the system evolves through time, the action of
the binary generates large amplitude density structures throughout the
circumbinary torus, extending outwards into the low density disk that
surrounds it. Likewise, similar structures propagate inwards from the
torus and, at various orbital phases, contact the circumstellar disks.
Some of the mass in these streams becomes a permanent part of the
circumstellar environment, while the remainder propagates outwards to
rejoin the torus again. In addition to mass accretion from the
circumbinary environment, mass contained in the circumstellar disks is
subject to the action of the same strong gravitational torques that
perturb the circumbinary material. It may therefore migrate radially
inward towards the star and, when it gets close enough, be accreted by
the star. 

In this section, we quantify the rates and timing of the accretion
into and out of the cirumstellar environment, and the masses of the
disks that result. As one aspect of our discussion, we will describe
simulations of the same physical configuration at two different
resolutions, differing by a factor of four in particle count. Through
the relationship in eq. \ref{eq:Balsara}, this difference will be
sufficient to change the numerical viscosity by a factor of $\sim2$. A
difference of this magnitude will in turn be sufficient to change the
accretion by a similar factor, if indeed resolution plays a
significant role in the accretion processes we model. We will discuss
such issues in section \ref{sec:stellar-env}.

\subsubsection{Accretion onto the circumstellar
disks}\label{sec:acc-onto-disks}

\begin{figure*}
\includegraphics[angle=-90,scale=0.65]{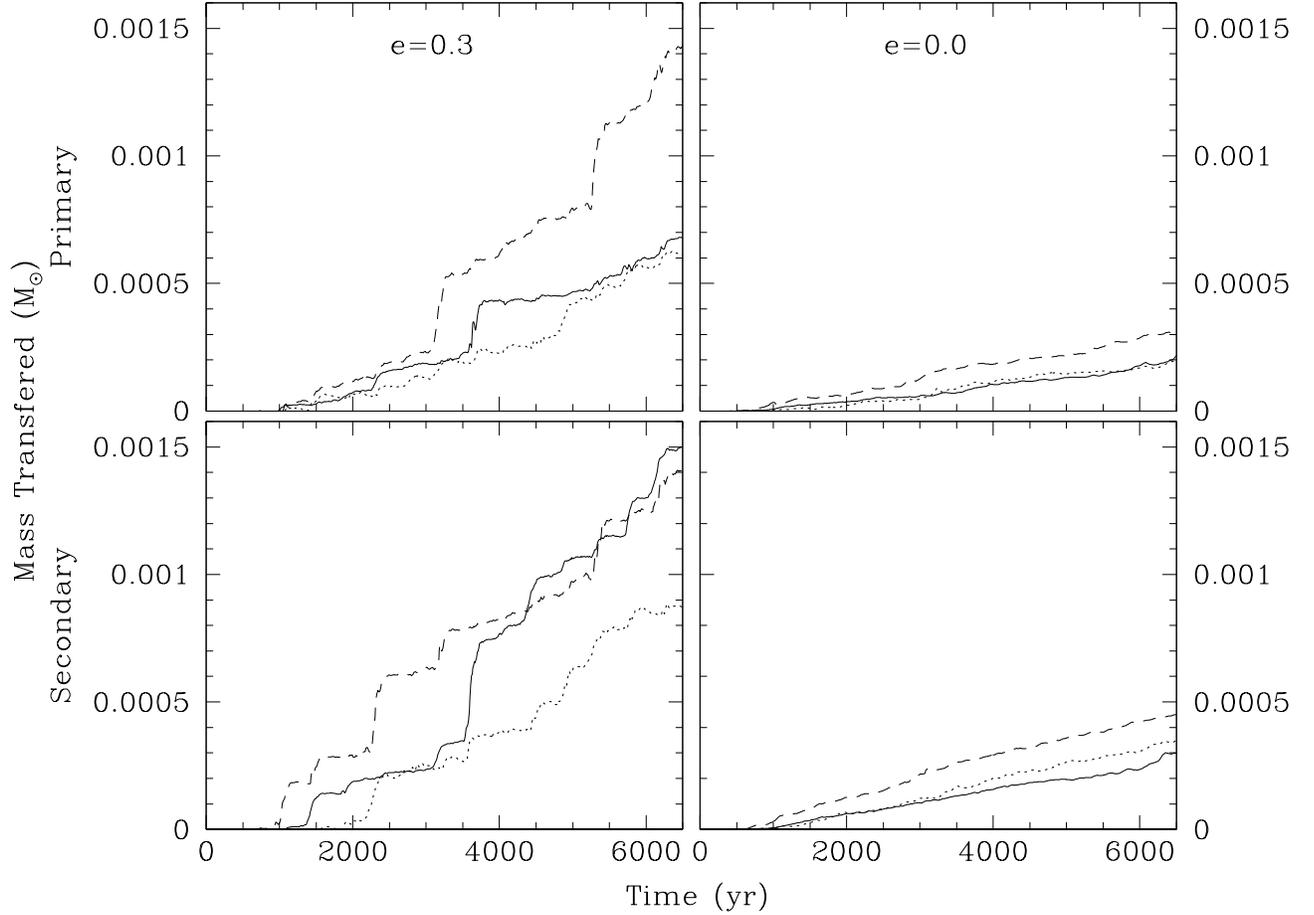}
\caption{\label{fig:masstransW}
The mass transfered into the circumstellar environment of the primary
(upper panels) and secondary (lower panels) stars in the binary system
with orbital separation $a=62$~AU. The left panels show the
simulations with assumed orbital eccentricity of $e=0.3$, and the
right panels show simulations with $e=0$. The solid, dotted and dashed
curves define the mass transfered for the high resolution torus/disk
system, its low resolution counterpart and the high resolution torus
only simulations, respectively.}. 
\end{figure*}

\begin{figure*}
\includegraphics[angle=-90,scale=0.65]{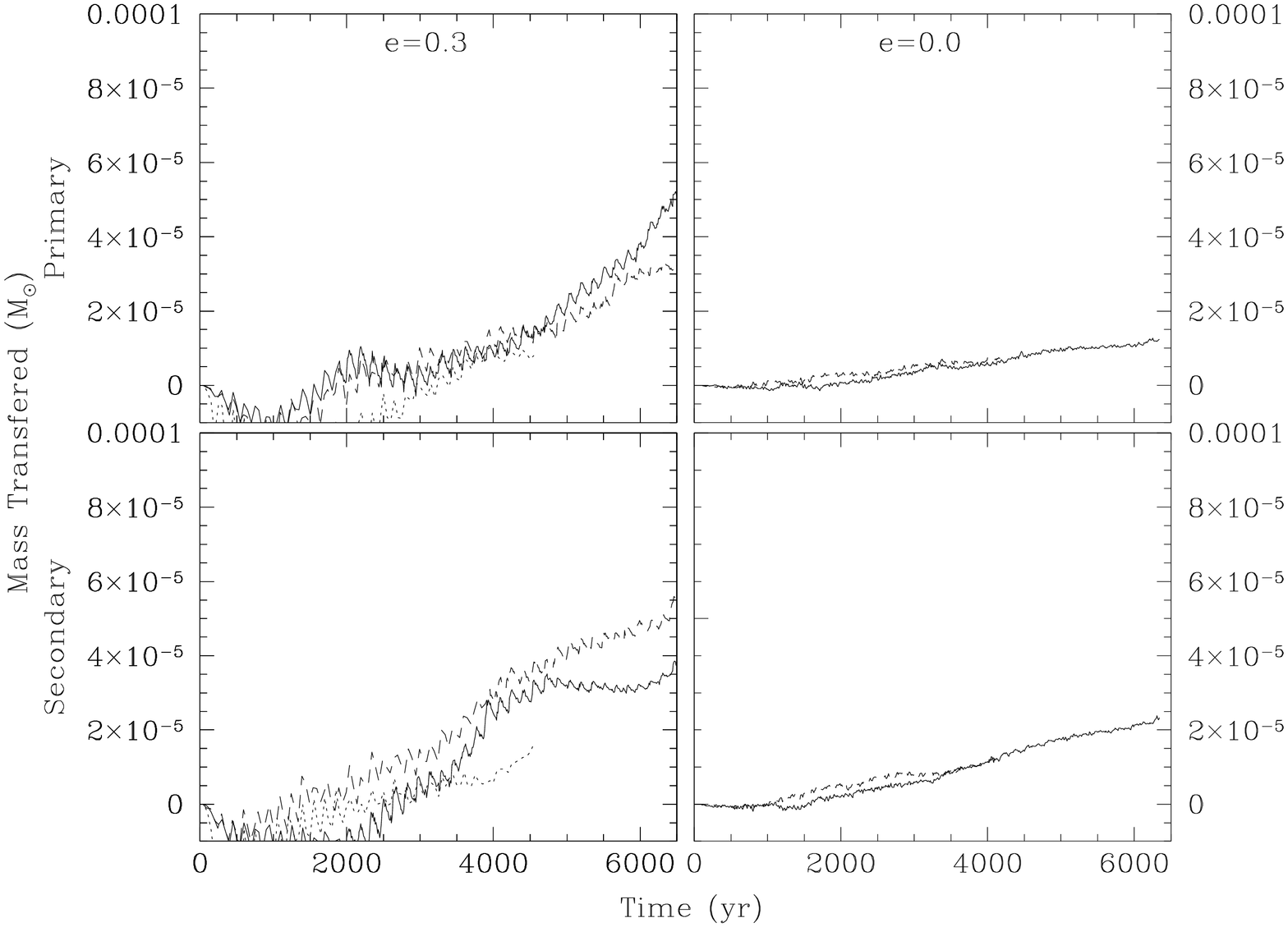}
\caption{\label{fig:masstransC}
Same as figure \ref{fig:masstransC}, but for the simulations with
orbital separation $a=32$~AU.}
\end{figure*}

Figures \ref{fig:masstransW} and \ref{fig:masstransC} show the mass
accreted by each of the circumstellar disks as a function of time for
the various simulations in our study. For purposes of defining the
``circumstellar disks'' for the primary and secondary stars, we define
an outer boundary to that environment to be 50\% larger than the
radius defined as the last stable orbit by \citet{HolWie99}. For each
of the wide binary simulations, the total mass transfered into each of
the circumstellar disks over the 6500~yr duration of the runs is
comparable in magnitude to the assumed initial disk masses themselves.
Accretion occurs preferentially onto the disk around the secondary
star for both the wide and close systems, and for both eccentric and
circular orbits. The torus-only models generate higher accretion rates
than do the torus/disk models, presumably because the higher mass
concentration in the inner circumbinary region is more readily
available for transfer inward.

For the close binaries, mass transfer occurs at rates some $\sim20$
times lower than occurs for corresponding simulations with wide
orbits, so that only a small fraction of the initial disk masses are
replaced over the lifetimes of the simulations. At early times in
these simulations, mass actually appears to be transfered out of the
circumstellar disks. This is, however, merely a consequence of the
initial spreading of the disks away from their initial condition and
the artificial limit we imposed to define the disk edge. The accretion
rates in these configurations are low enough to show small, apparent
oscillations in the calculated disk masses for the eccentric orbit
case (fig \ref{fig:masstransC}). In fact, these oscillations are more
likely due to the changes in the disk's spatial dimensions at
different phases of its orbit. Such `flexing' also causes a small
fraction of the disk mass to migrate outwards temporarily, beyond the
limiting radius we have defined for the disk edge, before moving again
into closer proximity to its host star.

Eccentric orbital motion enhances the time averaged accretion rates
over those observed for circular orbit systems, with rates a factor of
$\sim2-3$ higher in an eccentric orbit system than in an otherwise
identical circular orbit system. In addition, for the eccentric wide
binaries, accretion is highly episodic, with substantial mass transfer
occuring in bursts some 30-40~yr in duration and relatively little
transfer occuring at other times. These bursts reflect the periodic
gravitational perturbations of the torus by the stars as they travel
closer and further from its inner edge and excite spiral structures
there. As described in sections \ref{sec:wide-morph} and
\ref{sec:circstar-conf}, spiral structures generated in the
circumbinary disk by the action of the binary fall through the
otherwise cleared gap region outside the binary orbit. As they travel
through the circumstellar environment, a portion of their length may
encounter one or the other circumstellar disks and by accreted by it.
While similar features form in the simulations with $a=32$~AU (albeit
much weaker), the trajectories of the spiral structures through the
gap region do not intersect with the circumstellar disks due both to
the smaller spatial extent of the disks and to their greater distance
from the torus itself.

Due to the evident variability in the magnitude of the accretion
events from orbit to orbit, we have not attempted a detailed analysis
of their exact timing relative to the binary's orbital phase. We note
however that the bursts do vary from orbit to orbit, as does the
instantaneous accretion rate and total mass accreted during one event.
Typically, the events begin at orbital phases prior to orbital
periapse and reach their maximum intensity near periapse, declining
again as the stars recede from each other. The timing does not appear
to be strongly related to the binary motion itself, but rather to the
time required for material originating outside the circumstellar
environments to travel inwards far enough to be accreted into one or
the other circumstellar disk. 

\subsubsection{Accretion onto the stars}\label{sec:acc-onto-stars}

\begin{figure}
\includegraphics[angle=-90,scale=0.65]{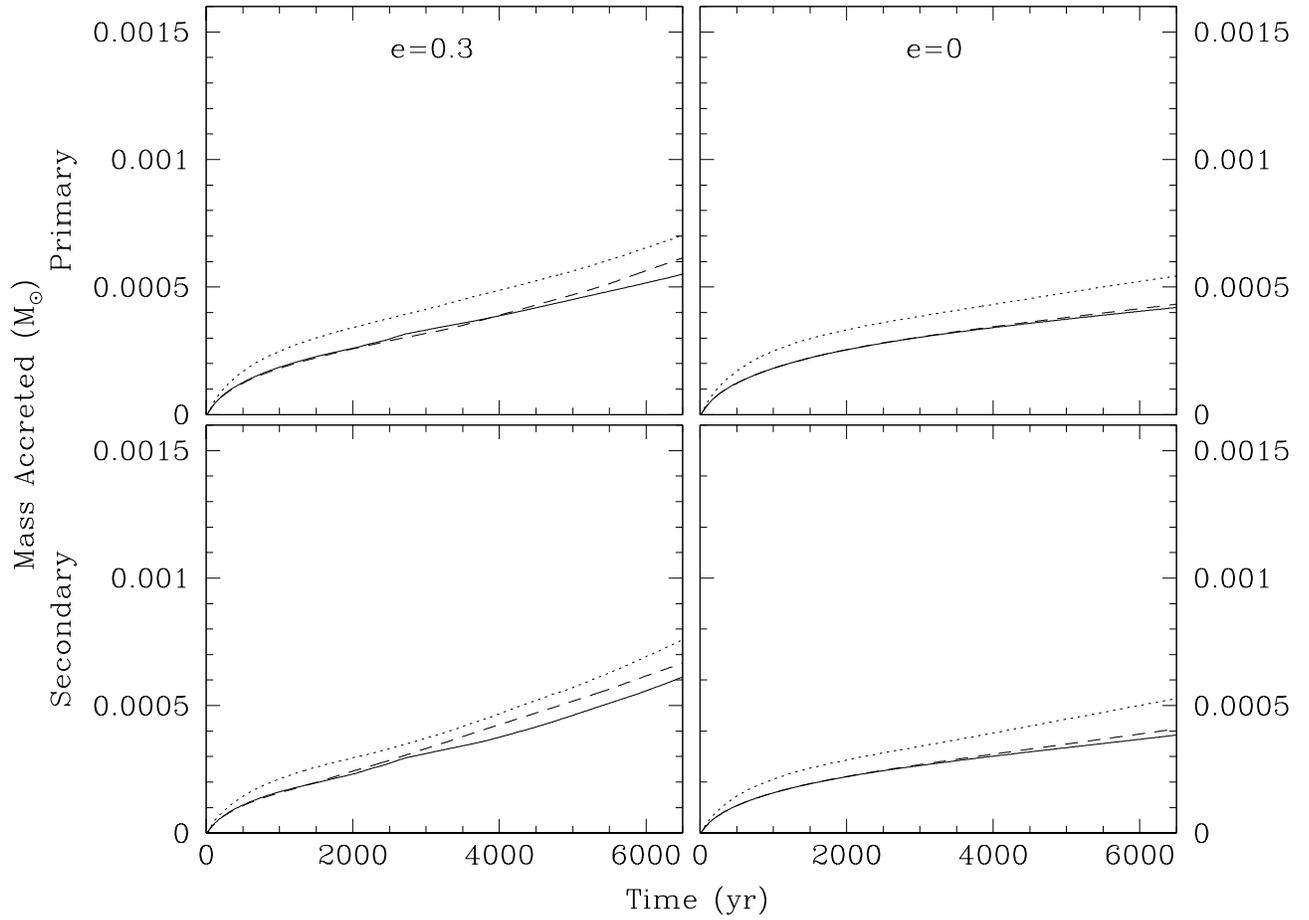}
\caption{\label{fig:massaccW}
Total mass accreted by the primary (top) and secondary (bottom) stars
for the series of models with binary separation $a=62$~AU and
eccentricity $e=0.3$ (left) and $e=0$ (right). The solid, dotted and
dashed curves correspond to the high and low resolution disk$+$torus
models and to the high resolution torus-only models, respectively.}
\end{figure}

\begin{figure*}
\includegraphics[angle=-90,scale=0.65]{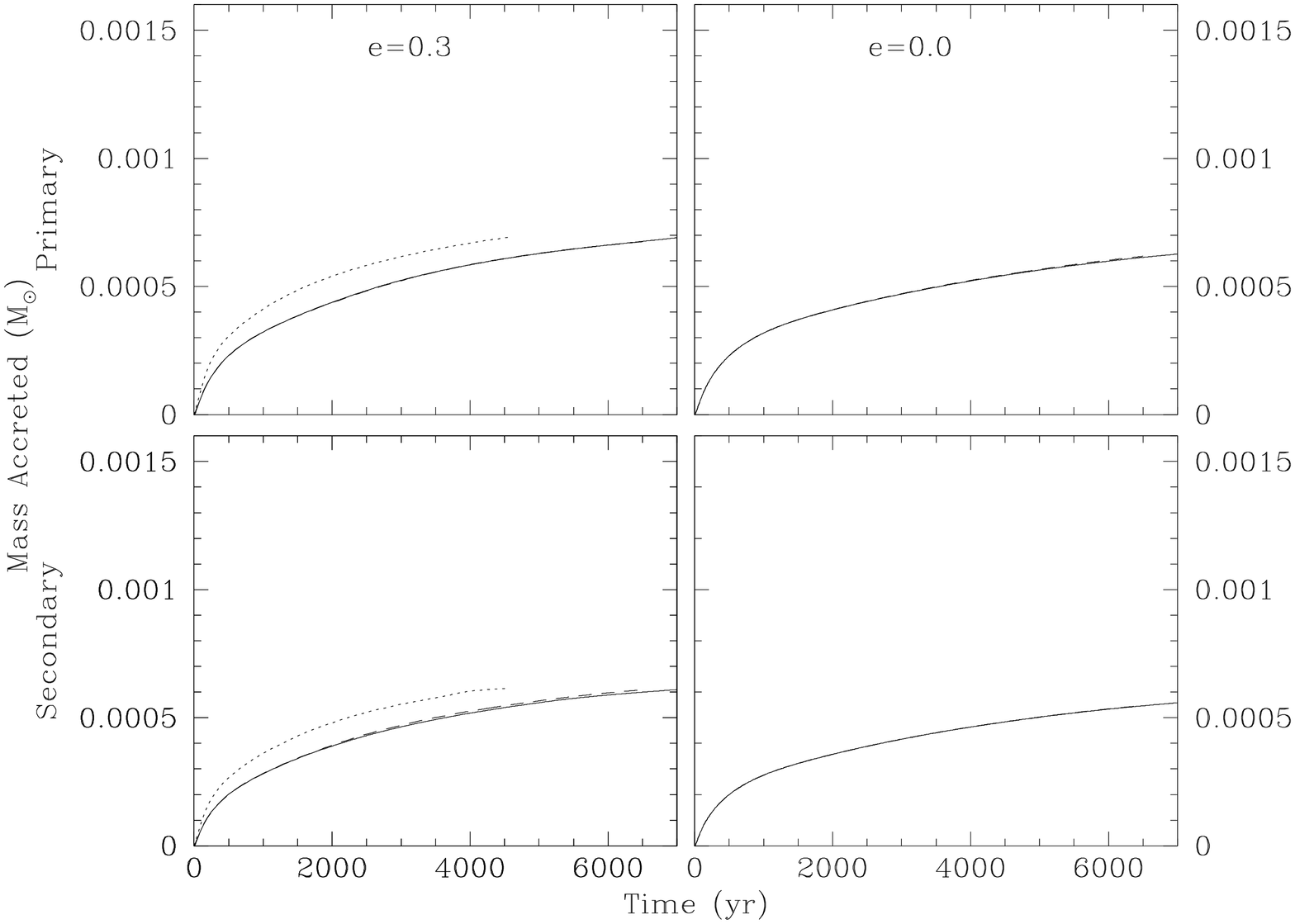}
\caption{\label{fig:massaccC}
Same as figure \ref{fig:massaccW}, but for the $a=32$~AU series of
simulations.}
\end{figure*}

In addition to mass transfer into the circumstellar environment from
the torus, mass leaves the circumstellar disks and accretes onto the
stars. Figures \ref{fig:massaccW} and \ref{fig:massaccC} show the
total mass accreted onto the stars from their disks as functions of
time, for the wide and close series of simulations, respectively. In
contrast to the episodic accretion patterns of mass entering the
circumstellar disks, accretion onto the stars exhibits no
periodicities or outburts, apart from an initial transient period
during which the disk material loses its memory of its initial
condition and moves towards a more physically appropriate
distribution.

For the wide binary simulations, accretion rates after the initial
transient fall between $6-9\times10^{-8}$\msun/yr for the $e=0.3$
series, and approximately half those rates for the $e=0$ series. The
eccentric orbit models each exhibit a slight increase in rates at late
times, while the $e=0$ models remain steady at near constant rates.
Given the larger accretion rates {\it into} the circumstellar disks in
the eccentric orbit simulations, increases in the accretion rates {\it
out of} those same disks is a natural consequence as the larger
material volume evolves along the same evolutionary path. Exactly the
opposite effect is true of the close orbit models. In those cases, the
initially rapid mass accretion transient drains a substantial fraction
of the mass from the disks. Because this material is not replaced as
rapidly (see section \ref{sec:acc-onto-disks}, above), the disks
become depleted of material (see section \ref{sec:diskmasses} below)
and the accretion onto the stars continues to slow down for the entire
duration of the simulations.

In all of our simulations, the resolution of the circumstellar disks
is extremely coarse. Of the nearly $1.9$ million particles employed in
our high resolution simulations, only about 20000 are initially
allocated to the circumstellar disks and, as we shall see in section
\ref{sec:diskmasses} below, that number does not increase substantially
over the life of the runs. Because of this low resolution, the viscous
dissipation originating from numerical sources is quite large. In
order to ensure that our conclusions regarding the accretion rates
through the disk are valid, we must therefore demonstrate that those
accretion rates are not sensitive to the numerical viscosity inherent
in the simulations themselves rather than to physical processes
present in the actual system we are modeling.

To this end, we ran identical physical models at two different
resolutions, and have included both the low and high resolution
results in figures \ref{fig:massaccW} and \ref{fig:massaccC}. In every
case, apart from a larger rate of accretion during the initial
transient, the cumulative mass accreted onto the stars for the low
resolution variant follows a curve with an essentially identical slope
to its high resolution counterpart. We conclude that our results are
not sensitive to the high numerical dissipation caused by the limited
resolution of the circumstellar environment. Our conclusions based on
these results are therefore also insensitive to the resolution.

\subsubsection{The disk masses as functions of
time}\label{sec:diskmasses}

\begin{figure}
\includegraphics[angle=-90,scale=0.65]{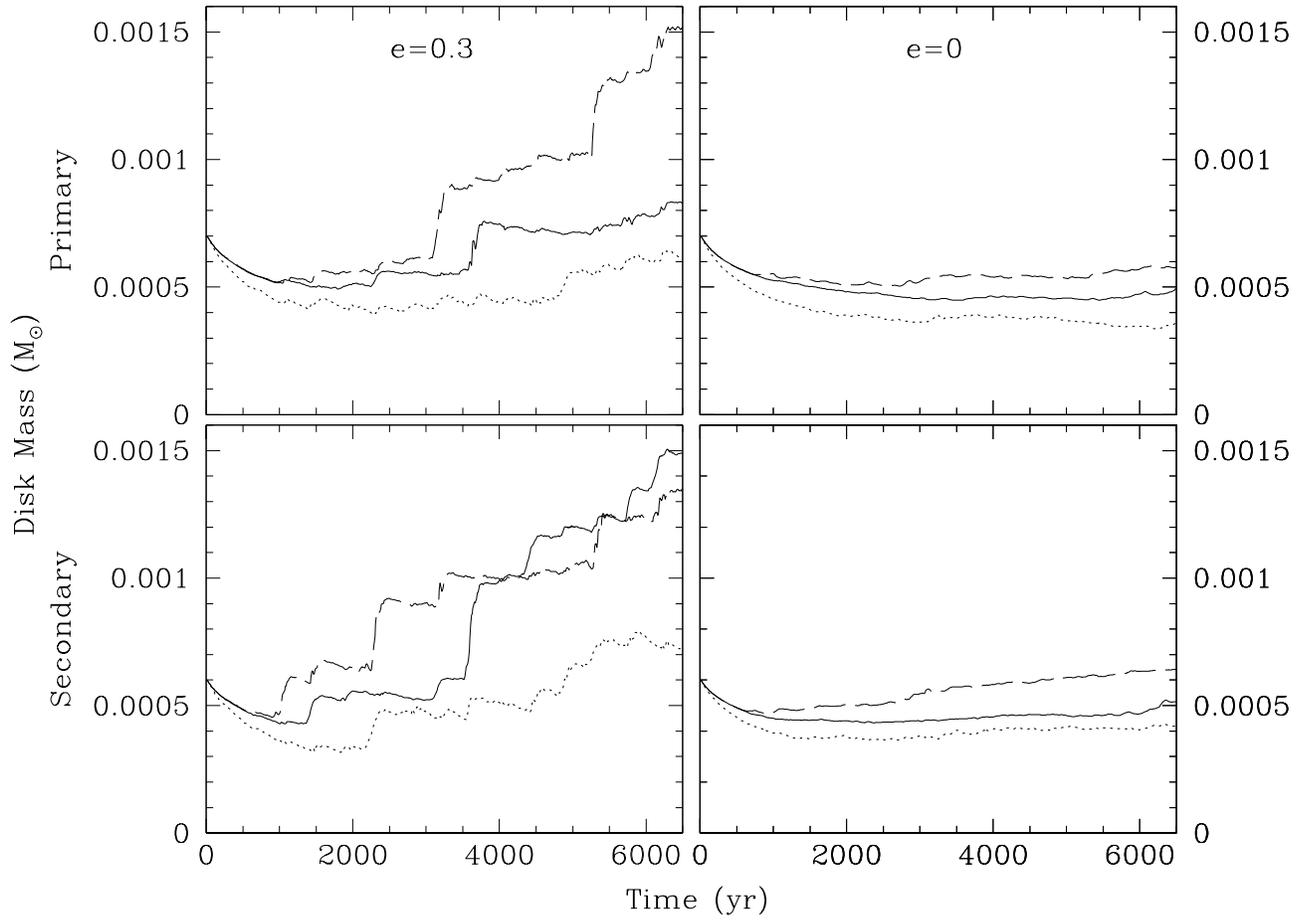}
\caption{\label{fig:massdiskW}
The masses of the circumprimary (top) and circumsecondary (bottom)
disks for the $e=0.3$ (left) and $e=0$ series of simulations with
$a=62$~AU. The solid, dotted and dashed curves correspond to the high
low resolution disk$+$torus models and to the high resolution
torus-only models, respectively.}
\end{figure}

\begin{figure}
\includegraphics[angle=-90,scale=0.65]{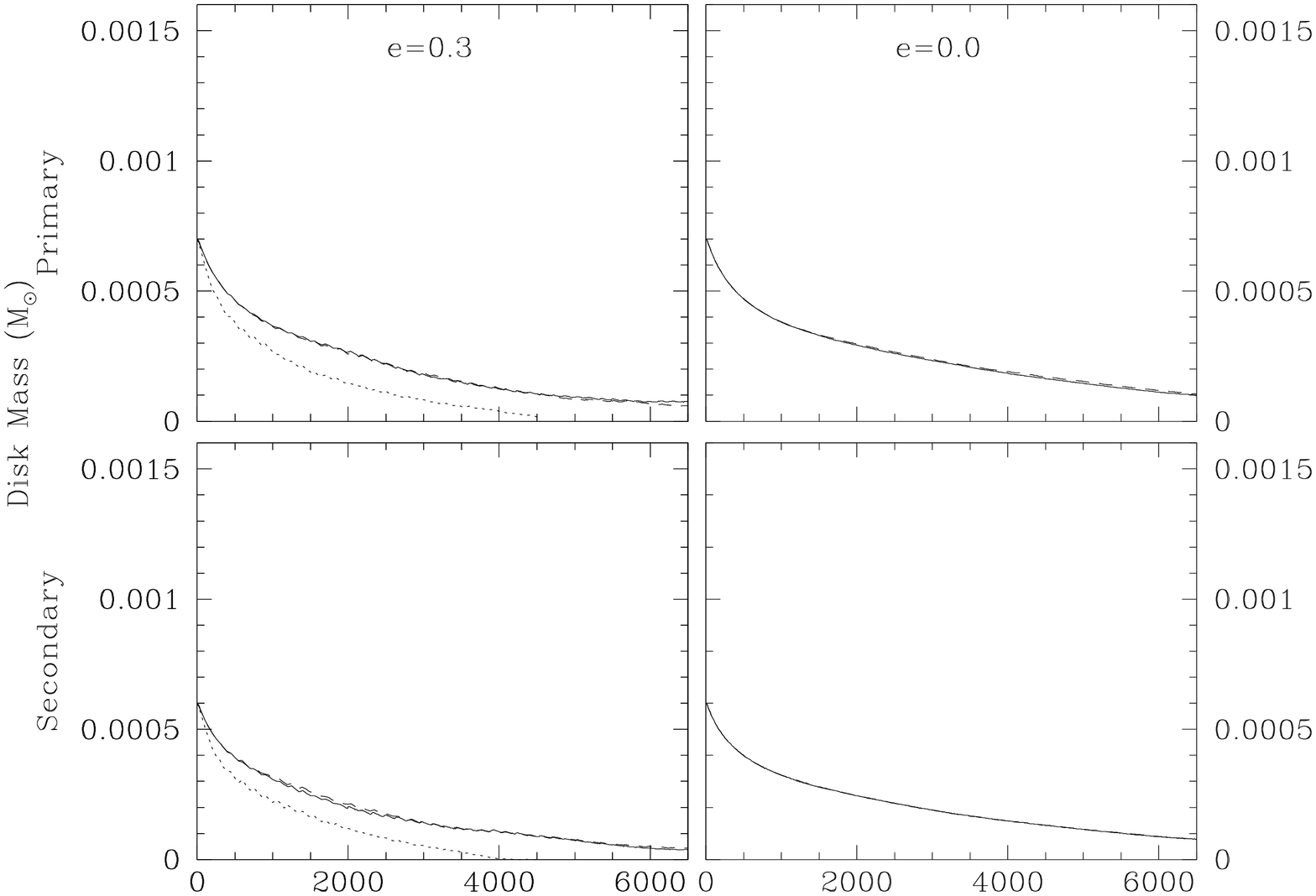}
\caption{\label{fig:massdiskC}
Same as figure \ref{fig:massdiskW}, but for simulations with
$a=32$~AU.}
\end{figure}

Although quantifying the mass accretion into and out of the
circumstellar disks as functions of time are important measures of the
system, by themselves they are incomplete because they do not specify
the mass contained in the disk at any given time. Figures
\ref{fig:massdiskW} and \ref{fig:massdiskC} show the mass of the disks
orbiting the primary and secondary stars as a function of time, for
the full set of wide and close simulations in our study. In each case,
the most significant changes in the disk masses over the first
$\sim1000-1500$~yr are due to the transient redistribution of disk
material as it loses memory of its initial condition. Thereafter, and
as the circumbinary material begins to exhibit activity, disks in the
wide, eccentric orbit models exhibit periodic accretion events, which
correspond to the orbital motion of the binary. In contrast, the disk
masses in simulations where the orbital eccentricity is zero settle
into steady states with little variation over time.  

For the close orbit models, the masses of the circumstellar disks
continues to decrease for both the $e=0.3$ and $e=0$ series of
simulations over the entire duration of the runs. After the 6500~yr
lifetime of the simulations, the masses of both circumstellar disks
has fallen to values an order of magnitude or more lower than their
initial masses. We anticipate that the disks would have been entirely
accreted within a comparatively short time of additional evolution.
Further evolution was not possible however, because the number of
particles remaining in the circumstellar disks decreased to such an
extent that instabliities developed in the time step controls for the
simulations (due to the poor resolution of the disks). Little forward
progress of the simulations was possible under such conditions, and we
therefore terminated them. Because of the larger mass accretion during
the initial transient, our lower resolution runs could only be run for
about two orbits of the circumbinary torus, much earlier than was
required for the higher resolution realizations. 

\section{Comparisons of features in our simulations with
observations}\label{sec:compare}

So far, we have described the behavior of simulations with initial
conditions specified to be very similar to \ggtaua, but we have not
yet discussed in any depth, direct comparisons of our results to it.
Other systems similar to \ggtaua\ also exist, to which the results of
our simulations may be suitably generalized as well. Of these, one of
the most similar is UY~Aur, with the recent discussions of
\citet{tang14} providing the most comprehensive description of the
system to date. The similarities between it and the \ggtaua\ system
are striking, both in the data themselves, and in the difficulties
experienced in trying to reconcile them with a single self-consistent
model of the system. Although the UY~Aur binary has a wider separation
and its disks much lower masses, both systems exhibit a remarkably
similar inventory of features otherwise. Such features include
structures in the circumbinary and circumstellar environments,
streamers and stellar orbit planes which may be misaligned with the
circumbinary disk if, as \citet{tang14} propose, the system is
actually a higher order multiple rather than a binary. Here, we
discuss comparisons of our results with observations of \ggtaua\ and
make a number of conclusions about the system and its evolution, with
an implicit understanding that the results will be applicable to a
larger class of hierarchical multiple systems of similar character.   

\subsection{Features in the circumbinary torus}\label{sec:torus-features}

The observed circumbinary torus of \ggtaua\ contains a number of
features \citep{krist02,krist05}, including spokes, gaps, variations
in brightness and torus width at different position angles along the
torus. Other work (e.g. GDS99), using high resolution interferometry,
has shown that the edges of the torus are no more than $\sim10$~AU in
radial extent. Despite these observations, detailed characterization
of the structures and attribution to specific causes is difficult
because the features themselves approach size scales comparable to the
resolution limits of the telescopes. While similar constraints apply
to our simulations, the spatial scale at which we are constrained is
far smaller, thereby permitting both comparisons to be made, as well
as, to some extent, predictions of types of features that may be
revealed by still higher resolution observations.

\begin{figure*}
\includegraphics[angle=0,scale=0.63]{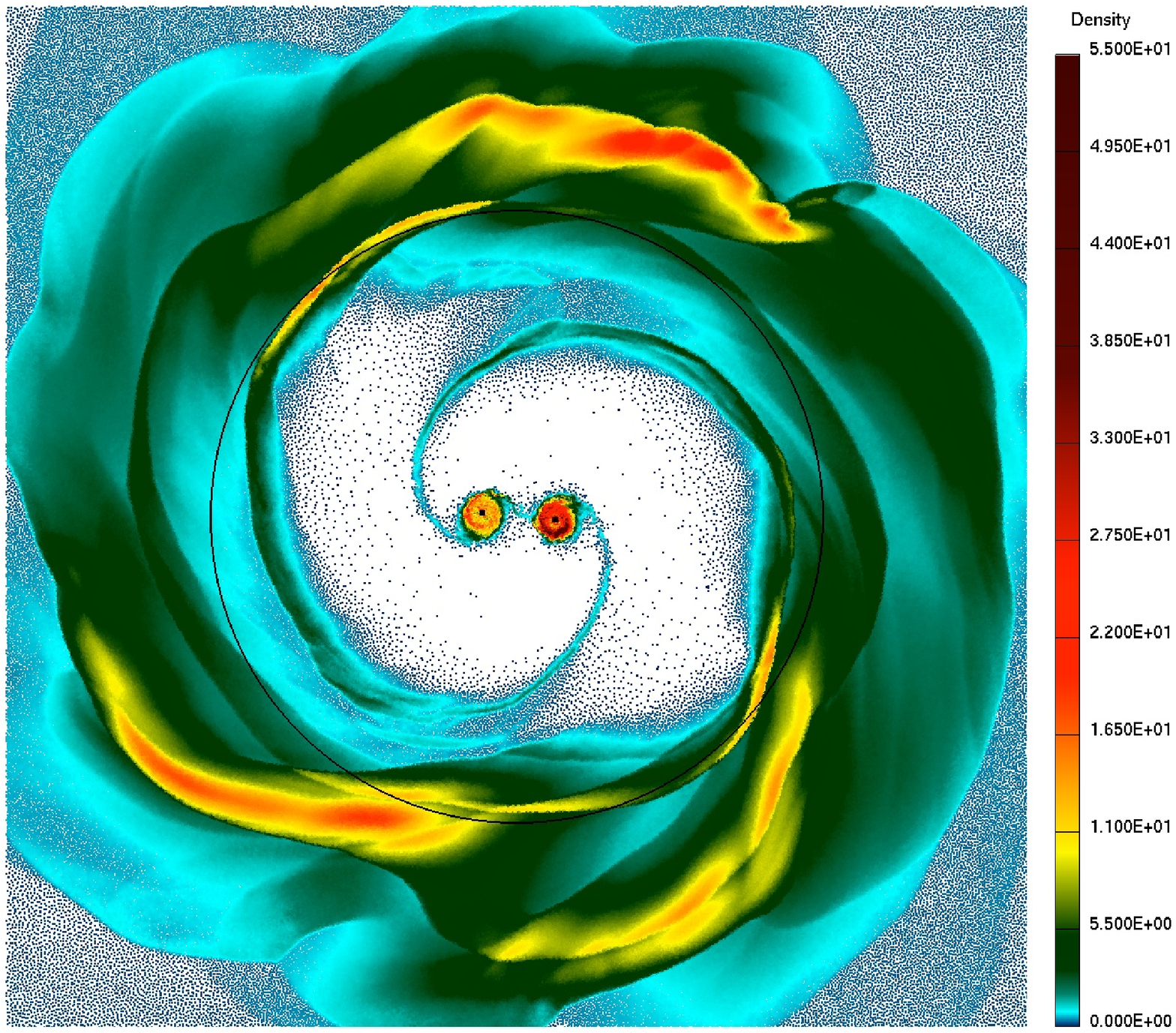}
\includegraphics[angle=0,scale=0.63]{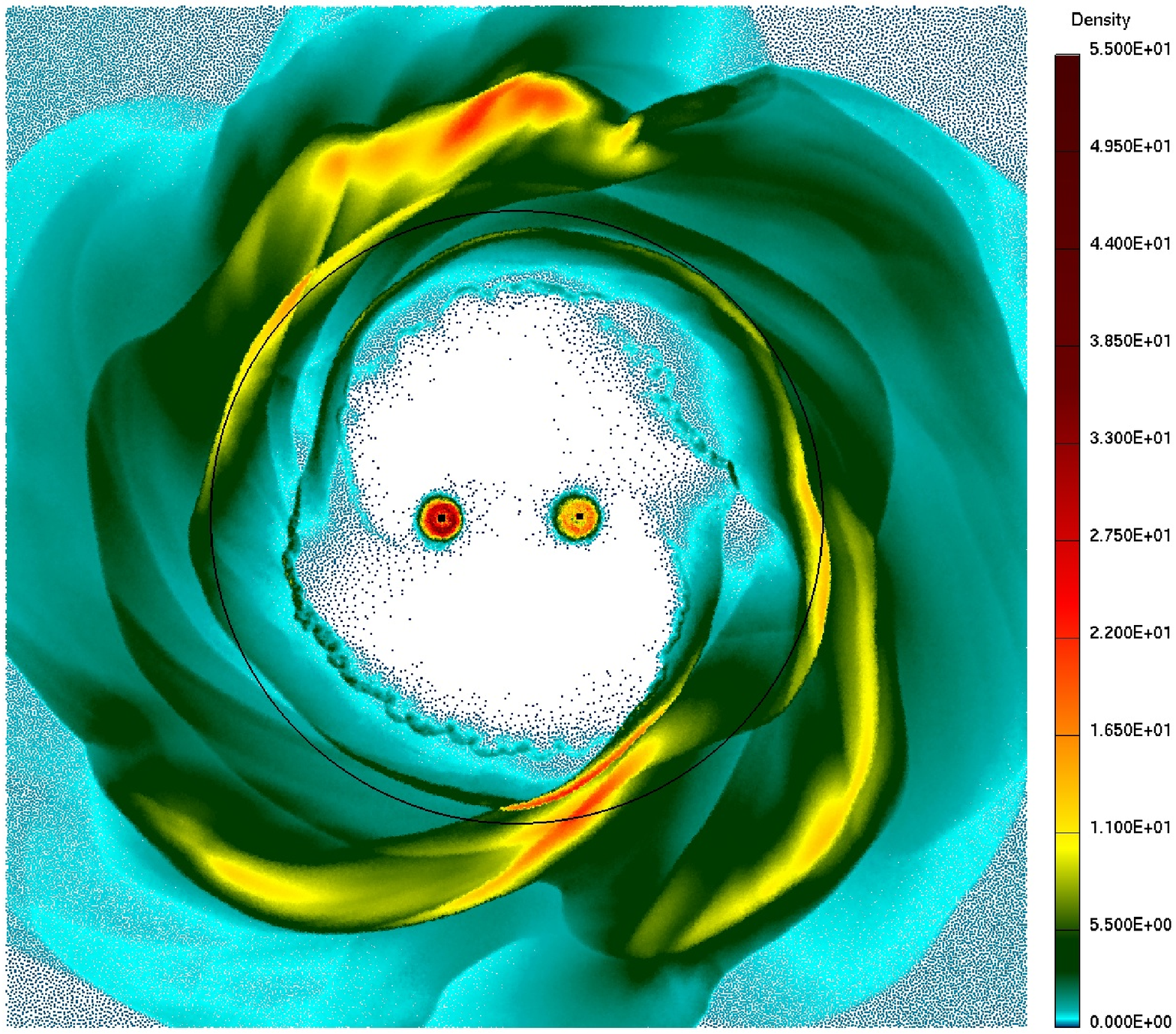}
\caption{\label{fig:wide-ecc-closeup} (bottom) A blowup image of the
system at the start of the 12th binary orbit (at apoapse, identical to
the 9th frame of figure \ref{fig:wide-ecc-manyorb}) and at the
immediately preceding periapse passage (top).}
\end{figure*}

Figure \ref{fig:wide-ecc-closeup} shows a blowup of one panel of the
mosaic in figure \ref{fig:wide-ecc-manyorb}, along with a snapshot of
the system at the preceding periapse passage of the binary. Two
streamers are clearly visible in the gap region between the
circumstellar material and torus, particularly in the periapse image,
which extend outwards to the torus and join with it. The streamers
remain distinguishable as coherent structures for well over a full
revolution around the system and, as we will discuss in more detail in
section \ref{sec:evo-circumstar}, they do not persist as stable
structures over the course of the binary's or torus' orbit. Instead,
they travel through the gap region, with a small portion impacting the
circumstellar disks, but with the largest fraction of material
contained in the streamer simply traveling outwards again, where it
rejoins the torus. Just such behavior has occured in the time leading
up to the apoapse snapshot, where the streamers no longer extend all
the way inwards to the stars, but instead appear effectively as the
torus' inner boundary. At this time, portions of each streamer also
exhibit small scale structure, particularly at the lower left and
upper right portion of the gap, due to the fact that as they traveled
these components have been stretched to the largest extent.

Even after reentering the torus, the structures themselves do not lose
coherence. Instead, they continue to propagate as distinct entities,
interacting with other streamers already present, which were generated
during previous orbits of the binary. As is clear from these images,
as well as similar panels in figure \ref{fig:wide-ecc-manyorb}, the
resulting torus is far from a smooth, azimuth invariant entity. The
combined density structures continue to stretch and deform as they
evolve further. In some portions of the torus, the remmants of
multiple streamers converge to form closely connected filaments. In
others, few such filamants are found. Each of these features is
defined by very distinct edges, which appear both as its inner and
outer boundaries and also as fine structure entirely within the main
body of the torus. Eventually, the features reach the outer edges of
the torus, where they effectively define its outer edge.

The torus' internal structure as seen in our simulations invites
comparison with the various structure observed in the \ggtaua\ system.
Most easily compared are the streamers which appear in the gap region
itself. Although many observations are at the limits of what can be
observed, so that their interpretations are not without ambiguities,
some, more recent observations, such as those of \citet{beck12} and
\citet{pietu11} have begun to remove those uncertainties. For example,
\citet{beck12} interpret their observations of emissions from hot gas
as evidence for streamers of material infalling onto circumstellar
material, which has been shock heated as it impacted the circumstellar
disks. Long wavelength observations of \citet{pietu11} reinforce this
interpretation, showing dust emission extending inwards from the torus
to the circumstellar environments where the hot gas resides. 
Correspondence between the streamers in our simulations and these
features in the observed system are easily made, particularly for the
images shown near periapse, for which the streamers extend inwards all
the way to the circumstellar disks. Some portions are also visible as
a small `bar' in the region extending between the binary components
themselves. The short time scales over which the features in our
simulations exist before dissipating back into the torus, also
correspond well with the locations where the hot material is observed,
in locations where its presence cannot be explained except as a
dynamically transient phenomenon. Finally, the presence of high
temperature material in the outer portions of the disks (see figure
\ref{fig:disk-temps}, above), shows that the accretion streams do,
indeed, generate exactly the observed behavior pattern. 

Within the torus, the distribution of spiral structure as a function
of azimuth is highly inhomogeneous. For example, in the periapse
image, while some comparatively low density structures are visible in
the image both to the left and right of the stars, far higher density
structures (yellow and red in the images) are visible above and below
them. As discussed in sections \ref{sec:wide-morph} and
\ref{sec:close-morph}, such structures are visible over the entire
duration of the simulations, and change their appearance as system
evolves. Although the orbital timescales in the torus are thousands of
years, the time scales for changes are far shorter, as evidenced by
the marked differences between the two images in figure
\ref{fig:wide-ecc-closeup}, which are separated in time by
$\sim200$~yr. From the earlier image to the later, the large amplitude
spiral arms have both traveled forwards in their orbits, and changed
in shape. For example, the large spiral arm near the top of each image
is longer and more dense in the top image than in the bottom.
Similarly, the spiral arm appearing in the bottom left quadrant of
each image is of higher density and angular extent in the top image
than in the bottom.

The features in our simulations are qualitatively similar to the
descriptions of depressions, filaments and gaps in the HST images
discussed in \citet{krist02} and \citet{krist05}. Unfortunately,
although the features are quite clear in our simulations, the
interpretation of the features in the observations themselves less so,
due simply to the fact that they are near the detection limits of the
telescope, where they are difficult to distinguish from various
sources of noise. The features in our simulations are natural
consequences of the dynamical activity generated by the motion of
binary and here, we make tentative identifications of these features
with those reported in the \ggtaua\ system. Further observations would
substantially bolster the case for such identifications, but would
require substantially more finely resolved images that are presently
available. We anticipate that any such improved observations, repeated
over the course of multi-decade timescales, may also begin to resolve
variability of the systems' appearance in time, though such an
observing program would be an extremely challenging endeavor.

\begin{figure}
\includegraphics[angle=-90,scale=0.60]{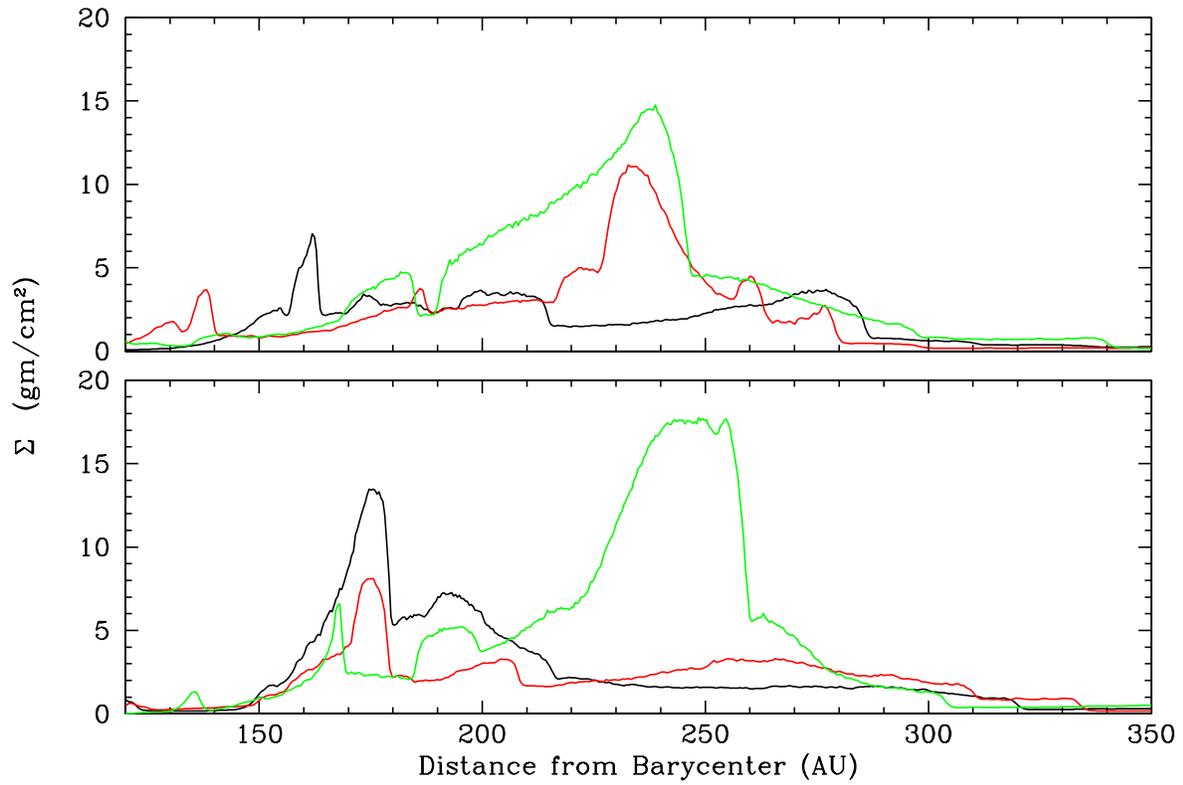}
\caption{\label{fig:dens-lineouts}
The surface densities along three rays extending from the system
barycenter along azimuth angle 0\degr\ (black), 45\degr\ (red) and
90\degr\ (green) in figure \ref{fig:wide-ecc-closeup}, as measured
with zero extending directly to the right in those images. The top and
bottom panels correspond to the top and bottom panels in figure
\ref{fig:wide-ecc-closeup}.} 
\end{figure}

Finally, the streamers, and their coherence over long periods of time,
provide a natural explanation for the observed sharply defined edges
of the \ggtaua\ torus. In figure \ref{fig:dens-lineouts}, we show the
radial profiles of the surface density in the torus along three rays
extending outwards from the system barycenter, each derived from the
surface density mappings described in section \ref{sec:quantitative},
below. The high density regions in the profiles correspond to slices
through the various spiral structures visible in figure
\ref{fig:wide-ecc-closeup}, and their sharply defined edges to their
inner and outer extents along the ray. For example, the green curves,
corresponding to two rays extending upwards in the two images each
intersect a spiral arm whose outer edge is found just inside 250~AU in
the top panel, and just inside 260~AU in the bottom. At those
locations, the surface density decreases abruptly from
$12-15$~gm/cm$^2$ to $\sim5$~gm/cm$^2$ over a spatial scale of
$\lesssim5$~AU. Similar features are present at various other
locations in the profiles, with those in the radially inner portions
corresponding to streamers located within the gap intersecting the
rays, while those at larger distances corresponding to larger scale
spiral disturbances within the torus itself. In some cases, the inner
edge appears more sharply defined, as is the case for the feature at
$\sim230$~AU along the 45\degr\ ray in the upper panel. In others, the
outer edge is sharper, as discussed above for the 90\degr\ rays.
Features with lower surface densities and lower contrasts are also
present in the profiles, extending to distances as far as
$\sim320-350$~AU from the system barycenter.

Each of the profiles include density contrasts of up to factors of
$\sim3-4$, over spatial scales of $\lesssim5$~AU in extent along the
ray. These scales are comparable to or smaller than the $\sim10$~AU
widths derived for the inner and outer ring edge width by
\citet{gds99} and \citet{pietu11} for the \ggtaua\ system. Each of
these edge width derivations are based on observations that approached
the resolution limits of the telescopes themselves so that, in
principle at least, the widths might be expected to be significantly
narrower. On the other hand, the smoothing lengths of the SPH
particles located in these regions is $\lesssim1$~AU, and we conclude
that both the spiral structures and their edges are well resolved in
our simulations. Therefore, given the results of our simulations, we
expect that future observations will fully resolve the edge structure
at size scales of a few AU.

In section \ref{sec:circumbin} we described the process by which we
defined initial conditions for the torus and disk and noted that
determination of self consistent initial conditions (i.e. rotation
velocities consistent with the implied pressure and gravitational
forces) were difficult to specify self consistently at the interface
between the two components. The sharp density gradients produced
correspondingly large pressure gradients that generated difficulties
in defining self consistent orbit velocities near the interface. To
sidestep the inconsistencies, we employed the `workaround' of widening
the interface in our initial condition over a radial region some
$2\times$ the observed width of the interface. Given the the sharp
interfaces described this section, a much more palatable physical
model for interpreting the data is that the sharpness of the inner and
outer edges of the torus is not a feature of a cylindrically symmetric
distribution of material in rotational equilibrium, but is more likely
simply to be due to the spiral structures generated by the actions of
the binary on the torus and of self gravitating disturbances.

\subsection{The meaning of fragmentation in the
torus}\label{sec:fragmentation}

In the discussion above, we noted that in many of our simulations the
circumbinary torus fragmented into one or more massive objects. We did
not however, discuss the significance of those objects, either in the
context of their significance from a numerical standpoint, or in the
context of what they might mean for the \ggtaua\ system. We now
consider such questions.

\subsubsection{Numerical criteria required for valid simulations}

Of primary importance in evaluating our results is to answer the
question of whether or not the fragments are simply artifacts of our
numerical methods or of deficiencies in the physical assumptions used
to model the system. Our first task, then, is to determine whether or
not the simulations obey known requirements for resolving
fragmentation, such as those described in \citet[][hereafter
N06]{N06}. The most pertinent of these criteria is the requirement
that the fragments be resolved by a sufficiently large number of
particles, both as they form and afterwards. A quantification of
`sufficiently large' is that the locally value of the `Toomre mass',
as defined in N06, be resolved by no fewer than $\sim6\times$ the
number of SPH neighbor particles, which corresponds to about 200
particles in 2D. We find that the conditions within the fragments
produced in our simulations are such that the Toomre mass reaches its
smallest values of $\sim10$~\mj\ in the highest density portions of a
fragment. Given the torus mass and the particle counts in our
simulations, the mass of a single SPH particle is $\sim0.02$\me. The
ratio of these two masses is approximately 10$^5$--far higher than the
required number of SPH neighbors, and we conclude that the
fragmentation is well resolved by our simulations.

Additional considerations for checking the sensitivity of our
simulations to various numerical and physical parameters are found in
a series of papers by Meru \& Bate. In \citet{mb10,mb11a}, they find
that, in addition to the result of \citet{LR04} that disks fragment if
their cooling timescale is shorter than $\sim1-2$ orbit periods
(depending on the adiabatic exponent, $\gamma$), fragmentation is also
sensitive to a variety of other physical parameters. In addition to
the cooling timescale, they examine the sensitivity of fragmentation
in disk simulations to the opacity in models with radiative cooling,
and to the mass and temperature profiles in the disk. They determine
an empirical relationship between the cooling timescale required for
fragmentation and the mass distribution and also note that disks
heated by external radiation, from e.g. stellar irradiation (as are
those in our models), tend to be less likely to fragment due to the
additional heating. 

Later however, in \citet{mb11b}, they realized that their previous
results were sensitive not only to changes in the various physical
models, but also to the number of SPH particles used to resolve the
disks in their simulations. They concluded that in many cases, their
simulations were not converged numerically, and questioned whether any
numerical simulations that included radiative transport could possibly
converge to a result which accurately modeled fragmentation in
circumstellar disks. After a further investigation \citep{mb12}, they
revise their initial conclusion to note that fragmentation can be
resolved, but depends strongly on the magnitude of the artificial
viscosity employed. Are these conditions important for the simulations
described in our work? Since our models also include an approximate
radiative cooling model, they too may be affected by a similar
deficiency, and we therefore explore that possibility here.  

We note first, that Meru \& Bate did not discuss the application of
any of the N06 resolution criteria to their results. Given this fact,
we begin with such a comparison in order to understand whether the N06
conditions are more or less restrictive than the convergence criteria
they describe. The condition most directly available for comparison in
the various Meru \& Bate works is the N06 requirement that the
vertical structure in 3D disks be resolved by at least $\sim4$~SPH
smoothing lengths per disk scale height, at the disk midplane. Figure
10 of \citet{mb12} shows this quantity, and we observe that the
azimuth averaged values of the ratio $h/H$ (smoothing length over
scale height--the inverse of the quoted N06 criterion), are larger
than the required 1/4 for all but the outer portions of the two
million particle run shown. Based on this observation, we conclude
that their simulations with two million particles are at best
marginally resolved in their outer regions, while the inner regions
and those runs with fewer particles are insufficiently resolved. The
direct consequence, as shown in N06, is that the mass densities in the
disk midplane in those runs will under-shoot their physically correct
(i.e. fully resolved) values by a factor of as much as 30\% when
$H/h\sim2$. This condition will be true for all of the simulations in
\citep{mb10,mb11a} and many in their later works \citep{mb11b,mb12} as
well.

The effect such errors will induce in other physics based quantities
or the overall outcome of a given simulation is difficult to assess
directly. It is noteworthy however that the sensitivities they
describe are much less pronounced above the 2 million particle
threshhold \citep{mb12} at which the vertical structure criterion of
N06 is marginally satisfied in the outer parts of their disks. We
conclude that the trends in their simulations are likely to be
manifestations of numerical inaccuracies similar to those underlying
the N06 vertical structure resolution requirement. This
conclusion, and the arguments on which it is based, is also shared by
\citet{LC11}. In addition, those authors argue that one possible
origin for the resolution sensitivity may lie in resolution dependent
changes in the thermal energy input supplied by the bulk artificial
viscosity, even when that contribution is small overall. They
emphasize however that the case for such a mechanism is ambiguous,
because some simulations, at high resolution and for which their model
predicts fragmentation, actually do not fragment. That absence is then
taken as evidence for physical mechanisms to suppress fragmentation.
We have discussed the importance of two such mechanisms in our work
(radiative heating and stirring generated by the binary interactions),
and so discount the resolution criterian in Meru \& Bate, in favor of
satisfying the N06 Toomre mass criterion directly, as the best
indicator that numerically induced fragmentation is absent in our
simulations.

\subsubsection{Physical significance} 

Given that our simulations do not contain numerical deficiencies which
generate fragmentation, we now discuss the significance of the
relevant physical models and their contribution to fragmentation.
Fragmentation occurs when an imbalance betweeen gravitational and
pressure forces develops which favors gravitational forces in some
volume so that it begins to collapse, typically on timescales of a few
orbits \citep{PPV_Durisen}. The imbalance develops when either
insufficient internal energy is absorbed by the disk via radiative
heating, or is generated in it by dissipation of kinetic energy, or
when cooling permits internal energy to escape more quickly than it
can be replaced. Each of these processes decrease the gas pressure
through the equation of state, and each are strongly time dependent in
character, varying as the disk evolves.

All of our simulations exhibit strong spiral structures generated by
the combination of self gravitating instabilities and the tidal
interactions with the binary. Some also produce fragments on short
timescales compared to the system lifetime. We conclude that our
simulations model systems lie very near their fragmentation boundary,
with some configurations being slightly more susceptible to
fragmentation and others slightly less so. In contrast, and although
marked structure exists, the \ggtaua\ system is not known to contain
any well defined fragments within its torus, even though the observed
torus conditions imply a strongly self gravitating structure. This
means that the balance between heating and cooling must be very finely
held, because the instabilities are neither fully suppressed, nor do
they grow to sufficient amplitude to generate fragements. Further,
this balance has been sustained over a long time, in spite of the
sensitive time dependence noted in \citet{PPV_Durisen}, since the
system lifetime is long compared to the $\sim$few orbit timescale
required for fragmentation to occur. Given that our simulations model
configurations defined to be very similar to \ggtaua, such that
similar physical behaviors are expected in both, we conclude that the
balance between heating and cooling is much more finely held in
reality than is possible for our models to reproduce. 

Nevertheless, two important, additional conclusions about the \ggtaua\
system are possible based on our results. First, we recall our
conclusions above that large scale structures seen in all of our
simulations owe their origin not only to the tidal action of the
binary but also to self gravitating instabilities in the torus.
Although no clear pattern emerges from our simulations that would
indicate which configurations are more likely to generate fragments in
the \ggtaua\ system, their presence in systems with many different
binary orbital parameters indicate that the conditions are generally
favorable for fragmentation. We can therefore conclude that the
\ggtaua\ system itself is very near its fragmentation boundary. Given
our results however, we cannot predict whether or not such an outcome
actually will occur.

Second, we conclude that the combined energy input from the radiative
heating from the stars and the stirring action of the binary are
responsible for a large fraction of the thermal energy input into the
torus, and thereby play an important role in its long term stability.
This is demonstrated by comparing the simulations described in section
\ref{sec:heatoff}. In the first, we model a single star of the same
mass and luminosity as the combined masses and luminosities of the
binary stars in our other simulations, but with the same torus. In the
second, we model a binary while omitting the stellar radiative heating
source. The torus is marginally stable with an embedded binary that
includes heating, producing none or at most one or two fragments
within the several orbit time frame of the simulations. In contrast,
removing either the radiative heating or the tidal stirring causes the
torus to become violently unstable, producing nearly 40 fragments in
less than a single torus orbit. Such an outcome clearly demonstrates
both the unphysical nature of the evolution where those processes are
absent and, at the same time, their importance where they are present.

\subsection{The orbital parameters of the \ggtaua\
binary}\label{sec:binary-orbit}

GDS99 report an inclination of 37$\pm$1 degrees for the \ggtaua\ torus
relative to the line of sight. This value is derived from the
observations by deprojecting the observed eccentric appearance of the
circumbinary torus into a circular shape. Given this inclination,
attempts to fit the available astrometric data for the binary's orbit
presently remain somewhat ambiguous. If an orbit coplanar with a
circular torus is assumed, then the best fit orbital parameters are
$a\sim35$~AU and $e\sim0.3$ \citep{MDG02,BEDU05,koehler11}.
Unfortunately, these parameters prove to be somewhat unsatisfactory
from the point of view that the semi-major axis is smaller than is
expected based on dynamical arguments and the position of the inner
edge of the torus.

Recognizing this shortcoming, the same authors and others fit for the
orbital parameters under the relaxed assumption that the planes of the
stellar orbit and the torus may be misaligned. In this case,
\citet{BEDU06} determined best fit orbit parameters of $a\sim62$~AU,
$e\sim0.35$ and a mutual inclination of $i\sim21-24$~degrees. They
also report a second family of fits with mutual inclinations of nearly
90 degrees, but find that the disk in such systems is either disrupted
entirely or becomes very thick. Following similar arguments regarding
alignment, the later analysis of \citet{koehler11} found the `most
plausible' orbit parameters to be $a\sim60$~AU and $e\sim0.44$, with a
relative inclination between the two planes of $\sim 25$ degrees. 

Under both sets of assumptions however, the fitted orbital parameters
still remain unsatisfactory based on the dynamical argument that a
substantial misalignment between the stars and torus would quickly
generate a warp in the torus, while no such feature is observed. Given
the analysis of section \ref{sec:quantitative}, we believe that the
data and the analyses of them may be fully reconciled if we relax the
assumption that the \ggtaua\ torus is circular, rather than the
assumption that the torus and binary orbit planes are inclined with
respect to each other. Under this alternate assumption, deprojection
of the torus' actual eccentric shape into a circular shape results in
an erroneous determination of its inclination. Then, the perceived
misalignment is simply a consequence of the incorrect determination of
the torus inclination, rather than a property of the system itself.

This alternate assumption will be plausible if the mutual inclination
determined from the fit is comparable to the inclination derived from
misinterpreting an elliptical torus as a circular torus inclined at an
angle, $i$, with respect to the line of sight. For a given ellipse
which is interpreted as a circle, the inferred inclination is related
to the true eccentricity, $e$, via the equation
\begin{equation}
\cos(i) = \sqrt{1 - e^2}. 
\end{equation}
Therefore, given a torus with eccentricity $e=0.3$, its inclination
would be perceived to be $i\approx17$ degrees. Similarly, a torus with
$e=0.1$ or $e=0.6$ would be perceived to have an inclination of
$\approx6$ or $\approx39$ degrees, respectively. The same error is
made if some other plane defines the torus' absolute orientation,
rather than the plane of the sky. In the case of the \ggtaua\ system,
this plane can be specified as the plane defining the orbit of the two
stars in the binary. 

The mutual inclinations derived in each of the analyses above fall
well within the range of `false' inclinations which could be
attributed to misinterpreting the eccentric shape of the torus itself.
On this basis, we conclude that the ambiguity in orbit parameters can
be resolved in favor of the larger semi-major axis. This conclusion
restores consistency between the observations of the torus'
dimensions--specifically the location of its inner edge--and the
dynamical results of \citet{ArtLu94}, which predict a wide binary
orbit based on the location of Lindblad resonances that drive tidal
truncation.

It is further reinforced by the fact that only the circumstellar disks
in our $a=62$~AU simulations remain massive enough to remain
consistent with the masses derived from observations. The disk masses
in all of our simulations with $a=32$~AU decrease quickly to much
smaller values due to the combination of losses through accretion onto
the stars and the lack of replenishing material accreting onto them
from the circumbinary environment. Observations indicating the
presence of accretion streams reinforces our conclusion further, for
two reasons. First, such streams are essentially non-existent in our
$a=32$~AU simulations, but ubiquitous in our $a=62$ simulations,
particularly in the $e=0.3$ run. Second, assuming an $a=62$~AU
semi-major axis, the \ggtaua\ system must be relatively close to its
periapse passage, given its apparent separation. This is important
because our simulations indicate that accretion streams extending
inward to the circumstellar disks occur only around this time,
becoming weaker or entirely absent at other orbital phases. Thus, the
simulations and observations suggest the same consistent model for the
system itself.

We note that an important consequence of adapting this assumption is
that deprojection of the apparent shape of the \ggtaua\ torus becomes
a much less accurate method of determining the system's actual
inclination. The difficulty is compounded by the fact that the
eccentricity of the torus varies widely and its orientation is not
correlated with the orientation of that of the binary's orbit plane.
Other constraints will be required in order to disambiguate the torus'
actual inclination and its actual eccentricity.

\subsection{Implications for the Circumstellar
Environment}\label{sec:stellar-env}

Observations of the \ggtaua\ system show conclusive evidence both for
circumstellar material and for accretion into the circumstellar
environment. Both hot and cold material are observed
\citep{dutrey14,beck12,pietu11} in a region extending some tens of AU
from the stars, in both circumstellar disks and in accretion streams.
Unfortunately, the observations extend over only a very short time
window in comparison to a full orbit of the binary, or even a full
periapse passage. They therefore define only a snapshot of the system
rather than a full evolutionary picture of its behavior over time.
Given the similarities we have already noted between our wide orbit
simulations and those observed in \ggtaua, it will be useful to apply
the comparisons first to make interpretations of the current state of
the system in the context of an evolutionary model, and then to make
predictions of its behavior as it evolves in the future.

Above, we concluded that the degeneracy in the orbital motion data for
the stars in the \ggtaua\ system can be resolved in favor of an orbit
with $a\sim62$~AU. This conclusion is further reinforced by the
presence of significant amounts of material in the circumstellar
environment of \ggtaua\ because mass accretes into the circumstellar
environments at rates sufficient to sustain and even grow the disks
{\it only} in our wide orbit simulations. Little accretion occurs in
the close orbit runs. Instead, mass transfer into the circumstellar
disks is very limited in those models and the disks initially assumed
to be present quickly accrete onto the stars, almost entirely emptying
them of material. These conditions are inconsistent with the
observations of features extending tens of AU from the stars in the
\ggtaua\ system. The presence and sizes of the circumstellar disks are
also consistent with the interactions by each component of the system
on the other, which lead to tidal truncation of the circumstellar
disks at sizes comparable to the dynamical stability limit described
by \cite{HolWie99} and shown numerically by \cite{ArtLu94}. The
episodic accretion seen in our eccentric binary simulations is quite
reminiscent of the same feature in the work of \citet{ArtLu96}. As in
their work, we find that accretion occurs preferentially onto the
secondary component and that, due to this preference, the mass
ratio between the components will tend towards equality over long
periods. The same conclusion was drawn in \citet{YC15}, for the long
term, time averaged accretion rates from massless, circumbinary disks
with isothermal evolution assuming temperatures both warmer and colder
than in our study.  

Given that the \ggtaua\ system is in a wide orbit, the fits provided
in \citet{koehler11} show that the stars are approaching their
periapse passage, rather than receding from it. Examination of our
simulations (e.g. panels 6-8 of figure \ref{fig:wide-ecc-1orb}) show
that at this time the circumstellar environment should include a
significant amount of material originating in the accretion streams
and wrapping around the disks and extending outwards to the torus. A
similarly important feature of the system is that the circumstellar
accretion disks should exhibit spiral structures and other
asymmetries, driven by the tidal interactions of the binary
components. Both classes of features are consistent with the
observations. To these interpretations of the system's short term
state, we add the longer term prediction that, over the duration of
the periapse passage, as much as a few tenths of a Jupiter mass will
accrete onto the circumstellar disks of the \ggtaua\ system,
preferentially onto the secondary component, as shown in figure
\ref{fig:masstransW}.

An important feature of the \ggtaua\ system that our simulations do
not reproduce is that the mass determined for the disk around the
secondary component is substantially less than that around the primary
component \citep[as discussed, e.g., by][]{pietu11}. In our
simulations, the reverse is true--mass accretes preferentially onto
the disk around the secondary, and its mass increases to $\sim2$ times
that of the disk around the primary (figure \ref{fig:massdiskW}). 
Should we interpret this inconsistency as an indication that our
models should be ruled out as inconsistent with the observations?
Fortunately for our models, \ggtaua\ itself provides the means to
reconcile simulation and observation: it simply is not the binary we
have modeled. Instead, recent interferometric observations of
\citet{difolco14} have shown that the GG~Tau~Ab component is itself a
binary consisting of two M stars separated by $\sim4.5$~AU. Unlike the
system we modeled, the circumstellar disks in this system will be
tidally truncated to radii no larger than about 1--1.5~AU each. The
shorter dynamical times inherent such small disks will lead to
correspondingly reduced disk lifetimes and masses, consistent with the
observations, even in the presence of rapid replenishment from the
accretion streams. Given this confounding factor, the
inconsistency between our simulations and the data is to be expected,
rather than being evidence for their invalidity. 

After its accretion into one or the other circumstellar disk, mass
continues to migrate through the disk, ultimately accreting onto the
star around which it orbited. Accretion rates onto the stars in
\ggtaua\ have been determined by \citet{GULL98} (who derive a rate of
$\sim1.8\times 10^{-8}$\msun/yr) and \citet{HAR95} (who derive a rate
of $\sim2\times 10^{-7}$\msun/yr) and are in very good agreement with
the rates inferred from our simulations, of a few $\times
10^{-8}$\msun/yr. Each of these works consider only the accretion onto
both stars as a sum, rather than onto the individual components. Our
results, summed over both stars, fall roughly half way between the two
observational determinations, supporting the conclusion that our 
simulations accurately model the \ggtaua\ system. Our rates, in turn,
are about an order of magnitude higher than those computed by
\citet{GUKLE}. While we have not investigated the origin of the
differences between our results and theirs, we note that our initial
setup was significantly different from theirs, as was our numerical
method.

The accretion rates correspond, in numerical terms, to rates of 1-2
particles per year, or several hundred per binary orbit. Given this
level of discreteness, such values suggest that we should consider the
accretion in our simulations to be, at least in part, a phenomenon
governed by numerical processes rather than physics. Perhaps
surprisingly however, the rates do not appear to be strongly
correlated to resolution. Accretion driven primarily by numerical
viscosity for example, should be proportional to the particle
smoothing length, through equation \ref{eq:Balsara}, and in turn to
the square root (in 2D) of the number of particles. Instead, both the
high and low resolution variants of the same physical configuration
return similar patterns of accretion over time. We conclude, due to
this relative insensitivity, that the accretion is not driven
primarily by viscous processes derived from numerial effects, but
rather by other physical processes, such as gravitational torques.

Mass transfer into the circumstellar disks occurs at rates similar to
those for stellar accretion for wide systems but at much lower rates
in the close systems. Thus, concerns similar to those noted for the
stellar accretion must be addressed for these cases as well.
\citet{YBC15} studied the sensitivity of the mass transfer rates into
circumstellar environments to the resolution employed in 2D SPH
simulations. They found that rates were numerically converged in
simulations where the ratio of particle smoothing length to disk scale
height fell in the range $h/H\sim0.05-0.5$ near Roche radii of the
stars. This occured at accretion rates of a few hundred particles per
binary orbit. In our high resolution runs, we find that
$h/H\sim0.1-0.2$ in this region. Given the similarity in these
parameters between their work and ours, we conclude that the mass
transfer rates in our simulations are also converged. In contrast, the
streams in the close orbit systems are far less massive and are
correspondingly more poorly resolved. The comparatively poorer
resolution of the mass transfer will not be fatal to our conclusion
that the rates are much lower than the wide orbit runs, because the
simulations are based on identical initial configurations of the
circumbinary material. Thus, the weaker streams are simply a
consequence of the intrinsic differences between the physical
configuration of the stellar and circumstellar material rather than
any shortcoming in the numerical realization.

\subsection{The likelihood for planet formation}\label{sec:plan-form}

The results we present may be extended to considerations regarding the
likelihood for planet formation in \ggtaua\ or other multiple systems.
From an observational perspective for example, \citet{dutrey14} detect
the presence of both cold and hot material in the circumstellar
environment of \ggtaua\ and interpret the presence of the cold
material as evidence for possible planet formation in the system. From
a theoretical perspective however, such a conclusion requires not only
the presence of substantial amounts of material in circumstellar
disks, but also a mechanism for accumulating that matter into massive,
condensed objects on short timescales. Theory suggests that the most
likely candidates responsible for forming Jovian mass planets and low
mass brown dwarfs are either gravitational fragmentation of large
scale spiral structure \citep{PPV_Durisen} or coagulation of small
solid grains followed by later accretion of additional gas \citep[i.e.
`core accretion'--see][]{PPV_LS} in the disks of forming stellar
systems.

In N00, one of us concluded that planet formation in the circumstellar
disks of equal mass binary systems with a separation similar to that
in the \ggtaua\ system was unlikely. The conclusion was based on the
fact that temperatures in the disks were too high either to support
gravitational instabilities or to permit solid grain growth since
grains would be vaporized in large portions of the disks. It was based
on models of a system similar to L1551~IRS~5, a system much younger
than \ggtaua\ and for which the disks are still relatively massive. To
what extent does a similar conclusion hold for systems like \ggtaua,
which has evolved to a much later evolutionary epoch, and for which
the circumstellar disks are far less massive?

In answer to this question, we note that the temperatures seen in
figure \ref{fig:disk-temps} (section \ref{sec:circstar-conf}, above)
are similar to those reported in N00, while the disk masses are more
than an order of magnitude lower. The effect of similar temperatures
and smaller disk masses will make the disks even more stable against
the growth of self gravitating structures, as defined by the value of
the Toomre $Q$ parameter, than reported in N00. Planet formation by
way of gravitational fragmentation is therefore suppressed to a
correspondingly greater degree. Similarly, high disk temperatures are
significant for the core accretion model, because the budget of solids
entrained in the gas decreases where temperatures rise above the
vaporization temperatures of those solids. In the present instance,
the `ice line', at which planet formation via the core accretion model
is enhanced by the presence of additional solid material, will lie
near to or outside the actual boundaries of the disks. Water ices will
therefore not be present within the circumstellar disks of \ggtaua,
except perhaps at high altitudes above the midplanes, and planet
formation via core accretion will be inhibited. In addition, while the
accretion of new material into the circumstellar environment might be
expected to enhance the growth of dust and small solid bodies, the
fact that the overall accretion rate through the disk is high will
negate this benefit. Any accumulations of dust grains and small
planetesimals will quickly be swept along with the overall mass flow
through the disk and be lost onto the star. Therefore, based on the
same arguments presented in N00, we can extend the conclusion in N00
to the later stages of evolution typical of \ggtaua.

In extending the N00 conclusion, we are mindful of apparently
contradictory data embodied in the detections of planets in mature
binary systems \citep[see e.g.][]{DB07,Piskorz15,Ngo15}. In many
cases, these systems are characterized by semi-major axes similar to
that found in \ggtaua\, making the contradiction more acute. It can be
resolved however, by noting that the companion masses derived for
these objects (or implied for them through their effective
temperatures) are substantially lower than those reported for the
primaries and that few planets in binaries with near equal mass ratios
were found. Such systems are therefore distinct from the cases
considered here and our conclusion does not apply. Further
contradictions emerge however, when observations of young systems are
accounted for, such as is seen in \citet{Cheetham15}, who show that
$\sim2/3$ of young, close binary systems have lost their circumstellar
disks by the time they are 1-2Myr in age, without regard to stellar
mass ratios. They conclude that planet formation will be less likely
in binary's compared to single stars. Such contradictions suggest that
a comparative study of the two classes of systems (of equal and
unequal mass components) would be of great interest, to explore the
physical reasons for the dichotomy between the presence of planets in
unequal mass binary systems and their paucity in equal mass systems.

Given these dynamical constraints, we conclude that the cold material
observed by \citet{dutrey14} has yet to interact substantially with
the circumstellar environment or, for some portion, will soon return
to the circumbinary torus as parts of the accretion stream moves out
of the circumstellar environment. We cannot therefore confirm their
conclusion that its presence in the circumstellar environment is
evidence for planet formation. Nevertheless, the presence of both cold
and hot material will remain of considerable interest for evolutionary
models because together they define a reservior of nearly pristine
material which has either recently undergone, or will soon undergo,
its first significant heating event. It may therefore retain
signatures of the processes important during that event.

In a comparison study of single and moderate separation binary
systems, \citet{Petal08} conclude that the first stages of planet
formation--grain growth--will not be affected by the presence or
absence of a binary companion. They base their conclusions on the lack
of observed differences between the 10$\mu$m silicate features seen in
the spectra of disks around stars in single and multiple systems.
While the presence of silicates and other solids in disks is certainly
a requirement for the formation of planets, in and of itself, simple
presence is insufficient to differentiate between a system where
planet formation is more likely and a similar system where planet
formation is less likely. These features may instead trace material
which is unlikely to become important for planet formation because of
its short lifetime in the disks. We believe that such is the case in
the \ggtaua\ system. 

\section{Summary and Discussion}\label{sec:discussion}

Having completed our study of systems configured to be similar to
\ggtaua\ and discussed their physical significance in relation it, we
now summararize our results and suggest future studies that may
profitably build on ours.

\subsection{Summary of Our Main Results}\label{sec:summary}

We have modeled the evolution of a forming binary star system,
configured to resemble the \ggtaua\ system, using the SPH code VINE.
The system consists of two stars, each surrounded by a circumstellar
disk and in orbit around each other, with a circumbinary torus$+$disk
structure in orbit around the combined system. Our simulations include
the self-gravity of the disk material, external heating of the gas due
to radiation from two stars and radiative cooling from the disks. 

We follow the evolution of a suite of systems for $\sim$6500-7500~yr,
corresponding to about three orbits of the torus around the center of
mass. We find that strong spiral structures develop in the torus due
to the stirring action of the binary and to its own self gravity. The
spiral structures are well defined in space, are characterized by
edges of $\lesssim5-10$~AU in width and frequently extend more than
360\degr\ around the center of mass. We interpret the observations of
sharply defined structures in the \ggtaua\ torus as evidence for such
spiral structures. The spiral structures propagate outwards over time,
generating a net outward mass flux extending into the low density
circumbinary disk surrounding the torus. At large distances, they lose
coherence, and we interpret the presence of the disk component of the
circumbinary material as an excretion feature produced as the spiral
structures transport mass to that region. In our torus-only
configurations, about 15\% of the mass of the torus moves outwards
over the $\sim6000$~yr duration of our simulations--about half of the
observed mass of the present disk (30\% of the torus mass). The rates
are much lower in the torus$+$disk initial configuration, and we
expect that on a longer timescales, as the disk component gains mass,
a nearly stationary state is reached between the outwards mass
transport into the disk, and the mass escaping from it, through its
outer edge or other mechanisms. 

Structures created at different times frequently interact with one
another and, late in the evolution of several simulations, fragment
into one or two clumps with initial masses comparable to brown dwarfs.
In a model where we replaced the binary with a single star of the same
mass, some 38 clumps formed in less than a single orbit of the torus.
We interpret this result as an indication of the importance of the
binary's stirring action as a contributor to the thermal energy input
of the torus, and as a strong indication of the importance of self
gravity in the \ggtaua\ torus. The torus changes shape and orientation
as the spiral structures propagate through it. We fit its shape to an
ellipse whose eccentricity varies between $0\lesssim e\lesssim0.6$ and
an orientation which precesses on timescales comparable to the torus
orbit period. We point out a degeneracy in the interpretation of the
data in the calculation of the torus' inclination and its
eccentricity, if the torus is actually eccentric rather than circular
in shape. Given this degeneracy, we resolve the question of the
supposed misalignment between the \ggtaua\ torus and the binary's
orbit, by interpreting the data as a coplanar system in which the
torus is eccentric, rather than circular.

We simulate systems configured with semi-major axes of either
$a=62$~AU (`wide') or $a=32$~AU (`close'), and with assumed orbital
eccentricity of either $e=0$ or $e=0.3$. Although torus structure
appears in all configurations, only wide configurations generate
substantial streams of material extending inward to the circumstellar
disks. A small fraction of the material in these streams accretes into
the circumstellar environment, with the remainder returning to the
torus. In wide binary configurations, and over the time span of our
simulations, mass comparable to the disks themselves is accreted onto
both primary and secondary components. Mass accretes preferentially
onto the secondary at steady rates in $e=0$ configurations, but is
episodic in $e=0.3$ configurations with highest rates occuring near
binary periapse. In contrast to the wide configurations, the streams
in close binary configurations contain much less material and the rate
of accretion into the circumstellar environment is two orders of
magnitude smaller. Accretion onto the stars is steady in all cases, at
rates sufficient to drain the disks almost entirely of material within
the time span of our simulations, if they are not otherwise
replenished. 

The net mass transport into and out of the disks in the wide binary
configurations, produces disk masses which remain steady near their
initial values in $e=0$ cases, but double over the same time span in
$e=0.3$ cases. In close binary configurations however, the net mass
inflow and outflow is insufficient to replenish the disks before they
accrete onto the stars. Replenishment of disk material from the torus
in wide orbit configurations can therefore significantly extend the
lifetime of the circumstellar disks. Mass replenishment is a
necessary, but not sufficient, condition for planet formation in the
circumstellar disks. We find that other conditions, such as material
temperatures, are not compatible with planet formation in the
circumstellar disks, and conclude that planet formation there is
unlikely.

Given the results of our simulations, we interpret the observations of
the \ggtaua\ system in favor of the wide, eccentric binary orbit
solutions of \citet{BEDU06} and \citet{koehler11}, for which
parameters of $a\sim62$~AU and $e\sim0.3$ are derived. In each case,
the fits require that the torus and binary orbit be inclined with
respect to each other. As noted above, we resolve this inconsistency
by interpreting the supposed mutual inclination as an eccentricity in
the torus. We also favor these parameters over the fits otherwise
prefered by each of the previous works, namely those with $a\sim32$~AU
and $e\sim0.3$, because the circumstellar disks in such systems would
be accreted onto their respective stars within a few thousand years,
and because material streams in such systems would be far weaker than
are observed.

\subsection{Questions for the Future}\label{sec:questions}

The results presented in this work answer a number of questions
regarding the character of the \ggtaua\ system, while raising others,
and leaving others untouched. First among the questions raised by our
results are the questions of whether the detailed morphology of
features in the circumbinary torus as seen in our simulations are
actually present in the \ggtaua\ system. The current best resolution
observations of the torus provide tantalizing hints that such features
do exist, but a definitive statement that such features are present
must await higher resolution observations of quality similar to those
described by \citet{HLTauALMA}, for the HL~Tau circumstellar disk.
Another important question concerns planet formation in multiple
systems. Whether or not any theoretical mechanism presently exists to
explain the formation of giant planets in binary systems, the fact
remains that at least a few binary systems, such as $\gamma$ Cephei
\citep{NEU07}, GI86 \citep{QUE00} and HD~41004 \citep{ZUCK04}, do
harbor planets. Therefore some formation mechanism does in fact exist.
While we find that accretion into the circumstellar disks occurs
rapidly enough so that they can survive for the comparatively long time
scales needed to form planets, other conditions, such as temperatures,
remain quite unfavorable. What are the mechanisms still missing from
our models that permit such objects to form? Finally, our simulations
model the evolution of the `A' component of the full \ggtau\ system
over a time span extending over only a tiny fraction of its formation
time scale and in only two spatial dimensions. We neglected the full
dynamical effects expected to be present in the system, insofar as our
results include neither the distant binary `B' component of the \ggtau\
system, nor the newly discovered tight binary nature of the \ggtaua~b
component. Even so, we find large scale morphological changes even over
this short time span and restricted dimensionality and physical system.
Given the vigor of the activity over such a short time scale, we would
expect activity of similar scale over to occur over longer time spans
as well, with correspondingly large consequences on the system
morphology. To what extent will 3D effects also play a role in the
evolution? What will be the end state configuration of the \ggtau\
system as a whole? Will the components eventually break apart? Merge?
Future investigations extending the work presented here will be
required in order to answer these questions.

\acknowledgements
We wish to thank the anonymous referee for comments which improved
our manuscript. Portions of this work were carried out under the 
auspices of the National Nuclear Security Administration of the U.S.
Department of Energy at Los Alamos National Laboratory under Contract 
No. DE-AC52-06NA25396, for which this is publication LA-UR-16-23283.

{}


\begin{thebibliography}{}

\bibitem[Alexander \& Ferguson(1994)]{AF94} Alexander, D. R.,
Ferguson, J. W., 1994, \apj, 437, 879

\bibitem[ALMA Partnership(2015)]{HLTauALMA} ALMA Partnership, 2015,
astro-ph:1503.02649


\bibitem[Artymowicz \& Lubow(1994)]{ArtLu94} Artymowicz, P.,
Lubow, S. H., 1994, \apj, 421, 651

\bibitem[Artymowicz \& Lubow(1996)]{ArtLu96} Artymowicz, P.,
Lubow, S. H., 1996, \apj, 467, L77

\bibitem[Balsara(1995)]{B95} Balsara, D., 1995, J. Comp.
Phys, 121, 357

\bibitem[Beck \etal(2012)]{beck12} Beck, T., Bary, J. S., Dutrey, A.,
Pi\`etu, V., Guilloteau, S. Lubow, S. H. Simon, M., 2012, \apj, 754,
72

\bibitem[Beust \& Dutrey(2005)]{BEDU05} Beust, H., \& Dutrey, A.,
2006, \aap, 439, 585.

\bibitem[Beust \& Dutrey(2006)]{BEDU06} Beust, H., \& Dutrey, A.,
2006, \aap, 446, 137.

\bibitem[Bodenheimer \etal(2000)]{PP4_BBKB} Bodenheimer, P.,
Burkert, A., Klein, R. I., Boss, A. P., 2000, in Protostars and
Planets IV, ed. Mannings, V., Boss, A. P., \& Russell, S. S.,
University of Arizona Press: Tucson, p. 675 

\bibitem[Cheetham \etal(2015)]{Cheetham15} Cheetham, A. C., Kraus, A.
L., Ireland, M. J., Cieza, L., Rizzuto, A. C., Tuthill, P. G., 2015,
\apj, 813, 83

\bibitem[Chiang \& Goldreich(1997)]{CG97} Chiang, E. I., Goldreich,
P., 1997, \apj, 490, 368

\bibitem[Close \etal(1998)]{UY-Aur} Close, L. M., Dutrey, A., Roddier,
F., Guilloteau, S., Roddier, C., Northcott, M., Menard, F., Duvert,
G., Graves, J. E., Potter, D., 1998, \apj, 499, 833

\bibitem[Cohen \& Kuhi(1979)]{CoKu79} Cohen, M., Kuhi, L. V., 1979,
\apjs, 41, 743

\bibitem[Desidera \& Barbieri(2007)]{DB07} Desidera, S., Barbieri, M.,
2007, \aap, 462, 345

\bibitem[di~Folco \etal(2014)]{difolco14} di~Folco, E., Dutrey, A.,
L~Bouquin, J.-B., Lacour, S., Berger, J.-P., K\"ohler, R., Guilloteau,
S., Pi\'etu, V., Bary, J., Beck, T., Beust, H., Pantin, E., 2014,
\aap, 565, L2

\bibitem[Duch\^ene \etal(2004)]{duchene04} Duch\^ene, G., McCabe, C.,
Ghez, A. M., Macintosh, B. A., 2004, \apj, 606, 969

\bibitem[Durisen \etal(2007)]{PPV_Durisen} Durisen, R. H., Boss, A.
P., Mayer, L., Nelson, A. F., Quinn, T., Rice, W. K. M., 2007 in
Protostars and Planets V, ed. Reipurth, B., Jewitt, D., Keil, K.,
University of Arizona Press: Tucson, p. 607 

\bibitem[Dutrey, Guilloteau \& Simon(1994)]{dutrey94} Dutrey, A.,
Guilloteau, S., Simon, M., 1994 \aap, 286, 149

\bibitem[Dutrey \etal(2014)]{dutrey14} Dutrey, A., di~Folco, E.,
Guilloteau, S. Boehler, Y., Bary, J, Beck, T., Beust, H., Chapillion,
E., Gueth, F., Hur\`e, J.-M. Pierens, A., Pi\'etu, V., Simon, M.,
Tang, Y.-W., 2014, Nature, 514, 600

\bibitem[G\"unter \& Kley(2002)]{GUKLE} G\"unther, R. \& Kley, W.
2002, \aap, 387, 550

\bibitem[Guilloteau, Dutrey \& Simon(1999)]{gds99} Guilloteau, S.,
Dutrey, A., \& Simon, M., 1999, \aap, 348, 570 (GDS99)

\bibitem[Gullbring \etal(1998)]{GULL98} Gullbring, E., Hartmann, L.,
Briceno, C., Calvet, N., 1998, ApJ, 492, 323

\bibitem[Haisch, Lada \& Lada(2001)]{HLL01} Haisch, K. E., Lada. E. A., 
Lada, C., J., 2001, \apjl, 553, 153

\bibitem[Holman \& Wiegert(1999)]{HolWie99} Holman, M. J., Wiegert, P.
A., \aj, 117, 621

\bibitem[Hartigan, Edwards \& Ghandour(1995)]{HAR95} Hartigan, P.,
Edwards, S., Ghandour, L. 1995, ApJ, 452, 736

\bibitem[Itoh \etal(2002)]{itoh02} Itoh, Y., \etal, 2002, PASJ, 54,
963

\bibitem[K\"ohler(2011)]{koehler11} K\"ohler, R., 2011, \aap, 530, 126

\bibitem[Krist, Stapelfeldt \& Watson(2002)]{krist02} Krist, J. E., 
Stapelfeldt, K. R., Watson, A. M., 2002, \apj, 570, 785

\bibitem[Krist \etal(2005)]{krist05} Krist, J. E., Stapelfeldt, K. R.,
Golimowski, D. A., Ardila, D. R., Clampin, M., Martel, A. R., Ford, H.
C., Illingworth, G. D. \& Hartig, G. F., 2005, \aj, 130, 2778

\bibitem[Leinert \etal(1991)]{leinert91} Leinert, Ch., Haas, M.,
Mundt, R., Richichi, A., Zinnecker, H., 1991, \aap, 250, 407

\bibitem[Lissauer \& Stevenson(2007)]{PPV_LS} Lissauer, J. J.,
Stevenson, D. J., 2007, in Protostars and Planets V, ed. Reipurth, B.,
Jewitt, D., Keil, K., University of Arizona Press: Tucson, p. 591 

\bibitem[Lodato \& Rice(2004)]{LR04} Lodato, G., Rice, W. K. M., 2004,
\mnras, 351, 630

\bibitem[Lodato \& Clarke(2011)]{LC11} Lodato, G., Clarke, C. J.,
2011, \mnras, 413, 2735

\bibitem[Mathieu \etal(2000)]{PP4_MGJS} Mathieu, R. D., Ghez, A. M.,
Jensen, E. L. N., Simon, M., 2000, in Protostars and Planets IV, ed.
Mannings, V., Boss, A. P., \& Russell, S. S., University of
Arizona Press: Tucson. p. 703

\bibitem[McCabe, Duch\^ene \& Ghez(2002)]{MDG02} McCabe, C., Duch\^ene,
G., Ghez, A. M., 2002, \apj, 575, 974

\bibitem[Meru \& Bate(2010)]{mb10} Meru, F., Bate, M. R., 2010,
\mnras, 406, 2279

\bibitem[Meru \& Bate(2011a)]{mb11a} Meru, F., Bate, M. R., 2010a,
\mnras, 410, 559

\bibitem[Meru \& Bate(2011b)]{mb11b} Meru, F., Bate, M. R., 2011b,
\mnras, 411, L1

\bibitem[Meru \& Bate(2012)]{mb12} Meru, F., Bate, M. R., 2012,
\mnras, 427, 2022

\bibitem[Monin \etal(2007)]{MONAL} Monin, J. L., Clarke, C. J., Prato,
L., McCabe, C.,  2007, in Protostar and Planets V, ed. Reipurth, B.,
Jewitt,  D., Keil, K., University of Arizona Press: Tucson

\bibitem[Murray(1996)]{Murray96} Murray, J., 1996, \mnras, 279, 402

\bibitem[Nelson \etal(1998)]{DynI} Nelson, A.F., Benz, W., Adams, F.C.,
Arnett, W.D., 1998, \apj, 502, 342 

\bibitem[Nelson \etal(2000)]{DynII} Nelson, A. F., Benz, W.,
Ruzmaikina, T. V., 2000, \apj, 529, 357 (\dyntwo)

\bibitem[Nelson(2000)]{N00} Nelson, A. F., 2000, \apjl, 537, L65-68
(N00)

\bibitem[Nelson(2006)]{N06} Nelson, A. F., 2006, \mnras, 373, 1039
(N06)

\bibitem[Nelson, Wetzstein \& Naab(2009)]{vineII} Nelson, A. F.,
Wetzstein, M., Naab, T.,  2009, \apjs, 184, 326

\bibitem[Neuhauser \etal(2007)]{NEU07} Neuhauser, R., Mugrauer, M.,
Fukagawa, M., Torres, G., Schmidt, T., 2007, A\&A 462, 777

\bibitem[Ngo \etal(2015)]{Ngo15} Ngo, H., \etal, ESS meeting \#3,
\#403.04, November 2015, BAAS, 47, \#6

\bibitem[Pascucci \etal(2008)]{Petal08} Pascucci, I., Apai, D.,
Hardegree-Ullman, E. E., Kim, J. S., Meyer, M. R., 2008, \apj, 673, 477

\bibitem[Pi\'etu \etal(2011)]{pietu11} Pi\'etu, V., Gueth, F.,
Hily-Blant, P., Schuster, K-F, Pety, J., 2011, \aap, 528, A81

\bibitem[Piskorz \etal(2015)]{Piskorz15} Piscorz, D., Knutson, H. A.,
Ngo, H., Muirhead, P. S., Batygin, K., Crep, J. R., Hinkley, S.,
Morton, T. D., 2015, \apj, 814, 148

\bibitem[Pollack, McKay \& Christofferson(1985)]{pmc85} Pollack, J. B.,
McKay, C. P., Christofferson, B. M., 1985, Icarus, 64, 471

\bibitem[Prato \& Simon(1997)]{PRASI} Prato, L., and Simon, M.,
1997, ApJ 474, 455

\bibitem[Press \etal(1992)]{NumRec}  Press, W. H., Teukolsky, S. A.,
Vetterling, W. T., Flannery, B. P., 1992 Numerical Recipes, Cambridge
University Press, Cambridge

\bibitem[Pringle(1991)]{pringle91} Pringle, J. E., 1991, \mnras, 248,
754

\bibitem[Queloz \etal(2000)]{QUE00} Queloz D., Mayor M., Weber L.,
Blecha A., Burnet M., Confino B., Naef D., Pepe F., Santos N., Udry
S., 2000, \aap, 354, 99

\bibitem[Roddier \etal(1996)]{roddier96} Roddier, C., Roddier, F.,
Northcott, M. J., Graves, J. E. and Jim, K., 1996, \apj, 463, 326

\bibitem[Rodriquez \etal(1998)]{L1551_Rod} Rodriguez, L. F.,
D'Alessio, P., Willner, D. J., Ho, P. T. P., Torrelles, J. M., Curiel,
S., G\'omez, Y., Lizano, S., Pedlar, A., Cant\'o J., Raga, A. C.,
1998, Nature, 395, 355

\bibitem[Semenov \etal(2003)]{semenov03} Semonov, D., Henning, Th.,
Helling, Ch., Ilgner, M., Sedlmayr, E., 2003, \aap, 410, 611

\bibitem[Shakura \& Sunyaev(1973)]{SS73} Shakura, N. J. \& Sunyaev, R. A.,
1973, \aap, 24, 337

\bibitem[Shu \etal(1987)]{SAL87} Shu, F. H., Adams, F. C. \& Lizano, S.,
1987, \araa, 25, 23

\bibitem[Silber \etal(2000)]{silber00} Silber, J., Gledhill, T.,
Duch\^ene, G., M\'enard, F., 2000, \apjl, 536, 89

\bibitem[Simon \& Guilloteau(1992)]{sg92} Simon, M., Guilloteau, S.,
1992, \apj, 397, L47

\bibitem[Stapelfeldt \etal(1998)]{HKtau-c} Stapelfeldt, K. R., Krist,
J. E., M\`enard, F., Bouvier, J., Padgett, D. L., Burrows. C. J.,
1998, \apjl, 502, L65

\bibitem[Tamazian \etal(2002)]{tamazian02} Tamazian, V. S., Docobo,
J. A., White, R. J., Woitas, J., 2002, \apj, 578, 925

\bibitem[Tang \etal(2014)]{tang14} Tang, Y.-W., Dutrey, A.,
Guilloteau, S., Pietu, V., Di~Folco, E., Beck, T., Ho, P. T. P,
Boehler, Y., Gueth, F., Bary, J., Simon, S., 2014, \apj, 793, 10

\bibitem[Wetzstein \etal (2009)]{vineI} Wetzstein, M., Nelson, A.
F., Naab, T., Burkert, A., 2009, \apjs, 184, 298

\bibitem[White \etal(1999)]{Whiteetal99} White, R. J., Ghez, A. M.,
Reid, I. N., Schultz, G., 1999, \apj, 520, 811

\bibitem[Young, Baird \& Clarke(2015)]{YBC15} Young, M. D., Baird, J.
T., Clarke, C. J., 2015, \mnras, 447, 2907

\bibitem[Young \& Clarke(2015)]{YC15} Young, M. D., Clarke, C. J.,
2015, \mnras, 452, 3085

\bibitem[Zucker \etal(2004)]{ZUCK04} 
Zucker S., Mazeh T., Santos N. C., Udry S., Mayor M., 2004,
A\&A 426, 695


\end{thebibliography}
\end{document}